\documentclass[twocolumn,showpacs,preprintnumbers,amsmath,amssymb,superscriptaddress,nofootinbib,english]{revtex4-1}
\pdfoutput=1
\usepackage{times,amsmath,amsfonts,amssymb,epstopdf}
\usepackage{graphicx}
\usepackage{dcolumn}
\usepackage{bm}
\usepackage{epsfig}
\usepackage{graphicx}
\usepackage{hyperref}
\usepackage[usenames]{color}
\usepackage{url}
\usepackage[normalem]{ulem}
\usepackage[T1]{fontenc}

\def\be{\begin{equation}}
\def\ee{\end{equation}}
\def\ben{\begin{eqnarray}}
\def\een{\end{eqnarray}}
\def\ba{\begin{array}}
\def\ea{\end{array}}

\newcommand{\bq}{\begin{eqnarray}}
\newcommand{\eq}{\end{eqnarray}}
\newcommand{\bes}{\begin{subequations}}
\newcommand{\ees}{\end{subequations}}

\begin{document}
\newcommand{\half}{{\textstyle\frac{1}{2}}}
\allowdisplaybreaks[3]
\def\triangledown{\nabla}
\def\grad3{\hat{\nabla}}
\def\a{\alpha}
\def\b{\beta}
\def\g{\gamma}\def\G{\Gamma}
\def\d{\delta}\def\D{\Delta}
\def\ep{\epsilon}
\def\et{\eta}
\def\z{\zeta}
\def\t{\theta}\def\T{\Theta}
\def\l{\lambda}\def\L{\Lambda}
\def\m{\mu}
\def\f{\phi}\def\F{\Phi}
\def\n{\nu}
\def\p{\psi}\def\P{\Psi}
\def\r{\rho}
\def\s{\sigma}\def\S{\Sigma}
\def\ta{\tau}
\def\x{\chi}
\def\o{\omega}\def\O{\Omega}
\def\k{\kappa}
\def\pa {\partial}
\def\ov{\over}
\def\br{\\}
\def\ud{\underline}

\newcommand\lsim{\mathrel{\rlap{\lower4pt\hbox{\hskip1pt$\sim$}}
    \raise1pt\hbox{$<$}}}
\newcommand\gsim{\mathrel{\rlap{\lower4pt\hbox{\hskip1pt$\sim$}}
    \raise1pt\hbox{$>$}}}
\newcommand\esim{\mathrel{\rlap{\raise2pt\hbox{\hskip0pt$\sim$}}
    \lower1pt\hbox{$-$}}}
\newcommand{\dpar}[2]{\frac{\partial #1}{\partial #2}}
\newcommand{\sdp}[2]{\frac{\partial ^2 #1}{\partial #2 ^2}}
\newcommand{\dtot}[2]{\frac{d #1}{d #2}}
\newcommand{\sdt}[2]{\frac{d ^2 #1}{d #2 ^2}}    

\title{Spherical collapse in Galileon gravity: fifth force solutions, halo mass function and halo bias}

\author{Alexandre Barreira}
\email[Electronic address: ]{a.m.r.barreira@durham.ac.uk}
\affiliation{Institute for Computational Cosmology, Department of Physics, Durham University, Durham DH1 3LE, U.K.}
\affiliation{Institute for Particle Physics Phenomenology, Department of Physics, Durham University, Durham DH1 3LE, U.K.}

\author{Baojiu Li}
\affiliation{Institute for Computational Cosmology, Department of Physics, Durham University, Durham DH1 3LE, U.K.}

\author{Carlton M. Baugh}
\affiliation{Institute for Computational Cosmology, Department of Physics, Durham University, Durham DH1 3LE, U.K.}

\author{Silvia Pascoli}
\affiliation{Institute for Particle Physics Phenomenology, Department of Physics, Durham University, Durham DH1 3LE, U.K.}

\begin{abstract}

We study spherical collapse in the Quartic and Quintic Covariant Galileon gravity models within the framework of the excursion set formalism. We derive the nonlinear spherically symmetric equations in the quasi-static and weak-field limits, focusing on model parameters that fit current CMB, SNIa and BAO data. We demonstrate that the equations of the Quintic model do not admit physical solutions of the fifth force in high density regions, which prevents the study of structure formation in this model. For the Quartic model, we show that the effective gravitational strength deviates from the standard value at late times ($z \lesssim 1$), becoming larger if the density is low, but smaller if the density is high. This shows that the Vainshtein mechanism at high densities is not enough to screen all of the modifications of gravity. This makes halos that collapse at $z \lesssim 1$ feel an overall weaker gravity, which suppresses halo formation.  However, the matter density in the Quartic model is higher than in standard $\Lambda$CDM, which boosts structure formation and dominates over the effect of the weaker gravity. In the Quartic model there is a significant overabundance of high-mass halos relative to $\Lambda$CDM. Dark matter halos are also less biased than in $\Lambda$CDM, with the difference increasing appreciably with halo mass. However, our results suggest that the bias may not be small enough to fully reconcile the predicted matter power spectrum with LRG clustering data.

\end{abstract} 
%\pacs{98.80.Cq}
\maketitle

\section{Introduction}

The so-called $\Lambda$CDM model has been extremely successful over the past 15 years in accounting for most of the accumulated cosmological data \cite{Hinshaw:2012fq, Ade:2013zuv, Guy:2010bc, Suzuki:2011hu, Efstathiou:2001cw, Sánchez11122009, Percival:2009xn, Beutler:2011hx, Reid:2012sw, Reid:2009xm, Anderson:2012sa, Sanchez:2012sg}. In this model, general relativity (GR) describes gravity, most of the matter is in the form of cold-dark-matter, and a cosmological constant $\Lambda$ acounts for the missing "dark energy" that is responsible for accelerating the expansion of the universe. However, despite the overall observational success, the fact that the value of $\Lambda$ required to explain the acceleration is many orders of magnitude below any standard quantum field theory predictions is a major embarassment. This problem has motivated the proposal of a number of alternatives, one of which is the modification of gravity (see \cite{Clifton:2011jh} for an extensive review). The idea behind these models is that GR breaks down on cosmological scales, in such a way that it accelerates the expansion of the universe without requiring $\Lambda$.

The Galileon gravity model \cite{PhysRevD.79.064036, PhysRevD.79.084003, Deffayet:2009mn} is an example of one such model, and it has been receiving growing interest lately \cite{Barreira:2012kk, Barreira:2013jma, Gannouji:2010au, PhysRevD.80.024037, DeFelice:2010pv, Nesseris:2010pc, Appleby:2011aa, PhysRevD.82.103015, Neveu:2013mfa, Appleby:2012ba, Okada:2012mn, Bartolo:2013ws}. In this model, the modifications of gravity are determined by a scalar field $\varphi$ (dubbed the Galileon field) whose Lagrangian is invariant under the Galilean shift transformation, $\partial_\mu\varphi \rightarrow \partial_\mu\varphi + b_\mu$, where $b_\mu$ is a constant four-vector. In \cite{PhysRevD.79.064036}, it was shown that in four-dimensional flat space-time there are only five Lagrangian densities that are Galilean invariant and that lead to second order field equations of motion. These Lagrangian densities are named after the power with which $\varphi$ appears (see Eq.~(\ref{eq:action}) below): besides the linear ($\mathcal{L}_1$) and quadratic ($\mathcal{L}_2$) terms, which describe a model like quintessence with a linear scalar potential, there are also the cubic, quartic and quintic terms ($\mathcal{L}_3$, $\mathcal{L}_4$ and $\mathcal{L}_5$, respectively) that are responsible for the modifications of gravity. The model was subsequently generalized to curved space-times in \cite{Deffayet:2009mn}, where it was concluded that explicit couplings between Galileon derivative terms and curvature tensors are needed in the quartic and quintic Lagrangians to retain the equations of motion from becoming higher than second-order. The second order nature of the equations is crucial to avoid the propagation of Ostrogradsky ghosts \cite{Woodard:2006nt}, and makes the Galileon model a subset of the more general Horndeski theory \cite{Horndeski:1974wa}. In the Galileon model, the spatial gradient of the scalar field contributes to the modifications of gravity, which are often referred to as a fifth force. These spatial gradients have to be suppressed in regions near massive objects, if the model is to survive the stringent Solar System gravity tests that constrain the magnitude of a long-distance fifth force to be very small \cite{1990PhRvL..64..123D, 1999PhRvL..83.3585B, Will:2005va, Kapner:2006si}. In the Galileon model, this is achieved by a mechanism known as the Vainshtein effect \cite{Vainshtein1972393, Babichev:2013usa, Koyama:2013paa}, which relies on the presence of nonlinear couplings of the scalar field derivatives that appear in $\mathcal{L}_3$, $\mathcal{L}_4$ and $\mathcal{L}_5$. The general picture is that far away from massive bodies, where the density is low, the nonlinearities are not important and the Galileon field satisfies a linear Poisson-like equation. On the other hand, near massive bodies, where the density is high, the nonlinear terms become important and effectively suppress the spatial variations of the scalar field.

In previous work \cite{Barreira:2012kk, Barreira:2013jma}, we have modified the {\tt CAMB} \cite{camb_notes} and {\tt CosmoMC} \cite{Lewis:2002ah} codes to include the cosmology of the Covariant Galileon model \cite{PhysRevD.79.084003}. We have used these extended codes to place observational contraints on the cosmological parameter space, using data from the WMAP 9-yr results for the temperature power spectrum of the cosmic microwave background (CMB) \cite{Hinshaw:2012fq}, type Ia supernovae (SNIa) from the SNLS 3-yr sample \cite{Guy:2010bc} and baryonic acoustic oscillations (BAO) measurements from the 6df \cite{Beutler:2011hx}, SDSS DR7 \cite{Reid:2012sw} and BOSS \cite{Percival:2009xn} galaxy surveys. Our work, which was kept at the linear level in perturbation theory, showed that the Galileon model can fit the CMB data better than $\Lambda$CDM, mainly due to the possibility of having less power than $\Lambda$CDM on large angular scales, which is preferred by the WMAP 9-yr data and also by the recent Planck results \cite{Ade:2013zuv}. However, we have also pointed out a tension in the ability of the model to explain the observed large-scale structure in the galaxy distribution. The interpretation of this tension, however, is subject to knowing exactly on which scales linear perturbation theory hold, and also on how halo and galaxy bias apply in modified gravity theories compared to $\Lambda$CDM. The investigation of these two uncertanties requires one to go beyond linear theory. This can be particularly challenging in Galileon gravity because of the highly nonlinear nature of its equations.

In a recent paper \cite{Barreira:2013eea}, we took a first step towards understanding the nonlinear formation of structure in Galileon gravity by performing the first N-body simulations of the Cubic Galileon model using the {\tt ECOSMOG} code \cite{Li:2011vk, Li:2013nua}. In this model, the higher order nonlinear terms that arise from $\mathcal{L}_4$ and $\mathcal{L}_5$ are absent, which makes the equations simpler and allows them to be more easily solved by N-body codes. For this model, we have found that for scales $k \lesssim 0.1h/{\rm Mpc}$, the nonlinearities do not affect the linear theory prediction, which can therefore be used to further constrain the model. However, the uncertainties relating halo and galaxy bias remain to be addressed. Moreover, as we will see below, the Cubic Galileon model, contrary to the more general (Quartic and Quintic) models, has the more serious problem of not being able to provide a reasonable fit to the low multipoles of the CMB temperature power spectrum \cite{Barreira:2013eea}. The next simplest Galileon model one can study is the Quartic Galileon model, in which the higher order nonlinear terms that arise from $\mathcal{L}_5$ are absent. The numerical algorithm to simulate the equations of this model has been presented recently in \cite{Li:2013tda}. This algorithm was implemented in the {\tt ECOSMOG} code \cite{Li:2011vk} to obtain the first nonlinear matter and velocity power spectra predictions for the Quartic Galileon model. These first results confirm that the extra nonlinearity plays a very important role in determining the modifications to gravity. For instance, contrary to the case of the Cubic Galileon model, in the Quartic model, the Vainshtein mechanism can have a measurable impact on scales $k \lesssim 0.1h/{\rm Mpc}$. At least to our knowledge, N-body simulations of the Quintic Galileon model (which is the most general, but the most nonlinear as well) have never been performed.

In this paper, our goal is to study the spherical collapse of matter overdensities in the Quartic and Quintic Galileon models, and use the excursion set formalism \cite{1991ApJ...379..440B} to predict the halo mass function and halo bias. By doing this, one is not expected to reach the same level of accuracy as the N-body simulations. However, the theoretical framework of the excursion set formalism provides a neat and easy way to capture the main physical features of the models, which helps to build intuition about their phenomenology. Part of this paper provides, therefore, a complementary analysis to the simulation results of the Quartic Galileon model presented in \cite{Li:2013tda}. It is not our goal to draw precise quantitative conclusions from our results. Instead, we are more interested in discussing the physics of the model in a more qualitative point of view.

The outline of this paper is as follows. In Sec.~\ref{sec:model}, we present the Galileon gravity model, the background equations, and the relevant nonlinear equations derived assuming spherical symmetry in the quasi-static and weak-field limits. We also review the cosmology of the model parameters we will focus our study on. In Sec.~\ref{sec:solutions}, we look at the properties of the fifth force in these models by discussing the existence of physical solutions and its time and density dependence. In particular, we will demonstrate that the Quintic Galileon models that are compatible with current CMB data do not admit physical solutions for the fifth force in high density regions. In Sec.~\ref{sec:excursion}, we outline the main ideas of the excursion set formalism and present the relevant equations for the spherical collapse of the overdensities. We present our main results for the mass function and halo bias for the Quartic Galileon model in Sec.~\ref{sec:results}. We conclude in Sec.~\ref{sec:conclusion}

Throughout this paper we assume the metric convention $(+,-,-,-)$ and work in units in which the speed of light $c = 1$. Greek indices run over $0,1,2,3$ and we use $8\pi G=\kappa=M^{-2}_{\rm Pl}$ interchangeably, where $G$ is Newton's constant and $M_{\rm Pl}$ is the reduced Planck mass.

\section{The Galileon model}\label{sec:model}

In this section, we present the Galileon gravity model and the equations that we use to calculate the fifth force. We shall also summarize the model predictions for the cosmic expansion history, and the CMB temperature and linear matter power spectra.

\subsection{Action and field equations}

The action of the minimally coupled covariant Galileon model is given by

\bq\label{eq:action}
&& S = \int {\rm d}^4x\sqrt{-g} \left[ \frac{R}{16\pi G} - \frac{1}{2}\sum_{i=1}^5c_i\mathcal{L}_i - \mathcal{L}_m\right],
\eq
where $g$ is the determinant of the metric $g_{\mu\nu}$, $R$ is the Ricci scalar and $\mathcal{L}_m$ represents the matter content, which is minimally coupled to the metric and Galileon fields.
The model parameters $c_{1-5}$ are dimensionless constants and the five terms in the Galileon Lagrangian density, fixed by the Galilean invariance in flat spacetime, $\partial_\mu\varphi \rightarrow \partial_\mu\varphi + b_\mu$, are given by
\bq\label{L's}
\mathcal{L}_1 &=& M^3\varphi, \nonumber \\
\mathcal{L}_2 &=& \nabla_\lambda\varphi\nabla^\lambda\varphi,  \nonumber \\
\mathcal{L}_3 &=& \frac{2}{M^3}\Box\varphi\nabla_\lambda\varphi\nabla^\lambda\varphi, \nonumber \\
\mathcal{L}_4 &=& \frac{1}{M^6}\nabla_\lambda\varphi\nabla^\lambda\varphi\Big[ 2(\Box\varphi)^2 - 2(\nabla_\mu\nabla_\nu\varphi)(\nabla^\mu\nabla^\nu\varphi) \nonumber \\
&& -R\nabla_\mu\varphi\nabla^\mu\varphi/2\Big], \nonumber \\
\mathcal{L}_5 &=&  \frac{1}{M^9}\nabla_\lambda\varphi\nabla^\lambda\varphi\Big[ (\Box\varphi)^3 - 3(\Box\varphi)(\nabla_\mu\nabla_\nu\varphi)(\nabla^\mu\nabla^\nu\varphi) \nonumber \\
&& + 2(\nabla_\mu\nabla^\nu\varphi)(\nabla_\nu\nabla^\rho\varphi)(\nabla_\rho\nabla^\mu\varphi) \nonumber \\
&& -6 (\nabla_\mu\varphi)(\nabla^\mu\nabla^\nu\varphi)(\nabla^\rho\varphi)G_{\nu\rho}\Big],
\eq
in which $M^3\equiv M_{\rm Pl}H_0^2$, where $H_0$ is the present-day Hubble expansion rate. In this model, the modifications to gravity are driven by derivative interactions of $g_{\mu\nu}$ and $\varphi$ (a proccess known as {\it kinetic gravity braiding} \cite{Deffayet:2010qz, Babichev:2012re}). These interactions arise through the coupling of covariant derivatives, and through the couplings to the Ricci scalar $R$ and the Einstein tensor $G_{\mu\nu}$ in $\mathcal{L}_4$ and $\mathcal{L}_5$. The latter two are necessary to prevent the equations of motion from having higher than second-order derivatives of the metric and the Galileon field in curved spacetimes, such as the one described by the Friedman-Robertson-Walker (FRW) metric \cite{PhysRevD.79.084003}. Such terms, however, break the Galilean shift symmetry. We will discuss later the implications of these couplings to curvature. 

We will consider the case in which the acceleration is due only to kinetic energy of the Galileon field and therefore we will set $c_1 = 0$. In this case, the action contains only derivatives of the scalar field, and as a result, the exact value of $\varphi$ is irrelevant for the physics of the model. The modified Einstein equations and the Galileon equation of motion are obtained by varying the action of Eq.~(\ref{eq:action}), with respect to $g_{\mu\nu}$ and $\varphi$, respectively. We do not show them in this paper, since they are lengthy and have been presented elsewhere (\cite{PhysRevD.79.084003, Appleby:2011aa, Barreira:2012kk}).

\subsection{Background equations}

We will work with the perturbed FRW metric in the Newtonian gauge

\bq\label{metric}
{\rm d}s^2 = \left(1 + 2\Psi\right){\rm d}t^2 - a(t)^2\left(1 - 2\Phi\right)\gamma_{ij}{\rm d}x^i{\rm d}x^j,
\eq
where $a$ is the cosmic scale factor and $\gamma_{ij} = \rm{diag}\left[1, 1, 1\right]$ (where $i, j  \in \{1,2,3\}$) is the spatial sector of the metric, which is taken here to be flat. The fields, $\varphi$, $\Psi$ and $\Phi$, are assumed to be functions of time and space.  In the equations below $\varphi = \bar{\varphi}(t) + \delta\varphi(t, \vec{x})$, where $\delta\varphi$ is the field perturbation and an overbar indicates background averaged quantities. We will always use $\varphi$ to denote the scalar field, and the context should determine whether we refer to $\bar{\varphi}$ or $\delta\varphi$.

The background Friedmann and Galileon field equations are, respectively, given by

\bq\label{eq:friedmann}
3H^2  &=& 8\pi G\left(\bar{\rho}_m + \bar{\rho}_r\right) + \frac{1}{2}c_2\dot{\varphi}^2 + 6\frac{c_3}{H_0^2}H\dot{\varphi}^3 \nonumber \\
&& + \frac{45}{2}\frac{c_4}{H_0^4}H^2\dot{\varphi}^4 + 21\frac{c_5}{H_0^6}H^3\dot{\varphi}^5,\ \  
\eq
and 
\bq\label{eq:eom}
0 &=& c_2(\ddot{\varphi}+3H\dot{\varphi}) + \frac{c_3}{H_0^2}\left(12H\dot{\varphi}\ddot{\varphi}+6\dot{H}\dot{\varphi}^2+18H^2\dot{\varphi}^2\right)\nonumber\\
&& + \frac{c_4}{H_0^4}\left(54H^2\dot{\varphi}^2\ddot{\varphi}+36\dot{H}H\dot{\varphi}^3+54H^3\dot{\varphi}^3\right) \nonumber \\
&& + \frac{c_5}{H_0^6}\left( 45\dot{\varphi}^4H^4   +60\ddot{\varphi}\dot{\varphi}^3H^3  +45\dot{\varphi}^4\dot{H}H^2 \right),
\eq
in which $\bar{\rho}_m$ and $\bar{\rho}_r$ denote the background densities for matter (baryonic and cold-dark-matter) and radiation, respectively, $H=\dot{a}/a$ is the Hubble expansion rate and an overdot denotes the physical time derivative. In the above equations, as well as in the rest of the paper, the Galileon field $\varphi$ is given in units of $M_{\rm{Pl}}$, i.e., we have applied the transformation $\varphi/M_{\rm{Pl}} \rightarrow \varphi$.

For completeness, the background energy density and pressure of the Galileon scalar field are given by

\bq\label{eq:density-background}
\kappa\bar{\rho}_\varphi &=& \frac{1}{2}c_2\dot{\varphi}^2 + 6\frac{c_3}{H_0^2}\dot{\varphi}^3H  +\frac{45}{2}\frac{c_4}{H_0^4}\dot{\varphi}^4H^2     \nonumber \\
&&+ 21\frac{c_5}{H_0^6}\dot{\varphi}^5H^3, \\
\kappa\bar{p}_\varphi &=& \frac{1}{2}c_2\dot{\varphi}^2 + -2\frac{c_3}{H_0^2}\ddot{\varphi}\dot{\varphi}^2 \nonumber \\
&& + 3\frac{c_4}{H_0^4}\left[ -4\ddot{\varphi}\dot{\varphi}^3H - \dot{\varphi}^4\dot{H} - \frac{3}{2}\dot{\varphi}^4H^2\right] \nonumber \\
&& + \frac{c_5}{H_0^6}\left[-15\ddot{\varphi}\dot{\varphi}^4H^2 - 6\dot{\varphi}^5\dot{H}H - 6\dot{\varphi}^5H^3\right].
\eq

\subsubsection{Background tracker solution}

In general, Eqs.~(\ref{eq:friedmann}) and (\ref{eq:eom}) have to be solved numerically to determine the expansion rate and the background evolution of the Galileon field. However, one can make use of an attractor tracker solution of the background equations to obtain analytical expressions for the background quantities to make the perturbed equations easier to handle. In the Galileon model, the tracker solution is described by \cite{DeFelice:2010pv}

\bq\label{eq:tracker}
H\dot{\varphi} = {\rm constant} \equiv \xi H_0^2,
\eq
where $\xi$ is a dimensionless constant. In \cite{Barreira:2012kk} and \cite{Barreira:2013jma} it was shown that the models that follow the tracker solution are those that best fit data from SNIa, BAO and CMB.

Multiplying both sides of Eq.~(\ref{eq:friedmann}) by $H^2$, using Eq.~(\ref{eq:tracker}) to eliminate $\dot{\varphi}$ and dividing the resulting equation by $H^4_0$, we obtain

\bq\label{eq:friedmann2}
E^4 &=& \left(\Omega_{m0}a^{-3} + \Omega_{r0}a^{-4}\right)E^2 \nonumber \\
&+& \frac{1}{6}c_2\xi^2 + 2c_3\xi^3 + \frac{15}{2}c_4\xi^4 + 7c_5\xi^5,
\eq
in which $E\equiv H/H_0$, $\Omega_{m0} = \bar{\rho}_{m0}/\rho_{\rm{c0}}$ and  $\Omega_{r0} = \bar{\rho}_{r0}/\rho_{\rm{c0}}$, where $\rho_{\rm{c0}} = 3H_0^2/(8\pi G)$ is the critical energy density today. At the present day ($a = 1$ and $E = 1$),  Eq.~(\ref{eq:friedmann2}) gives

\bq\label{eq:xi}
\frac{1}{6}c_2\xi^2 + 2c_3\xi^3 + \frac{15}{2}c_4\xi^4 + 7c_5\xi^5 &=& 1-\Omega_{m0} - \Omega_{r0}, \nonumber \\
\eq
which can be used to determine the value of $\xi$ given $c_2, c_3, c_4, c_5$ and $\Omega_{m0}$. Combining Eqs.~(\ref{eq:xi}) and (\ref{eq:friedmann2}), we get
\bq
E^4 &=& \left(\Omega_{m0}a^{-3} + \Omega_{r0}a^{-4}\right)E^2+1-\Omega_{m0} - \Omega_{r0},
\eq
which gives the Hubble expansion rate at $a$ analytically as
\bq\label{eq:tracker_H}
&&E(a)^2 = \frac{1}{2}\left[\left(\Omega_{m0}a^{-3} + \Omega_{r0}a^{-4}\right) \right. \nonumber \\
&&  \left. + \sqrt{(\Omega_{m0}a^{-3} + \Omega_{r0}a^{-4})^2 + 4(1-\Omega_{m0} - \Omega_{r0})}\right].
\eq
Finally, using Eq.~(\ref{eq:tracker}) we have

\bq\label{eq:tracker_galileon}
\dot{\varphi} &=& \xi H_0/E \ \ \ \Longrightarrow \ \ \ \varphi' = \xi / E^2,
\eq
where $' \equiv {\rm d}/{\rm d} N$, with $N = {\rm ln}(a)$.

\subsection{Spherically symmetric nonlinear equations}

We assume that $\delta\varphi$, $\Phi$ and $\Psi$ are spherically symmetric, under which case the nonlinear field equations simplify considerably. To make the problem tractable we shall also employ two other simplifying assumptions. The first one is the so-called quasi-static approximation which corresponds to the limit where the time derivatives of the perturbed quantities are negligible compared to their spatial derivatives. For instance, $ \partial_t\partial_t\Phi \ll \partial_r\partial_r\Phi$ or $\partial_t\partial_r\varphi \ll \partial_r\partial_r\varphi$ \footnote{Note that $\partial_r\varphi = \partial_r\delta\varphi$ is a perturbed quantity.}. In \cite{Barreira:2012kk, DeFelice:2010as}, it was shown that such an approximation typically works well in the Galileon model on length scales smaller than $k \sim 0.01 h/\rm{Mpc}$. The second simplifying assumption amounts to neglecting the terms that are suppressed by the scalar potentials, $\Phi$ and $\Psi$, and their first spatial derivatives, $\partial_i\Phi$ and $\partial_i\Psi$. This is known as the weak-field approximation where, for instance, $\left(1-2\Phi\right)\partial^i\partial_i\varphi \sim \partial^i\partial_i\varphi$ or $\partial_i\Phi\partial^i\Phi \ll \partial_i\partial^i\Phi$. This is plausible since these fields are typically very small ($\lesssim 10^{-4}$) on nonlinear scales. We will discuss the implications of these assumptions later in the paper.

Under the above approximations, the perturbed Poisson ($\delta G^0_0 = \kappa \delta T^0_0$), slip ( $\delta G^r_r = \kappa \delta T^r_r$) and Galileon field equations of motion follow, respectively,

\begin{widetext}
\bq
\label{eq:poisson0}
2\frac{1}{r^2}\left(r^2\Phi,_r\right),_r &=& -2\frac{c_3}{H_0^2}\dot{\varphi}^2\frac{1}{r^2}\left(r^2\varphi,_r\right),_r + \frac{c_4}{H_0^4}\left[6\frac{\dot{\varphi}^2}{a^2}\frac{1}{r^2}\left(r(\varphi,_r)^2\right),_r - 12H\dot{\varphi}^3\frac{1}{r^2}\left(r^2\varphi,_r\right),_r + 3\dot{\varphi}^4\frac{1}{r^2}\left(r^2\Phi,_r\right),_r\right] \nonumber \\
&& + \frac{c_5}{H_0^6}\left[-4\frac{\dot{\varphi}^2}{a^4}\frac{1}{r^2}\left((\varphi,_r)^3\right),_r + 12\frac{H\dot{\varphi}^3}{a^2}\frac{1}{r^2}\left(r(\varphi,_r)^2\right),_r - 15H^2\dot{\varphi}^4\frac{1}{r^2}\left(r^2\varphi,_r\right),_r \right. \nonumber \\ 
&& \ \ \ \ \ \ \ \ \ \ \left. + 6H\dot{\varphi}^5\frac{1}{r^2}\left(r^2\Phi,_r\right),_r - 6\frac{\dot{\varphi}^4}{a^2}\frac{1}{r^2}\left(r\varphi,_r\Phi,_r\right),_r\right]  + 8\pi G\bar{\rho}_m\delta a^2,\\
\label{eq:slip0}
 \frac{2}{r}\left(\Phi,_r - \Psi,_r\right) &=& \frac{c_4}{H_0^4}\left[\left(-4H\dot{\varphi}^3 - 12\ddot{\varphi}\dot{\varphi}^2\right)\frac{\varphi,_r}{r} - \dot{\varphi}^4\frac{\Phi,_r}{r} - 3\dot{\varphi}^4\frac{\Psi,_r}{r} + 2\frac{\dot{\varphi}^2}{a^2}\left(\frac{\varphi,_r}{r}\right)^2\right] \nonumber \\
&+& \frac{c_5}{H_0^6} \left[12\frac{\ddot{\varphi}\dot{\varphi}^2}{a^2}\left(\frac{\varphi,_r}{r}\right)^2 + 6\ddot{\varphi}\dot{\varphi}^4\frac{\Phi,_r}{r} - 6\left(\dot{H}\dot{\varphi}^4 + H^2\dot{\varphi}^4 + 4H\ddot{\varphi}\dot{\varphi}^3\right)\frac{\varphi,_r}{r} \right. \nonumber \\
&& \ \ \ \ \ \ \ \ \ \ - \left. 6H\dot{\varphi}^5\frac{\Psi,_r}{r} + 6\frac{\dot{\varphi}^4}{a^2}\frac{\varphi,_r}{r}\frac{\Psi,_r}{r}\right],\\
\label{eq:eom_sph0}
0 &=& -c_2\frac{1}{r^2}\left(r^2\varphi,_r\right),_r + \frac{c_3}{H_0^2}\left[\frac{4}{a^2}\frac{1}{r^2}\left(r(\varphi,_r)^2\right),_r - 4(\ddot{\varphi} + 2H\dot{\varphi})\frac{1}{r^2}\left(r^2\varphi,_r\right),_r - 2\dot{\varphi}^2\frac{1}{r^2}\left(r^2\Psi,_r\right),_r\right] \nonumber \\
&& +\frac{c_4}{H_0^4}\left[-\frac{4}{a^4}\frac{1}{r^2}\left((\varphi,_r)^3\right),_r + 12\frac{\ddot{\varphi} + H\dot{\varphi}}{a^2}\frac{1}{r^2}\left(r(\varphi,_r)^2\right),_r - \left(12\dot{H}\dot{\varphi}^2 + 24\ddot{\varphi}\dot{\varphi}H + 26H^2\dot{\varphi}^2\right)\frac{1}{r^2}\left(r^2\varphi,_r\right),_r \right. \nonumber \\
&& \left. + (12\ddot{\varphi}\dot{\varphi}^2 + 4H\dot{\varphi}^3)\frac{1}{r^2}\left(r^2\Phi,_r\right),_r - 12H\dot{\varphi}^3\frac{1}{r^2}\left(r^2\Psi,_r\right),_r - 4\frac{\dot{\varphi}^2}{a^2}\frac{1}{r^2}\left(r\varphi,_r\Phi,_r\right),_r + 12\frac{\dot{\varphi}^2}{a^2}\frac{1}{r^2}\left(r\varphi,_r\Psi,_r\right),_r\right] \nonumber \\
&& +\frac{c_5}{H_0^6}\left[-8\frac{\ddot{\varphi}}{a^4}\frac{1}{r^2}\left((\varphi,_r)^3\right),_r + 12\frac{\dot{H}\dot{\varphi}^2 + H^2\dot{\varphi}^2 + 2H\ddot{\varphi}\dot{\varphi}}{a^2}\frac{1}{r^2}\left(r(\varphi,_r)^2\right),_r - 12\frac{\dot{\varphi}^2}{a^4}\frac{1}{r^2}\left(\Psi,_r(\varphi,_r)^2\right),_r \right. \nonumber \\
&& \left. - 12\left(3H^2\ddot{\varphi}\dot{\varphi}^2 + 2\dot{H}H\dot{\varphi}^3 + 2H^2\dot{\varphi}^3\right)\frac{1}{r^2}\left(r^2\varphi,_r\right),_r - 24\frac{\ddot{\varphi}\dot{\varphi}^2}{a^2}\frac{1}{r^2}\left(r\varphi,_r\Phi,_r\right),_r + 24\frac{H\dot{\varphi}^3}{a^2}\frac{1}{r^2}\left(r\varphi,_r\Psi,_r\right),_r \right. \nonumber \\
&& \left. - 6\frac{\dot{\varphi}^4}{a^2}\frac{1}{r^2}\left(r\Psi,_r\Phi,_r\right),_r + 6\left(4H\ddot{\varphi}\dot{\varphi}^3 + \dot{H}\dot{\varphi}^4 + H^2\dot{\varphi}^4\right)\frac{1}{r^2}\left(r^2\Phi,_r\right),_r - 15H^2\dot{\varphi}^4\frac{1}{r^2}\left(r^2\Psi,_r\right),_r\right],
\eq
\end{widetext}
where $r$ is the comoving radial coordinate and $,_r \equiv {\rm d}/{\rm d}r$. We have checked that these equations (together with the remaining components of the Einstein equations, which we do not show for brevity) satisfy the independent conservation equations $\nabla_\nu \delta G^{\mu\nu} = \nabla_\nu \delta T^{\mu\nu} = 0$. In the last term in Eq.~(\ref{eq:poisson0}), $\delta = \rho_{m}/\bar{\rho}_m - 1$ is the matter density contrast of the spherical top-hat overdensity w.r.t. the cosmic mean density. In this paper, $\delta$ characterizes the density of the spherical halos throughout their entire evolution, and not only during the stages where it is small ($|\delta| \ll 1$).

Eqs.~(\ref{eq:poisson0}) and (\ref{eq:eom_sph0}) can be simplified by integrating over $\int{4\pi r^2dr}$. Doing so, and moving to the radial coordinate $\chi \equiv aH_0r$, we can write Eqs.~(\ref{eq:poisson0}), (\ref{eq:slip0}) and (\ref{eq:eom_sph0}) as

\begin{widetext}
\bq
\label{eq:poisson}\frac{\Phi,_{\chi}}{\chi} &=& \frac{\Omega_{m0}\delta a^{-3} + A_1\left(\varphi,_{\chi}/\chi\right) + A_2\left(\varphi_{\chi}/\chi\right)^2 + A_3\left(\varphi_{\chi}/\chi\right)^3}{A_4 + A_5\left(\varphi_{\chi}/\chi\right)}, \\
\label{eq:slip}\frac{\Psi,_{\chi}}{\chi} &=& \frac{B_0\left(\Phi,_{\chi}/\chi\right) + B_1\left(\varphi,_{\chi}/\chi\right) + B_2\left(\varphi,_{\chi}/\chi\right)^2}{B_3 - B_4\left(\varphi,_{\chi}/\chi\right)}, \\
\label{eq:eom_sph}0 &=& C_1\frac{\varphi,_{\chi}}{\chi} + C_2\left(\frac{\varphi,_{\chi}}{\chi}\right)^2 + C_3\left(\frac{\varphi,_{\chi}}{\chi}\right)^3  + C_4\frac{\Phi,_{\chi}}{\chi} + C_5\frac{\Psi,_{\chi}}{\chi}  + C_6\frac{\varphi,_{\chi}}{\chi}\frac{\Phi,_{\chi}}{\chi} + C_7\frac{\varphi,_{\chi}}{\chi}\frac{\Psi,_{\chi}}{\chi} + C_8\left(\frac{\varphi,_{\chi}}{\chi}\right)^2\frac{\Psi,_{\chi}}{\chi} \nonumber \\
&+& C_9\frac{\Phi,_{\chi}}{\chi}\frac{\Psi,_{\chi}}{\chi}.
\eq
\end{widetext}
The quantities $A_i$, $B_i$ and $C_i$ depend only on time and are given in the Appendix. One can use Eqs.~(\ref{eq:poisson}) and (\ref{eq:slip}) to eliminate $\Phi,_{\chi}$ and $\Psi,_{\chi}$ in Eq.~(\ref{eq:eom_sph}). The resulting equation is a sixth-order algebraic equation for $\varphi,_{\chi}/\chi$, which can be cast as

\begin{widetext}
\bq\label{eq:eom_sph_alg}
0 &=& \eta_{02}\delta^2 + \eta_{01}\delta + \left(\eta_{11}\delta + \eta_{10}\right)\left[\frac{\varphi,_{\chi}}{\chi}\right] + \left(\eta_{21}\delta + \eta_{20}\right)\left[\frac{\varphi,_{\chi}}{\chi}\right]^2 + \left(\eta_{31}\delta + \eta_{30}\right)\left[\frac{\varphi,_{\chi}}{\chi}\right]^3 + \eta_{40}\left[\frac{\varphi,_{\chi}}{\chi}\right]^4 + \eta_{50}\left[\frac{\varphi,_{\chi}}{\chi}\right]^5 \nonumber \\
& +& \eta_{60}\left[\frac{\varphi,_{\chi}}{\chi}\right]^6.
\eq
\end{widetext}
The coefficients $\eta_{ab}$ are given in terms of the functions $A_i$, $B_i$ and $C_i$ in Eqs.~(\ref{eq:poisson}), (\ref{eq:slip}) and (\ref{eq:eom_sph}). Their expression is very lengthy and for brevity we do not show them explicitly.

The strategy used to determine the total gravitational force is as follows. For every moment in time and for a given matter overdensity $\delta$ one has to solve the algebraic equation, Eq.~(\ref{eq:eom_sph_alg}), to determine the gradient of the Galileon field inside the overdensity. Note that in the case of a top-hat profile, this gradient will be proportional to the radial coordinate, just like in GR. Having obtained the solution for $\varphi,_{\chi}/\chi$, one can then plug it into Eqs.~(\ref{eq:poisson}) and (\ref{eq:slip}) to determine the total gravitational force (GR + fifth force), which is given by $\Psi,_{\chi}$. 

In the following, it will be convenient to measure the impact of the fifth force in terms of an effective gravitational constant $G_{\rm{eff}}$. The latter is determined by the ratio of the total force to the normal gravity contribution:

\bq\label{eq:Geff}
\frac{G_{\rm{eff}}}{G}(a, \delta) = \frac{\Psi,_{\chi}/\chi}{\Psi,_{\chi}^{\rm GR}/\chi} = \frac{\Psi,_{\chi}/\chi}{\Omega_{m0}\delta a^{-3}/2}.
\eq
In the Galileon model, $G_{\rm{eff}}$ is in general time and density dependent, but it is constant within a top-hat density profile.

\subsection{Model parameters}\label{sec:model-parameters}

\begin{table}
\caption{Parameters of the Galileon models studied in this paper. $\Omega_{r0}$, $\Omega_{b0}$, $\Omega_{c0}$, $h$, $n_s$, and $\tau$ are, respectively, the present day fractional energy density of radiation ($r$), baryons ($b$) and cold dark matter ($c$), the dimensionless present day Hubble expansion rate, the primordial scalar spectral index and the optical depth to reionization. The scalar amplitude at recombination $A_s$ refers to a pivot scale $k = 0.02 {\rm Mpc}^{-1}$. The universe is spatially flat in this model. The parameters $c_2$, $c_3$, $c_4$, $c_5$ are the dimensionless constants that appear in the action Eq.~(\ref{eq:action}) and $\rho_{\varphi, i} / \rho_{m,i}$ is the ratio of the Galileon and total matter ($m$) energy densities at $z_i$. We also show the value of $\chi^2 = -2{\rm log}P$ (where $P$ is the posterior probability obtained from the data from the WMAP 9yr results \cite{Hinshaw:2012fq}, the SNLS 3yr sample \cite{Guy:2010bc} and the BAO measurements from the 6dF Galaxy Survey \cite{Beutler:2011hx}, from the SDSS DR7 \cite{Percival:2009xn} and from the SDSS-III BOSS \cite{Anderson:2012sa}), the Galileon field time derivative at $z_i$, the age of the Universe and the present day value of $\sigma_8$. Only in this table, the subscript "$_i$" refers to quantities evaluated at $z = z_i = 10^6$. }
\begin{tabular}{@{}lccccccccccc}
\hline\hline
\\
Parameter  & Quintic Galileon& \ \ Quartic Galileon& \ \ 
\\
\hline
\\
$\chi^2$                          &   $7989.97$                                   			         &\ \ $7995.60$ & \ \ 
\\
$\Omega_{r0}{h}^2$                  &     $4.28\times10^{-5}$          				&\ \ $4.28\times10^{-5}$ & \ \ 
\\
$\Omega_{b0}{h}^2$                     & $0.02178$           				&\ \ $0.02182$ & \ \ 
\\
$\Omega_{c0}{h}^2$                       & $0.125$          				&\ \   $0.126$ & \ \ 
\\
${h}$                                                    & $0.735$            			&\ \   $0.733$ & \ \ 
\\
$n_s$                                                   & $0.947$             			           &\ \    $0.945$ & \ \ 
\\
$\tau$                                                   & $0.0680$                  			&\ \   $0.0791$ & \ \
\\
${\rm log}\left[ 10^{10}A_s \right]$    & $3.127$			&\ \  $3.152$ & \ \ 
\\
${\rm log}\left[\rho_{\varphi, i} / \rho_{m,i}\right]$   &$-6.51$      &\ \   $-37.39$ & \ \ 
\\
$c_2 / c_3^{2/3}$                                                    &$-3.59$                               &\ \    $-4.55$ & \ \
\\
$c_3$                                                                        &$10$           &\ \    $20$ & \ \
\\
$c_4/ c_3^{4/3}$                                                  &$-0.199$                                 &\ \    $-0.096$ & \ \
\\
$c_5/ c_3^{5/3}$                                                  &$0.0501$                                 &\ \    $0$ (fixed) & \ \
\\
\hline
\\
$\dot{\bar{\varphi}}_i c_3^{1/3}$		&$2.31\times10^{-14}$			  		&\ \  $1.54\times10^{-20}$ & \ \
\\
Age (Gyr)					  	&$ 13.778$	&\ \ $13.770$ & \ \
\\
$\sigma_8(z = 0)$					  &$0.975$		&\ \ $0.998$ & \ \
\\
\hline
\hline
\end{tabular}
\label{table:table-max}
\end{table}

\begin{figure*}
	\centering
	\includegraphics[scale=0.48]{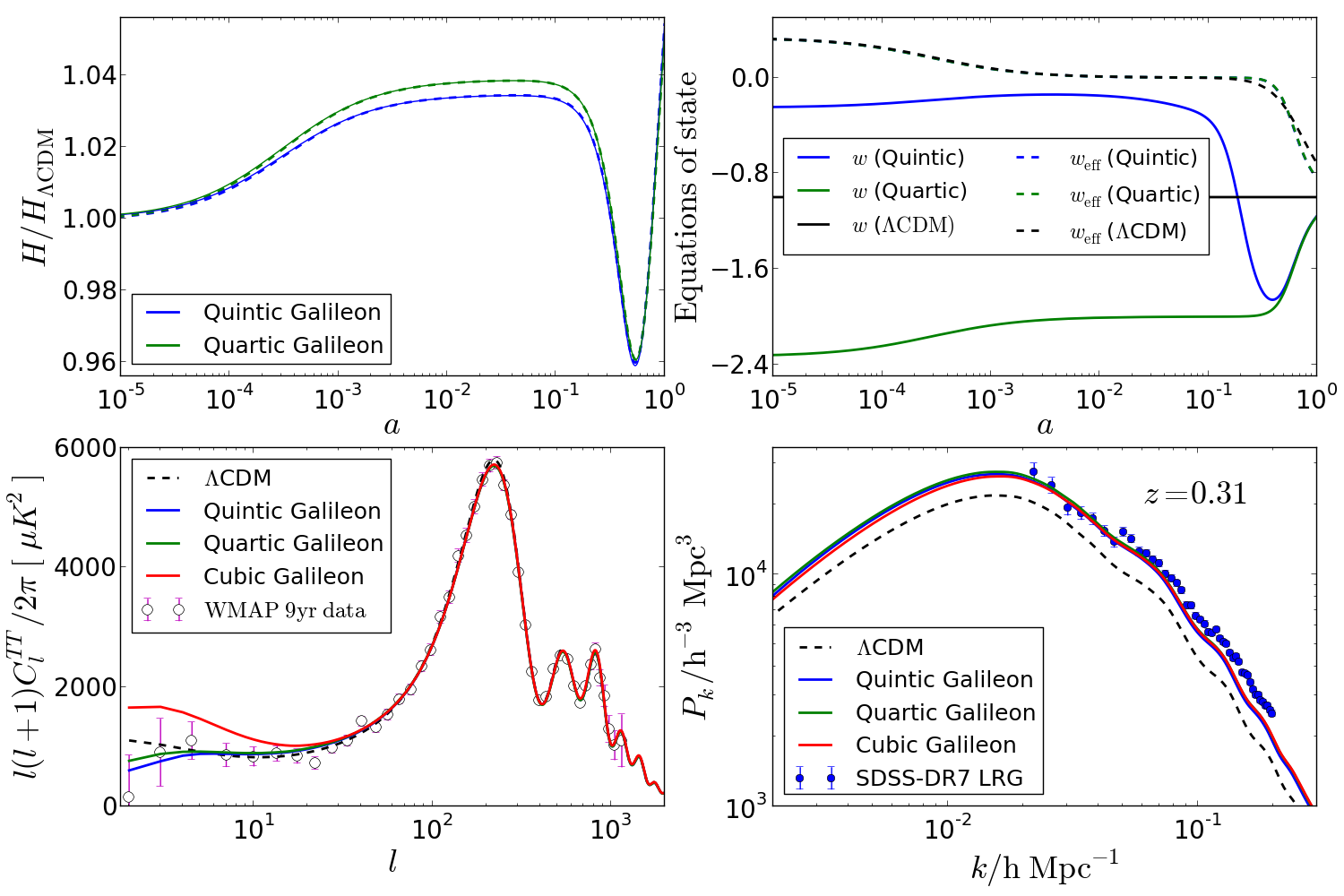}
	\caption{(Top left) Time evolution of the expansion histories of the Quintic (blue) and Quartic (green) Galileon models, plotted as $H/H_{\Lambda {\rm CDM}}$. The solid lines represent the full numerical solution, whereas the dashed lines show the tracker solution of Eq.~(\ref{eq:tracker_H}). (Top right) Time evolution of the Galileon field (solid) and cosmological (dashed) equations of state, $w$ and $w_{\rm eff}$, respectively, for the $\Lambda$CDM (black), Quartic (green) and Quintic (blue) Galileon models. (Bottom left) CMB temperature fluctuations angular power  spectra, as function of the multipole moments, for the $\Lambda$CDM (dashed black), Cubic (solid red), Quartic (solid green) and Quintic (solid blue) Galileon models.  Also shown are the data points with errorbars of the WMAP 9-yr results \cite{Hinshaw:2012fq}. (Bottom right) Linear matter power spectrum, as function of scale $k$, for the $\Lambda$CDM (dashed black), Cubic (solid red), Quartic (solid green) and Quintic (solid blue) Galileon models. The power spectrum is shown for $z = 0.31$, which is the mean redshift of the Luminous Red Galaxies of the SDDS DR7 used to estimate the host halo spectrum shown as the data points with errorbars \cite{Reid:2009xm}.}
\label{fig:w-h-cl-pk}\end{figure*}

Throughout the paper, whenever we refer to the Quintic and Quartic Galileon models, we will be referring to the models with the parameters given in Table~\ref{table:table-max}. These are model parameters that provide a reasonably good fit to a combination of data made up of the WMAP 9yr results \cite{Hinshaw:2012fq}, SNIa from the SNLS 3yr sample \cite{Guy:2010bc} and the BAO measurements from the 6dF Galaxy Survey \cite{Beutler:2011hx}, from the SDSS DR7 \cite{Percival:2009xn} and from the SDSS-III BOSS \cite{Anderson:2012sa}. These parameters were obtained with our modified versions of the {\tt CAMB} \cite{camb_notes} and {\tt CosmoMC} \cite{Lewis:2002ah} codes \cite{Barreira:2012kk, Barreira:2013jma}.

The time evolution of the expansion rate, the Galileon equation of state parameter $w_{\varphi} = \bar{p}_{\varphi}/\bar{\rho}_{\varphi}$ and the effective cosmological equation of state  $w_{\rm{eff}} = \left(\bar{\rho}_r/3 + w_{\varphi}\bar{\rho}_{\varphi}\right)/\left(\bar{\rho}_r + \bar{\rho}_m + \bar{\rho}_{\varphi}\right)$, are shown in Fig.~\ref{fig:w-h-cl-pk} for the Quintic and Quartic Galileon models. In the top left panel, we show both the numerical solution (solid) and the analytical expression (dashed) for the expansion rate (Eq.~(\ref{eq:tracker_H})). One can see the very good agreement between the two at all the epochs shown. This is because in both of these models, the tracker is reached before the epoch when dark energy starts to play a measurable role in the dynamics of the universe. At earlier times, radiation and matter dominate, and hence the expansion rate is not sensitive to the evolution of the Galileon field and whether or not it is on the tracker. In the top right panel of Fig.~\ref{fig:w-h-cl-pk}, one sees that this tracker solution is characterized by an equation of state $w_{\varphi} < -1$. Moreover, the Quintic Galileon model is attracted to the tracker much later than the Quartic model, which follows the tracker dynamics from $a<10^{-5}$. This is a consequence of the much lower energy density of the Galileon field at $z = 10^6$ in the Quartic Galileon compared to the Quintic, which favours the tracker to be reached at much earlier epochs.

%The time when the Galileon field joins the tracker solution is essentially determined by the primordial value of its background energy density, $\bar{\rho}_{\varphi, i}$. In particular, the smaller $\bar{\rho}_{\varphi, i}$, the sooner the field will evolve according to the tracker \cite{DeFelice:2010pv, DeFelice:2010as}. In \cite{Barreira:2013jma}, we have shown that the CMB data, in particular through the ISW effect, constraints the inital density to be smaller than a given critical value.  As can be seen in the top right panel of Fig.~\ref{fig:w-h-cl-pk}, this tracker solution is characterized by an evolution where $w_{\varphi} < -1$. Also, the Quintic Galileon model merges the tracker much later than the Quartic model, which joins the tracker at $a<10^{-5}$. 

In the bottom panels of Fig.~\ref{fig:w-h-cl-pk}, we show the predicted power spectra for the CMB temperature fluctuations (bottom left) and for the linear clustering of matter (bottom right). We show the predictions of the Quintic and Quartic Galileon models, as well as the Cubic Galileon model that best fits the CMB, SNIa and BAO data (see \cite{Barreira:2013eea}) and the $\Lambda$CDM model with the WMAP 9-year parameters \cite{Hinshaw:2012fq}. With respect to the CMB data, one sees that the Quartic model, contrary to the Cubic model, is able to provide a fit similar to that of the Quintic Galileon model. In \cite{Barreira:2013jma}, we showed that the latter can fit the WMAP 9-yr data better than standard $\Lambda$CDM, being just slightly disfavoured when the low-redshift SNIa and BAO data is also taken into account. With respect to the Quintic Galileon model, the Quartic Galileon is disfavoured by $\Delta\chi^2_{\rm Quartic} \sim - 6$. This is much smaller than the corresponding difference for the Cubic Galileon case $\Delta\chi^2_{\rm Cubic} \sim - 17$ \cite{Barreira:2013eea} \footnote{Note that in these $\chi^2$ differences we are not taking into account the fact that the different models have different numbers of free parameters.}. These differences in the fits of the different Galileon models are mostly determined by the Integrated Sachs-Wolfe (ISW) effect, which is sensitive to the time variations of the lensing potential $\phi = \left(\Phi + \Psi\right)/2$. In \cite{Barreira:2012kk}, we showed that $\phi$ can have a nontrivial time and scale dependence (see Fig.~4 of \cite{Barreira:2012kk}). In particular, depending on the choice of the Galileon parameters $c_i$, the lensing potential can grow or decay very rapidly, or display a milder time evolution. The latter cases are those preferred by the data since they contribute less to the ISW power on large angular scales (low-$l$). In the case of the Cubic Galileon model one has that $c_4 = c_5 = 0$, which gives the model less flexibility to produce milder time evolutions in $\phi$. Thus, the Cubic model does not fit the  low-$l$ data of the CMB as well as the Quartic and Quintic Galileon models. Note that in the $\Lambda$CDM model, the lensing potential decays at late times, which is why this model predicts more ISW power that the Quartic and Quintic Galileon models.

Contrary to the CMB predictions, the Cubic, Quartic and Quintic Galileon models predict very similar power for the linear clustering of matter. In particular, all models show a general enhancement of the clustering power with respect to the standard $\Lambda$CDM prediction, on all scales. However, there are a number of uncertainties associated with clustering measurements that prevent a direct comparison with the data. Firstly, there is the uncertainty related to the validity of linear perturbation theory, whose assessment is less obvious in modified gravity theories due to the nonlinear screening mechanisms. To determine the regime of validity of linear theory one usually needs to resort to N-body cosmological simulations. In \cite{Barreira:2013eea}, we showed that in the case of the Cubic Galileon model, the simulation results recover the linear perturbation theory prediction on scales $k \lesssim 0.1 h/{\rm Mpc}$. These are scales where linear theory is usually expected to be a good approximation. On the other hand, N-body simulations of the Quartic model \cite{Li:2013tda} find that the nonlinear Vainshtein mechanism can have a measurable impact (although small) on the growth of structure for $k \lesssim 0.1 h/{\rm Mpc}$. Nevertheless, even if the Vainshtein screening is found to be negligible above a given length scale, there is still a second important uncertainty that is related to the bias of dark matter halos and galaxies. Although it might be reasonable to expect that the bias would result in an overall enhancement of the clustering power of high-mass halos, the exact value of the bias and its mass and scale dependence are not clear in modified gravity theories. It is therefore important to have a better understanding of the bias in models like the Galileon before making a robust comparison with the current and future data. One of the goals of this paper is to take a first step in this direction.

\section{Fifth force solutions}\label{sec:solutions}

In Eq.~(\ref{eq:Geff}), we have parametrized the modifications to gravity (the fifth force) as a rescaling of the effective gravitational constant, which is time and density dependent. The process of determining the total force involves solving a nonlinear algebraic equation, Eq.~(\ref{eq:eom_sph_alg}), which in general has more than one branch of real solutions. Therefore, care must be taken in making sure that the physical branch exists and is correctly identified. We discuss these issues next.

\subsection{Quintic Galileon}

\begin{figure*}
	\centering
	\includegraphics[scale=0.445]{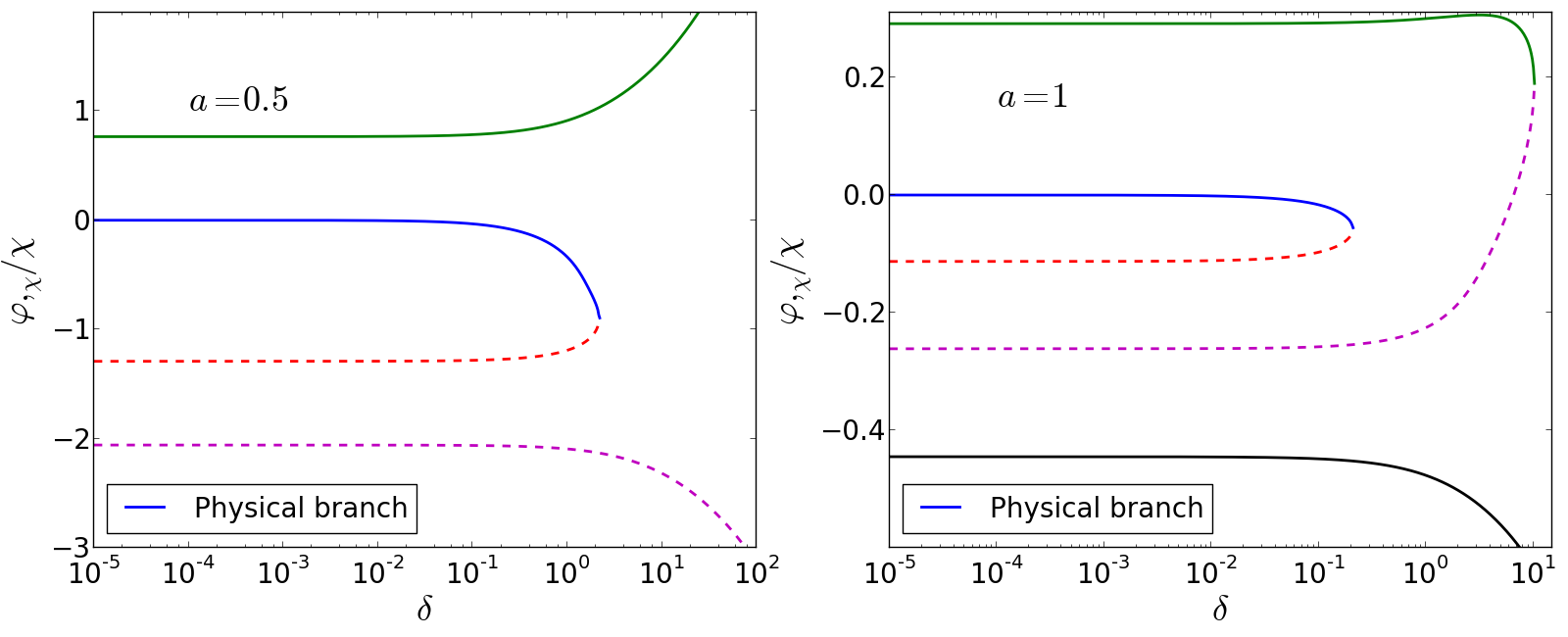}
	\caption{Evolution of the branches of solutions of Eq.~(\ref{eq:eom_sph_alg}) as function of the density constrast $\delta$, for $a = 0.5$ (left panel) and $a = 1$ (right panel). The physical branch corresponds to the solid blue line for which $\varphi,_{\chi}/\chi\left(\delta=0\right) = 0$. For visualization purposes, in the $a = 1$ panel we do not show the branch $\varphi,_{\chi}/\chi\left(\delta=0\right) = -11.477$, which is far below the scale of the plot and has little importance for the discussion.}
\label{fig:quintic-branches}\end{figure*}

In the case of the Quintic Galileon model, Eq.~(\ref{eq:eom_sph_alg}) has six branches of solutions, which in general can be either complex or real. We require the physical branch to be real and to satisfy: 

\bq\label{eq:physical-branch}
\frac{\varphi,_{\chi}}{\chi}\left(\delta \rightarrow 0\right) \rightarrow 0.
\eq
This is the solution that exhibits the physical behavior that there should be no fifth force if there are no density fluctuations sourcing it. We must ensure that this solution exists at every moment in time, and for every value of $\delta \geq -1$. 

However, as we will show next, the Quintic Galileon model equations do not satisfy this requirement. To better understand why this happens, one can differentiate Eq.~(\ref{eq:eom_sph_alg}) w.r.t. $\delta$ to obtain a differential equation for $\varphi,_{\chi}/\chi$:

\begin{widetext}
\bq\label{eq:eom_sph_diffeq}
\frac{\rm{d}}{\rm{d}\delta}\left[\frac{\varphi,_{\chi}}{\chi}\right] = \frac{-2\eta_{02}\delta - \eta_{01} - \eta_{11}\left[\frac{\varphi,_{\chi}}{\chi}\right] - \eta_{12}\left[\frac{\varphi,_{\chi}}{\chi}\right]^2 - \eta_{31}\left[\frac{\varphi,_{\chi}}{\chi}\right]^3}{\eta_{11}\delta + \eta_{10} + 2\left(\eta_{21}\delta + \eta_{20}\right)\left[\frac{\varphi,_{\chi}}{\chi}\right] + 3\left(\eta_{31}\delta + \eta_{30}\right)\left[\frac{\varphi,_{\chi}}{\chi}\right]^2 + 4\eta_4\left[\frac{\varphi,_{\chi}}{\chi}\right]^3 + 5\eta_5\left[\frac{\varphi,_{\chi}}{\chi}\right]^4 + 6\eta_6\left[\frac{\varphi,_{\chi}}{\chi}\right]^5}.
\eq
\end{widetext}
Just to illustrate our point, it suffices to consider the equations at $a = 0.5$  and $a = 1$ (we have checked that our conclusion holds for other epochs too). When $\delta = 0$, Eq.~(\ref{eq:eom_sph_alg}) has four real roots $\left\{-2.059,\ -1.292,\ 0,\ 0.765\right\}$ at $a = 0.5$, whereas at $a = 1$ there are six real roots $\left\{ -11.477,\ -0.445,\ -0.261,\ -0.113,\ 0,\ 0.291 \right\}$. These can be used as the initial conditions to solve Eq.~(\ref{eq:eom_sph_diffeq}) and evolve the different branches. The result is shown in Fig.~\ref{fig:quintic-branches}. The physical branch is the one that starts from zero at $\delta = 0$, but one sees that it cannot be evaluated beyond $\delta \approx 2$ and $\delta \approx 0.2$ at $a=0.5$ and $a=1$, respectively. At these values of $\delta$, the differential equation becomes singular because the physical branch becomes complex (and therefore unphysical), together with the branch represented by the dashed red line. The same thing happens for the (unphysical) branches represented by the solid green and dashed magenta lines at $a = 1$, although at different values of $\delta$. We have explicitly looked at Eq.~(\ref{eq:eom_sph_alg}) for cases near these critical values of $\delta$ to confirm that the breakdown of the differential equation is related to the absence of real roots. Moreover, we have also checked that the problem persists for different choices of the Galileon and cosmological parameters around the regions of parameter space preferred by the CMB, SNIa and BAO data \cite{Barreira:2013jma}.

The spherical collapse in the Galileon model has been also studied in \cite{Bellini:2012qn}. In the latter, the authors found that physical fifth force solutions exist both at low and high densities. In particular, by taking the limit $\delta \gg 1$, the authors derive the conditions for the existence of real solutions for Eq.~(\ref{eq:eom_sph_alg}). This assumes that the physical solution does not become complex for intermediate densities, which is what is shown not to happen in Fig.~\ref{fig:quintic-branches} of this paper. We point out that it is hard to directly compare the results of the two papers because of the different notation adopted to describe the tracker background evolution; also, contrary to \cite{Bellini:2012qn}, we focus on the parameters of the model that fit the current data.

At this point, one may wonder whether this problem can be avoided by relaxing the quasi-static and weak-field approximations used to derive Eqs.~(\ref{eq:poisson0}), (\ref{eq:slip0}) and (\ref{eq:eom_sph0}). However, note that Fig.~\ref{fig:quintic-branches} shows that the physical solution does not even exist in high density regions, where the terms that have been neglected are expected to be small, and hence our approximations are justified (we will return to this point in the next section).  Another way to try to circunvent the problem is to explore different choices of the Galileon and cosmological parameters. However, even if for a different choice of parameters one could find physical solutions for all $\delta$, such parameters would already be ruled out by the current CMB, SNIa and BAO data. For these reasons, our study of the spherical collapse in the Quintic model stops here!

\subsection{Quartic Galileon}\label{subsection:quartic-galileon}

When $c_5 = 0$, Eq.~(\ref{eq:eom_sph_alg}) becomes

\bq\label{eq:eom_sph_alg-quartic}
0 &=& \eta_{01}\delta + \left(\eta_{11}\delta + \eta_{10}\right)\left[\frac{\varphi,_{\chi}}{\chi}\right] +\eta_{20}\left[\frac{\varphi,_{\chi}}{\chi}\right]^2 \nonumber \\
&& + \eta_{30}\left[\frac{\varphi,_{\chi}}{\chi}\right]^3,
\eq
which is third order, and therefore admits analytical solutions given by the general expression

\bq\label{eq:analytic-solution}
\frac{\varphi,_{\chi}}{\chi} = -\frac{1}{3\eta_{30}}\left[\eta_{20} + \mu_k \Gamma + \frac{\Sigma_0}{\mu_k\Gamma}\right],\ k \in \left\{1,2,3\right\},
\eq
where

\bq
\Gamma &=& \left[\frac{\Sigma_1 + \sqrt{\Sigma_1^2 - 4\Sigma_0^3}}{2}\right]^{1/3}, \\
\Sigma_0 &=& \eta_{20}^2 - 3\eta_{30}\left(\eta_{11}\delta + \eta_{10}\right), \\
\Sigma_1 &=& 2\eta_{20}^3 - 9\eta_{30}\eta_{20}\left(\eta_{11}\delta + \eta_{10}\right) + 27\eta_{30}\eta_{01},
\eq
and the three branches of solutions (labelled by $k$) correspond to

\bq
\mu_1 = 1, \ \ \ \ \ \mu_2 = \exp\left[-i\pi/3\right], \ \ \  \ \mu_3 = \exp\left[i\pi/3\right].\ \ \ 
\eq
The physical branch, Eq.~(\ref{eq:physical-branch}), corresponds to the $k = 3$ solution, which is a complex number. As a result, $\Gamma$ must be complex as well and we can write it as

\bq
\Gamma = \Sigma_0^{1/2}\exp\left[i\theta/3\right],
\eq
with $\theta$ given by
\bq
{\rm cos}\ \theta = \frac{\Sigma_1/2}{\Sigma_0^{3/2}}, \ \ \ \ \ \ \theta \in \left[0, \pi\right].
\eq
Using these expressions, Eq.~(\ref{eq:analytic-solution}) can be written as

\bq\label{eq:analytical-solution-2}
\frac{\varphi,_{\chi}}{\chi} = -\frac{1}{3\eta_{30}}\left[\eta_{20} + 2\sqrt{\Sigma_0}\cos\left(\frac{\theta}{3} - \frac{2\pi}{3}\right)\right],
\eq
which allows us to analytically determine the magnitude of the effective gravitational strength ($G_{\rm eff}$) using Eq.~(\ref{eq:Geff}). 

\begin{figure*}
	\centering
	\includegraphics[scale=0.455]{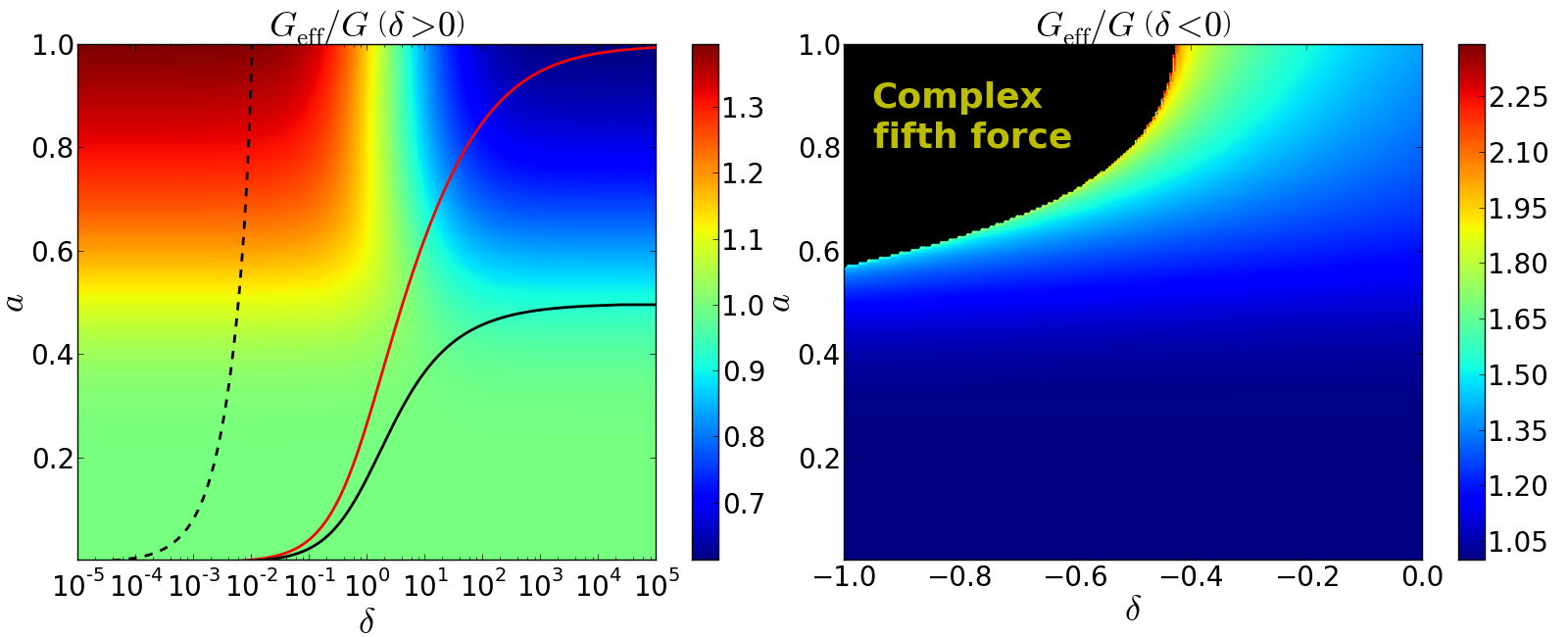}
	\caption{Time and density dependence of the effective gravitational strength $G_{\rm eff}$ of Eq.~(\ref{eq:Geff}) for $\delta > 0$ (left panel) and $\delta < 0$ (right panel). The colour scale bars at the right of each panel show the value of $G_{\rm eff}/G$. In the left panel the solid red and solid black lines represent the trajectory in $a-\delta$ space of a halo that collapses at $a = 1$ ($z = 0$) and $a = 0.5$ ($z = 1$) in the Quartic Galileon model, respectively. The dashed black line shows the trajectory of a large linear density region with density contrast $\delta = 0.01$ today. In the right panel, the region marked in black in the top left corner shows the values of $a$ and $\delta$ for which the solution of the fifth force becomes a complex number. To facilitate the visualization, note that the colour scale in the two panels is not the same.}
\label{fig:Geffmap}\end{figure*}

The value of $G_{\rm{eff}}$ as a function of the scale factor $a$ and density $\delta$ is shown in the colour map of Fig.~\ref{fig:Geffmap}, for the Quartic Galileon model. The left and right panels correspond to $\delta > 0$ and $\delta < 0$, respectively. For $\delta > 0$ we see that, contrary to the case of the Quintic Galileon model, there are physical solutions for sufficiently large values of the density contrast $\delta$. When $a \lesssim 0.5$ one has $G_{\rm{eff}}/G \approx 1$. At later times, however, $G_{\rm{eff}}/G$ progressively deviates from unity, and this happens in a density dependent way. In the linear regime ($\delta \ll 1$), $G_{\rm{eff}}$ increases with time, being roughly $40 \%$ larger than $G$ today.  However, for $\delta \gtrsim 1$, one sees that gravity becomes weaker with time ($G_{\rm{eff}}/G < 1$), or in other words, the fifth force becomes repulsive.  In particular, at the present day, the effective gravitational strength is reduced to $\sim 60\%$ of the standard gravity value.

The effects of the fifth force that modify $G_{\rm eff}$ in the Quartic model can be thought of as being two-fold. Firstly, one has the extra terms proportional to $\varphi,_{\chi}/\chi$, that add up to the total gravitational strength in Eqs.~(\ref{eq:poisson}) and (\ref{eq:slip}). Secondly, there are also the time-dependent coefficients $A_4$, $B_0$ and $B_3$ that multiply the standard gravity terms, and that arise via explicit couplings of the Galileon field to curvature. The effect of the screening can be seen by writting Eq.~(\ref{eq:eom_sph_alg}) in the limit where $\delta \gg 1$, 

\bq\label{eq:vainshtein-operation}
0 &\approx& \eta_{01} + \eta_{11} \left[\frac{\varphi,_{\chi}}{\chi}\right].
\eq
Here, one sees that in regions where the density is sufficiently high, the spatial gradient of the Galileon field, $\varphi,_{\chi}/\chi$, does not depend on $\delta$. The Vainshtein mechanism in the Quartic model works because

\bq\label{eq:vainshtein-operation1}
\left|\frac{\varphi,_{\chi}}{\chi}\right| = \left|\frac{\eta_{01}}{\eta_{11}}\right| \ll \left|\frac{\Psi,_{\chi}}{\chi}\right| \sim \delta\ \ \ \ \ \ \left(\delta \gg 1\right),
\eq
and increasing the density $\delta$ further does not increase the gradient of the Galileon field \footnote{In the case of the Cubic Galileon model one has $\varphi,_{\chi}/\chi \propto \sqrt{\delta}$ in high densities \cite{Barreira:2013eea}.}. However, the coefficients $A_4$, $B_0$ and $B_3$ depend only on the background evolution of the Galileon field, and will not be affected by the Vainshtein mechanism. This is why $G_{\rm eff}/G$ does not approach unity when $\delta \gg 1$ (c.f.~Fig.~\ref{fig:Geffmap}). This result has in fact been found to be generically possible in the framework of the most general second-order scalar tensor theory \cite{Babichev:2011iz, Kimura:2011dc}, which encompasses the Quartic Galileon model studied here.

The fact that the effective gravitational strength is time-varying if the density is high is an unpleasent novelty of the model. In fact, this may imply that the Quartic Galileon model is automatically ruled out by the local gravity tests that constrain the modifications to gravity to be very small. For instance, \cite{Babichev:2011iz, Kimura:2011dc} have claimed that Lunar Laser Ranging experiments \cite{Williams:2004qba} can place very strong constraints on models like the Quartic Galileon. It seems therefore reasonable to state that the survival of the Quartic Galileon model as a candidate for dark energy depends upon finding a cure for this apparent local time variation of $G_{\rm eff}$. One may invoke the validity of the quasi-static approximation in an attempt to ease this problem. For instance, if the time derivative of the Galileon field perturbation is not completely negligible, then its contribution to the coefficients $A_4$, $B_0$ and $B_3$ could help soften the time variation of $G_{\rm eff}$. However, we argue that this should not be the case. The successful implementation of the screening in the Quartic Galileon model means that the fluctuations of the Galileon field, $\delta\varphi$,  have to be much smaller than the metric perturbation, i.e., $\delta_\varphi \ll \Psi$ (c.f.~Eqs.~(\ref{eq:vainshtein-operation}) and (\ref{eq:vainshtein-operation1})). Since $\Psi$ is typically very small for collapsed objects like cluster- and galaxy-mass halos or the Sun ($\Psi \sim 10^{-6}$ to $10^{-4}$) \footnote{Near black holes, for instance, one can have larger metric perturbations $\Psi \sim 1$.}, then $\delta\varphi$ has to be even smaller. This justifies the use of the weak-field assumption for the Galileon perturbation, $\delta\varphi$. Consequently, for consistency, the time variation of $\delta\varphi$ has to be very small as well, $\dot{\delta\varphi} \ll \dot{\Psi} \ll \dot{\bar{\varphi}}$. The same reasoning applies to the Quintic model studied in the last subsection. In the remaining of the paper, we will focus on the cosmological interpretation of the results.

For $\delta < 0$, in Fig.~\ref{fig:Geffmap}, we see again that the modifications to gravity arise only for $a \gtrsim 0.5$, but here gravity can only become stronger. In addition, there are no physical solutions for the epochs and densities indicated by the black region in the top left corner of the right panel of Fig.~\ref{fig:Geffmap}. In particular, the fifth force becomes complex in the most empty voids ($\delta \sim -1$) for $a \gtrsim 0.6$; at $a = 1$, underdense regions where $\delta  \lesssim -0.4$ also do not admit a real fifth force. This is a problem that exists also in Cubic Galileon gravity models \cite{Barreira:2013eea}. This absence of real physical solutions for the fifth force is probably related to the fact that the quasi-static limit may not be a good approximation in low density regions. Nevertheless, in this work we are interested in studying the formation of halos (rather than voids), for which these low densities are irrelevant. 

\section{Excursion set theory in Galileon gravity}\label{sec:excursion}

In this section, we layout the main premises of excursion set theory \cite{1991ApJ...379..440B, Zentner:2006vw} and of the dynamics of the gravitational collapse of spherical overdensities in the Galileon model (see, e.g.~\cite{Gaztanaga:2000vw, Schaefer:2007nf, Martino:2008ae, Li:2011qda, Borisov:2011fu, Lam:2012fa, Li:2012ez, Clampitt:2012ub, Lombriser:2013wta, Kopp:2013lea} and references therein for applications of the spherical collapse model and excursion set theory to other modified gravity models).

\subsection{Basics of excursion set theory}

\subsubsection{Unconditional probability distribution and halo mass function}

The main postulate of excursion set theory is that dark matter halos form from the gravitational collapse of regions where the linear density contrast smoothed over some comoving length scale $R$,

\bq\label{eq:smoothed-density-field}
\delta_{\rm lin}\left(\bold{x}, R\right) &=& \int W\left(|\bold{x} - \bold{y}|, R\right)\delta_{\rm lin}\left(\bold{y}\right)\rm{d}^3\bold{y} \nonumber \\
&=& 4\pi\int k^2\tilde{W}\left(k, R\right)\delta_{\rm lin, k}e^{i\bold{k}\bold{x}}{\rm d}{k},
\eq
exceeds a certain critical density threshold $\delta_{\rm lin,crit}$ (to be defined below). Here $W\left(|\bold{x} - \bold{y}|, R\right)$ is the real space filter (or window) function of comoving size $R$, and $\tilde{W}\left(k, R\right)$ and $\delta_{\rm lin, k}$ are the Fourier transforms of $W\left(|\bold{x} - \bold{y}|, R\right)$ and $\delta_{\rm lin}\left(\bold{y}\right)$, respectively. We use the subscript "$_{\rm lin}$" to remind ourselves of the situations where the density contrast should be interpreted as being small ($|\delta| \ll 1$), i.e., in the linear regime.

The mass of the halo is given by

\bq\label{eq:mass-overdensity}
M = 4\pi \bar{\rho}_{m0}R^3/3.
\eq
For the same comoving radius $R$, the halo mass is different for models with different matter densities $\bar{\rho}_{m0}$. As the standard practice, we will assume that the probability distribution of the initial ($z_i = 300$)  linear density contrast $\delta_{\rm lin}(\bold{x})$ is a Gaussian with zero mean

\bq\label{eq:prob_gaussian}
{\rm Prob.}\left(\delta_{\rm lin},  S\right){\rm d}\delta_{\rm lin} = \frac{1}{\sqrt{2\pi S}}\exp\left[-\frac{\delta_{\rm lin}^2}{2S}\right]\rm{d}\delta_{\rm lin},
\eq
in which $S \equiv S(R)$ is the variance of the density contrast field on the scales of the size of the filter function $R$, and is given by

\bq\label{variance}
S(R) \equiv \sigma^2(R) = 4\pi\int k^2 P_k\tilde{W}\left(k, R\right){\rm d}k,
\eq
where $P_k$ is the linear matter power spectrum. Note that for a fixed model, the variables $R$, $M$ and $S$ are related to one another and will be used interchangeably throughout when referring to the scale of the halos.

In hierarchical models of structure formation, $S(R)$ is a monotonically decreasing function of $R$. Consequently, the probability that the density field on a region smoothed over a very large $R$ exceeds the critial density $\delta_{\rm{lin,crit}}$ is very small, since the variance is also very small. As one smooths the density field with decreasing $R$, the field $\delta_{\rm lin}\left(\bold{x}, R\right)$ undergoes a random walk with "time" variable $S$. In the excursion set theory language, $\delta_{\rm{lin, crit}}$ defines a "barrier" that the random walks cross, and the aim is to determine the probability distribution, $f(S){\rm d}S$, that the first up-crossing of the barrier occurs at $\left[S, S + {\rm d}S\right]$. In the particular case where the filter function is a top-hat in $k$-space, then the random walk of the density field will be Brownian. As we will see below, in the case of the Galileon model, the critical density for collapse does not depend on the scale $S$ considered. This is called a "flat barrier". In this case, $f(S)$ admits a closed analytical formula given by \cite{1991ApJ...379..440B}

\bq\label{eq:first-crossing}
f(S) = \frac{1}{\sqrt{2\pi}}\frac{\delta_c}{S^{3/2}}\exp\left[-\frac{\delta_c^2}{2S}\right],
\eq
where $\delta_c$ denotes the initial critical density, $\delta_{\rm{lin, crit}}$, for a spherical overdensity to collapse at a given redshift, linearly extrapolated to the present day, assuming $\Lambda$CDM linear growth factor\footnote{In the case of $\delta_c$, we will avoid writting the subscript $_{\rm lin}$ to ease the notation.}. This linear extrapolation is done only for historical reasons so that the values of $\delta_c$ we present in this paper can be more easily compared with previous work. Note also that, for consistency, one must compute the variance $S$ in Eq.~(\ref{variance}) using the initial power spectrum of the models, but evolved to $z = 0$ with the $\Lambda$CDM linear growth factor. We use the BBKS fitting formula \cite{Bardeen:1985tr}, whose accuracy in reproducing the $\Lambda$CDM and Quartic Galileon model $P_k$ at the initial time is more than sufficient for the purposes of the qualitative discussion we present here\footnote{Note that one can use $\Lambda$CDM to compute the matter power spectrum of the Galileon model at the initial time, but one has to use the parameters given in Table \ref{table:table-max}.}. We will follow the standard procedure of adopting a filter function that is a top-hat in real space, whose Fourier transform is given by

\bq
\tilde{W}\left(k, R\right) = 3\frac{\sin\left(kR\right) - kR\cos\left(kR\right)}{\left(kR\right)^3}.
\eq
Note that, strictly speaking, for this filter function the excursion set random walks are not Brownian, and as a result, there is some degree of approximation in taking Eq.~(\ref{eq:first-crossing}). On the other hand, this choice of filter function is that which is compatible with our definition of the mass of the smoothed overdense region (Eq.~(\ref{eq:mass-overdensity})).

In this paper the halo mass function is the comoving differential number density of halos of a given mass per natural logarithmic interval of mass. This quantity is obtained by associating $f(S){\rm d}S$ with the fraction of the total mass that is incorporated in halos, whose variances fall within $\left[S, S + {\rm d}S\right]$ (or equivalently, whose masses fall within $\left[M, M + {\rm d}M\right]$). The mass function observed at redshift $z$ is then given by

\bq\label{eq:mass-function}
&&\frac{{\rm d}n(M)}{{\rm d}{\rm ln}M}{\rm d}{\rm ln}M = \frac{\bar{\rho}_{m0}}{M} f(S){\rm d}S \nonumber \\
&&= \frac{\bar{\rho}_{m0}}{M}\frac{\delta_c}{\sqrt{2\pi S}}\left|\frac{{\rm d}{\rm ln}S}{{\rm d}{\rm ln}M}\right|{\rm exp}\left(-\frac{\delta_c^2}{2S}\right){\rm d}{\rm ln}M.
\eq
This is known as the Press-Schechter mass function \cite{1974ApJ...187..425P}. The redshift dependence is included into $\delta_c$ (c.~f.~Fig.~\ref{fig:dc}). In principle, one can distinguish the formation time from the observation time of the halos (see e.g.~\cite{Parfrey:2010uy}). For simplicity, in this paper we assume that these are the same, i.e., $z = z_{{\rm form}} = z_{{\rm obs}}$.

\subsubsection{Conditional probability distribution and halo bias}

Equations (\ref{eq:first-crossing}) and (\ref{eq:mass-function}) assume that the starting point of the excursion set random walk is the origin of the $\delta_{\rm lin}-S$ plane. The mass function computed using Eq.~(\ref{eq:mass-function}) gives the abundance of halos that have collapsed from the mean cosmological background. However, it is well known that the clustering of halos is biased towards the underlying clustering of dark matter, i.e., the number density of halos is different in different regions. Within the framework of excursion set theory, this is described by the so-called halo bias parameter $\delta_h$ \cite{Mo:1995cs}. The latter is determined by computing the abundance of halos that have formed from a region characterized by $S = S_0$ and $\delta_{\rm lin} = \delta_0$, and compare it with the abundance of the halos that have formed from the mean background ($S = \delta_{\rm lin} = 0$). It can be shown that $\delta_h$ is given by \cite{Mo:1995cs}

\bq\label{eq:halo-overabundance}
\delta_h = \left(1+\delta_{\rm env}\right)\frac{f(S|S_0, \delta_0){\rm d}S}{f(S){\rm d}S} - 1,
\eq
where $\delta_{\rm env}$ is the density contrast of the underlying dark matter region or environment where the halos are forming. $f(S|S_0, \delta_0)$ is the probability distribution that a random walk that starts (or passes through) $\left(\delta_0, S_0\right)$ crosses the critical barrier $\delta_c$ at $\left[S, S + {\rm d}S\right]$, and is given by

\bq\label{eq:conditional-first-crossing}
f(S|S_0, \delta_0) = \frac{1}{\sqrt{2\pi}}\frac{\delta_c-\delta_0}{\left(S-S_0\right)^{3/2}}\exp\left[-\frac{(\delta_c - \delta_0)^2}{2(S-S_0)}\right], \nonumber \\
\eq
for a flat barrier. Here, $\delta_0$ is the linearly extrapolated (with the $\Lambda$CDM linear growth factor to today) initial density of the underlying dark matter region, so that its density is $\delta_{\rm env}$, at a given redshift\footnote{Just like for $\delta_c$, we will avoid writting the subscript $_{\rm lin}$ in $\delta_0$ to ease the notation.}. From Eqs.~(\ref{eq:conditional-first-crossing}) and (\ref{eq:halo-overabundance}) one sees that dense regions can boost the clustering of halos, since the effective height of the barrier becomes lower $\left(\delta_c - \delta_0 < \delta_c\right)$. On the other hand, the clustering can also be suppressed if the mass of the halos is comparable to the mass available in the region specified by $S_0$. For example, halos with variance $S < S_0$ will not form because the random walks cannot cross the barrier before their starting point (this effect is known as halo exclusion).

One is often interested in the limit of very large regions with small density contrast ($S_0 \ll 1,\ 0 < \delta_0 \ll 1$), where the treatment simplifies considerably. In this case, we can Taylor expand $\delta_h$ as \cite{Fry:1992vr}:

\bq
\delta_h = \sum_{k=0}^{\infty}\frac{b_k}{k!}\delta^k \approx b_0 + b_1\delta_{\rm env} + \mathcal{O}\left(\delta_{\rm lin,env}^2\right),
\eq
where we have truncated the series at the linear term, as we are assuming low density regions (from here on $\delta_{\rm env}$ should be interpreted as a small linear overdensity). Since we taking the limit where the dark matter regions look like the mean background, $S_0, \delta_0 \rightarrow 0$, then $b_0 = 0$. The linear term $b_1$ is then the leading one, and is given by

\bq\label{eq:linear-bias}
b_1 &=& \frac{\rm{d}}{{\rm d}\delta_{\rm env}}\delta_h|_{\delta_{\rm env} = 0} \nonumber \\
&=& \frac{1}{f(S)}\left[f(S) + \left(\frac{\rm{d}\delta_0}{{\rm d}\delta_{\rm env}}\right)\frac{\rm{d}}{{\rm d}\delta_0}f(S| 0, \delta_0)|_{\delta_0 = 0} \right] \nonumber \\
&=& 1 + \left(\frac{\rm{d}\delta_0}{{\rm d}\delta_{\rm env}}\right)\frac{\delta_c^2/S - 1}{\delta_c} \nonumber \\
&=& 1 + g(z)\frac{\delta_c^2/S - 1}{\delta_c}.
\eq 
To find the expression of $g(z)$, one notes that
\bq\label{eq:linear-bias-Dfactor}
\delta_{\rm env} &=& \frac{D^{{\rm model}}(z)}{D^{{\rm model}}(z_i)}\delta_{\rm env, i} = \frac{D^{{\rm model}}(z)}{D^{{\rm model}}(z_i)} \frac{D^{\Lambda{\rm CDM}}(z_i)}{D^{\Lambda{\rm CDM}}(0)} \delta_0 \nonumber \\
&=& \frac{D^{{\rm model}}(z)}{D^{\Lambda{\rm CDM}}(0)} \delta_0,
\eq
where $\delta_{\rm env, i}$ is the initial density of the regions whose density today in a given model is $\delta_{\rm env}$. In Eq.~(\ref{eq:linear-bias-Dfactor}), $D^{\rm model}(z)$ is the linear growth factor of a given model and we have assumed that $D^{{\rm model}}(z_i) = D^{\Lambda{\rm CDM}}(z_i)$ (see next subsection). Thus, $g(z)$ is simply given by

\bq\label{eq:g}
g(z) \equiv \frac{\rm{d}\delta_0}{{\rm d}\delta_{\rm env}} = \frac{D^{\Lambda{\rm CDM}}(0)}{D^{{\rm model}}(z)}.
\eq

In Eq.(\ref{eq:linear-bias}), the model dependence is included in $g(z)$ and $\delta_c$. In particular, $g(z)$ accounts for the fact that different models have different values of $\delta_0$ to yield the same $\delta_{\rm env}$ at redshift $z$.

\subsection{Linear growth factor and spherical collapse dynamics}

The final ingredient to derive the mass function and the linear halo bias is to determine the threshold barrier $\delta_c$, and to specify the equation that governs the evolution of the linear overdensities (which determines $g(z)$, Eq.~(\ref{eq:g})). For scales inside the horizon, the latter can be written as

\bq\label{eq:linear-growthfactor0}
\ddot{\delta}_{\rm lin} + 2H\dot{\delta}_{\rm lin} - 4\pi G\bar{\rho}_m\delta_{\rm lin} = 0,
\eq
or equivalently, by changing the time varible to $N = {\rm ln} a$, as

\bq\label{eq:linear-growthfactor}
D'' + \left(\frac{E'}{E} + 2\right)D' - \frac{3}{2}\frac{G_{{\rm eff}}(a)}{G}\frac{\Omega_{m0}e^{-3N}}{E^2} &=& 0,
\eq
where the linear growth factor $D(a)$ is defined as $\delta_{\rm lin}(a) = D(a)\delta_{\rm lin}(a_i)/D(a_i)$. The initial conditions are set up at $z_i = 300$ using the known matter dominated solution $D(a_i) = D'(a_i) = a_i$ \footnote{Not to be confused with the initial times of Table \ref{table:table-max}.}. These initial conditions are the same for all the models we will study (c.f.~Eq.~(\ref{eq:linear-bias-Dfactor})). The linear growth factor obtained by solving Eq.~(\ref{eq:linear-growthfactor}) enters the calculation of the linear halo bias through $g(z)$.

Recall we have defined $\delta_c$ as the linearly extrapolated value (using the $\Lambda$CDM linear growth factor) of the initial density of the spherical overdensity for it to collapse at a given redshift. To determine this value, we consider the evolution equation of the physical radius $\zeta$ of the spherical halo at time $t$, which satisfies the Euler equation

\bq\label{eq:sph_col}
\frac{\ddot{\zeta}}{\zeta} - \left(\dot{H} + H^2\right) &=& -\frac{\Psi,_{\zeta}}{\zeta} = -H_0^2\frac{\Psi,_{\chi}}{\chi}\nonumber \\ 
&=& -\frac{G_{{\rm eff}}(a, \delta)}{G}\frac{{H_0^2}\Omega_{m0}\delta a^{-3}}{2},
\eq
where we have used Eq.~(\ref{eq:Geff}) in the last equality. Note that $\zeta = ar = \chi/H_0^2$, where $r$ is the comoving radial coordinate. Changing the time variable to $N$ and defining $y(t) = \zeta(t)/\left(aR\right)$, Eq.~(\ref{eq:sph_col}) becomes

\bq\label{eq:sph_col1}
y'' &+& \left(\frac{E'}{E} + 2\right)y' \nonumber \\
&+& \frac{G_{{\rm eff}}(a, y^{-3} - 1)}{G}\frac{\Omega_{m0}e^{-3N}}{2E^2}\left(y^{-3} - 1\right)y = 0, \nonumber \\
\eq
where we have used that $\delta = y^{-3} - 1$ invoking mass conservation \footnote{Explicitly: $\bar{\rho}_ma^3R^3 = \left(1+\delta\right)\bar{\rho}_mr^3 \Rightarrow \delta = \left(aR/r\right)^3 - 1 = y^{-3} - 1$.}. The initial conditions are then given by $y(a_i) = 1 - \delta_{\rm lin, i}/3$ and $y'(a_i) = \delta_{\rm lin,i}/3$ (here, $\delta_{\rm lin,i}$ is the initial linear density contrast). The value of $\delta_c$ is found by a trial-and-error approach to determine the initial density $\delta_{\rm lin, i}$ that leads to collapse ($y = 0$, $\delta \rightarrow \infty$) at redshift $z$, evolving this afterwards until the present day using the $\Lambda$CDM linear growth factor.

It is important to note that, despite the presence of the Vainshtein screening, the modifications to gravity incorporated into $G_{\rm eff}$ do not introduce any scale dependence in the dynamics of the collapse of the spherical overdensities. The reason for this is that the implementation of the Vainshtein mechanism does not depend on the size of the halo $R$, but only on its density. Consequently, the critical barrier $\delta_c$ is "flat", i.e., it is only time-dependent and not $S$-dependent. In fact, in the previous subsection we have already anticipated this result, which is the one for which Eqs.~(\ref{eq:first-crossing}), (\ref{eq:mass-function}) and (\ref{eq:conditional-first-crossing}) are valid. The situation is different, for instance, in models that employ the chameleon screening mechanism. In these cases, the fifth force is sensitive to the size of the halo, and the barrier can have a nontrivial shape \cite{Li:2011qda}.

\subsubsection{Limitations of the spherical top-hat profile description}

It is well known that the Sheth-Tormen mass function \cite{Sheth:1999mn, Sheth:1999su, Sheth:2001dp} fits $\Lambda$CDM N-body simulation results better than Eq.~(\ref{eq:mass-function}). The reason is because the Sheth-Tormen mass function is derived by assuming the ellipsoidal collapse of the overdensities, which is a more realistic description of the intrinsically triaxial proccesses of gravitational instability. In the excursion set picture, the ellipsoidal collapse translates into a mass dependent (i.e.~'non-flat') critical barrier. In this paper, we are only interested in a qualitative analysis and, therefore, the spherical collapse model is sufficient. However, even if one models the Galileon mass function with the standard Sheth-Tormen formulae, some complications may still arise. We comment on two such complications.

Firstly, the Sheth-Tormen mass function contains two free parameters ($a$ and $p$ in Eq.~(10) of \cite{Sheth:1999mn}), which were originally fitted against N-body simulations of $\Lambda$CDM models. The ellipsoidal collapse captures a number of departures from the spherical collapse, but the magnitude of such departures can be different for different models. As a result, one expects these two parameters to be different in Galileon gravity. Secondly, in the paradigm of hierarchichal structure formation, larger objects form by the merging or accretion of smaller objects. As a result, the assumption that the overdense regions remain a top-hat throughout all stages of the collapse may not be a good approximation, specially when it comes to capture the effects of the screening mechanism. For example, consider the formation of a very massive halo; then, in the case of the spherical top-hat collapse, the effects of the screening mechanism only become important in the late stages of the collapse, when the density of the region is sufficiently high. In reality, however, the screening mechanism should start to have an impact on the formation of this very massive halo much earlier, because the halo forms via the continuous merging/accretion of higher-density objects that has been affected by the screening since earlier times.

The investigation of the performance of the excursion set theory formalism in reproducing the simulation results of Galileon gravity models \cite{Barreira:2013eea, Li:2013tda} is the subject of ongoing work.

\section{Results}\label{sec:results}

\begin{table*}
\caption{Summary of the models for which we study the mass function and halo bias. We also show the collapse threshold $\delta_c$ at redshift zero for each of these models.}
\begin{tabular}{@{}lccccccccccc}
\hline\hline
\\
Model  &\ \ $\Omega_{m0}h^2$& \ \ $H(a)$ & $G_{\rm eff}/G$ & $\delta_c\left(z = 0\right)$&\ \ 
\\
\hline
\\
$\Lambda$CDM                    &\ \ $0.137$&\ \  $\Lambda$CDM &  \ \ $1$ & $1.677$&\ \ 
\\
$\rm{QCDM}$                         &\ \ $0.148$&\ \  Eq.~(\ref{eq:tracker_H}) &\ \   $1$& $1.565$&\ \ 
\\
Linear force Quartic Galileon        &\ \ $0.148$ &\ \ Eq.~(\ref{eq:tracker_H}) &  Eq.~(\ref{eq:Geff}) ($\delta \ll 1$) & $1.497$&\ \ 
\\
Full Quartic Galileon             &\ \ $0.148$ &\ \  Eq.~(\ref{eq:tracker_H}) &  \ \ Eq.~(\ref{eq:Geff}) & $1.594$&\ \ 
\\
\hline
\hline
\end{tabular}
\label{table:models}
\end{table*}

In this section we present our results for the halo mass function and halo bias. These will be shown for the WMAP9 best-fitting $\Lambda$CDM model \cite{Hinshaw:2012fq} (dashed black) and three variants of the Quartic Galileon model. The first one is the "full" Quartic Galileon (solid blue) model characterized by Eqs.~(\ref{eq:tracker_H}) and (\ref{eq:Geff}). The second model is a linear force Quartic Galileon model (solid green), in which $G_{\rm eff}/G$ is obtained by taking the limit where $|\delta| \ll 1$ (c.f.~Fig.\ref{fig:Geffmap}). Comparing these two models allows one to measure the effects of the $\delta$-dependence of $G_{\rm eff}$. The last variant is a model we call QCDM (solid red), in which the modifications to gravity are absent $G_{\rm eff}/G = 1$, but the expansion history and matter density are the same as in the other two variants. This model is useful to isolate the changes introduced by the modified gravitational strength, excluding those that arise through the different matter density and modified expansion rate. These models are summarized in Table \ref{table:models}.

\subsection{Evolution of the critical density $\delta_c$}

\begin{figure}
	\centering
	\includegraphics[scale=0.39]{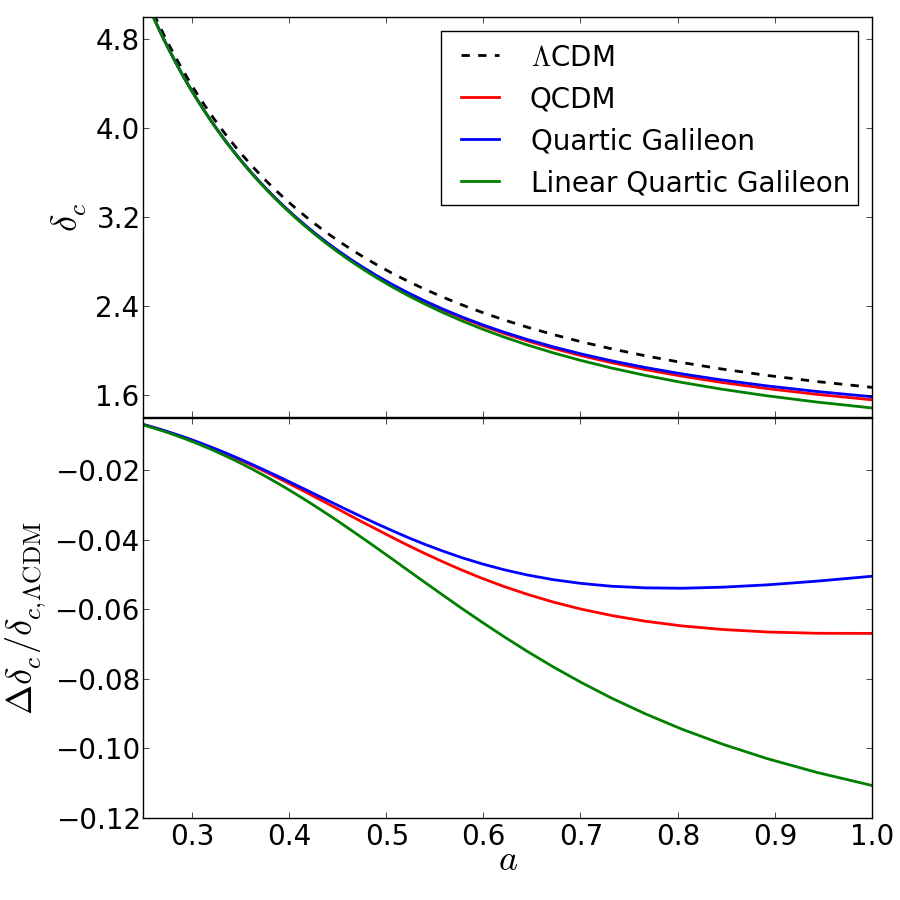}
	\caption{The top panel shows the time evolution of the linearly extrapolated value (assuming $\Lambda$CDM linear growth factor) of the initial critical density for the halo collapse to occur at scale factor $a$ for the $\Lambda$CDM (dashed black), QCDM (solid red), linear force Quartic model (solid green) and full Quartic Galileon model (solid blue). The bottom panel shows the difference relative to $\Lambda$CDM.}
\label{fig:dc}\end{figure}

Before presenting the predictions for the halo mass function and bias, it is instructive to look at the time dependence of $\delta_c$. This is shown in the top panel of Fig.~\ref{fig:dc}, and the bottom panel shows the difference with respect to the $\Lambda$CDM model. Comparing the $\Lambda$CDM and QCDM models, the differences are driven by the different matter densities and by the different expansion rates. The physical matter density, $\Omega_{m0}h^2$, is smaller in the $\Lambda {\rm CDM}$ than in the QCDM model (c.f.~Table \ref{table:models}), so that structure formation is enhanced in the latter. On the other hand, the expansion rate acts as a friction term that slows down structure formation. In Fig.~\ref{fig:w-h-cl-pk}, we saw that $H^{\Lambda {\rm CDM}} > H^{{\rm QCDM}}$ for $0.3 \lesssim a \lesssim 0.8$. During these times, the friction will be higher in $\Lambda$CDM, but lower for all other times. The net effect is that structure formation is suppressed overall in the $\Lambda {\rm CDM}$ model, which is why $\delta_c$ is larger: the initial critical densities have to be larger to account for the slower collapse. One also notes that the relative difference between these two models starts to flatten for $a \gtrsim 0.5$. This is due to the fact that, after this time, $H^{ {\rm QCDM}}$ starts to grow relative to $H^{\Lambda{\rm CDM}}$, which effectively brings the rate of the growth of structure closer together in the two models. 

The differences between the three variants of the Quartic Galileon model are driven only by the differences in $G_{\rm eff}$. In particular, in the linear force model, $\delta_c$ is smaller than in QCDM because gravity is stronger at late times ($a \gtrsim 0.5$) and the initial densities have to be smaller for the collapse to occur at the same epoch. On the other hand, $\delta_c$ is larger in the full Quartic Galileon model compared to QCDM, which means that the collapsing halo feels an overall weaker gravity. This is illustrated by the solid red in the left panel of Fig.~\ref{fig:Geffmap}, which represents the trajectory in $a-\delta$ space of a halo that collapses at the present day. Here, one sees that by the time the fifth force deviates from unity ($a \gtrsim 0.5$), the density of the halo is already sufficiently large for it to feel the negative fifth force ($G_{\rm eff}/G < 1$). It is interesting to note that this brings the full model predictions closer to $\Lambda$CDM because the resulting weaker gravity in the Quartic Galileon model compensates the faster growth driven by the higher matter density.

As we look back in time, the curves of the three Quartic model variants get closer to one another. This is expected because $G_{\rm eff}/G \approx 1$ in the three models for $a \lesssim 0.5$, and therefore there is nothing driving any differences. The solid black line in the left panel of Fig.~\ref{fig:Geffmap} shows the trajectory in $a-\delta$ space of a halo that collapses at $a = 0.5$ ($z = 1$), where one sees that it never crosses any region where $G_{\rm eff}/G$ significantly deviates from unity.

\subsection{Halo mass function}

\begin{figure*}
	\centering
	\includegraphics[scale=0.405]{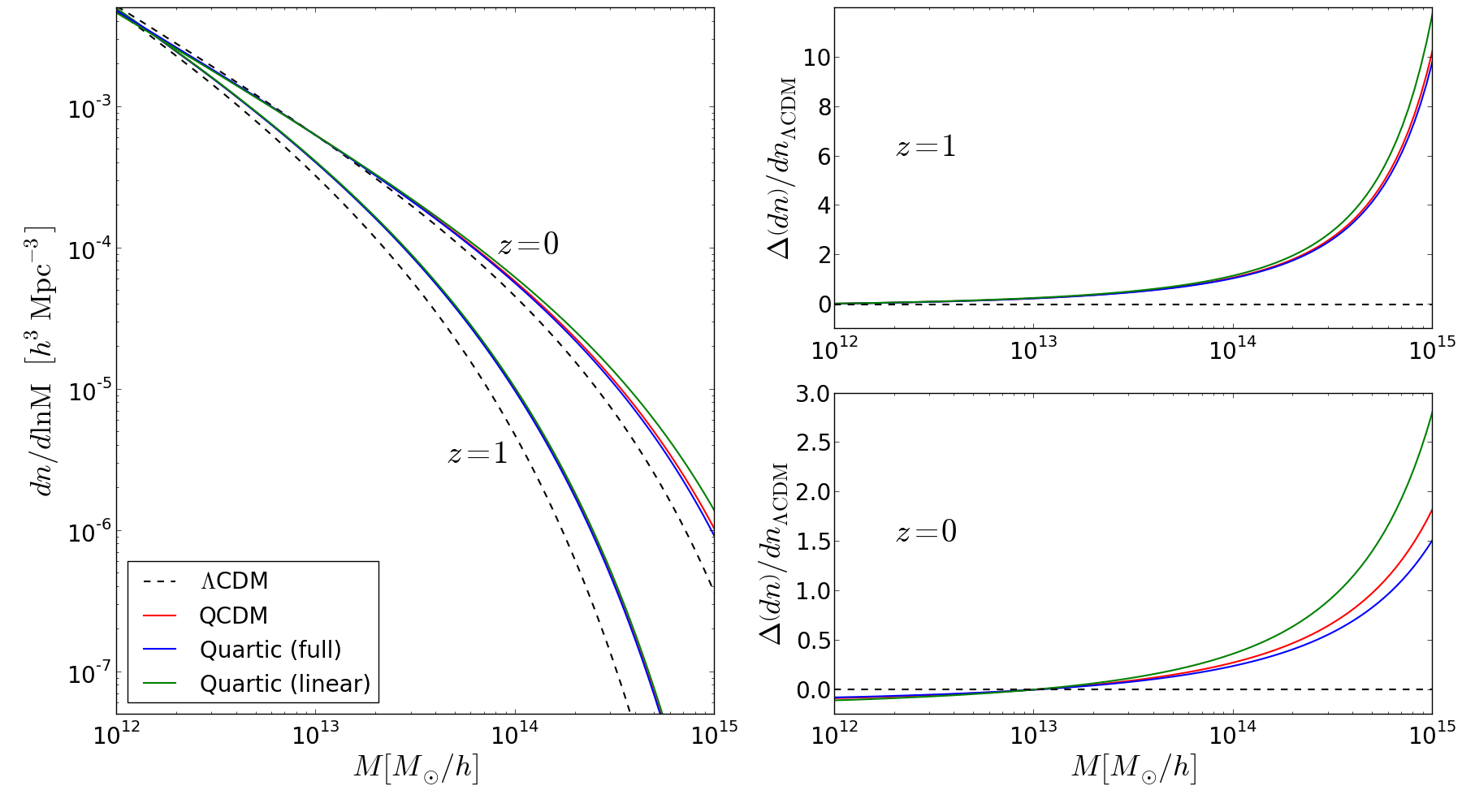}
	\caption{The left panel shows the halo mass function of Eq.~(\ref{eq:mass-function}) for the $\Lambda$CDM (dashed black), QCDM (solid red), linear force Quartic model (solid green) and full Quartic Galileon model (solid blue), for two different redshifts $z = 1$ and $z = 0$. The two panels on the right show the difference relative to $\Lambda$CDM for the two redshifts.}
\label{fig:mass-function}\end{figure*}

The left panel of Fig.~\ref{fig:mass-function} shows the mass function of Eq.~(\ref{eq:mass-function}) predicted for the models of Table \ref{table:models} at redshifts $z = 1$ and $z = 0$. All the models show the standard result that the number density of halos decreases with halo mass. Moreover, the number of the most massive halos progressively increases with time, while the number of lowest mass halos decreases (the latter effect is not seen due to the range
of abundances plotted). This is a result of hierarchical structure formation: with time, low-mass halos merge to form higher mass objects. The two panels on the right show the difference with respect to the $\Lambda$CDM model at each redshift. 

At $z = 0$, all the variants of the Quartic Galileon model predict more massive halos, but fewer low-mass halos compared to $\Lambda$CDM. This is expected because $\delta_c$ is smaller in all the Quartic variants (structure formation is enhanced), which favours the merging of smaller halos into bigger ones. The linear force model has the lowest value of $\delta_c$, and therefore is the model in which these differences to $\Lambda$CDM are more pronounced. In the excursion set language, the explanation is that lower values of $\delta_c$ shift the peak of the first-crossing distribution $f(S)$, Eq.~(\ref{eq:first-crossing}), towards lower $S$, or equivalently, towads higher $M$\footnote{In other words, if $\delta_c$ is lower then the random walks first up-cross the barrier sooner (low $S$), rather than later (high $S$).}. This enhances the abundance of high-mass halos, but suppresses at the same time the number of low-mass halos. The opposite happens in the case of the full Quartic Galileon model. In this case, the $\delta$-dependence of $G_{\rm eff}$ results in an overall weaker gravity for halos that form at $z > 1$, which increases $\delta_c$. As a result, one finds that there are fewer high-mass halos compared to QCDM; the overall weaker gravitational strength felt by the collapsing halos in the Quartic Galileon model compensates slightly the effects of the higher matter density.

The differences between the results for the three variants of the Quartic Galileon model become less pronounced as one looks back in time. This follows from the fact that $G_{\rm eff}/G \sim 1$ at sufficiently early times $a \lesssim 0.5$, and so the models become essentially undistinguishable.

\subsection{Halo bias}

\begin{figure*}
	\centering
	\includegraphics[scale=0.405]{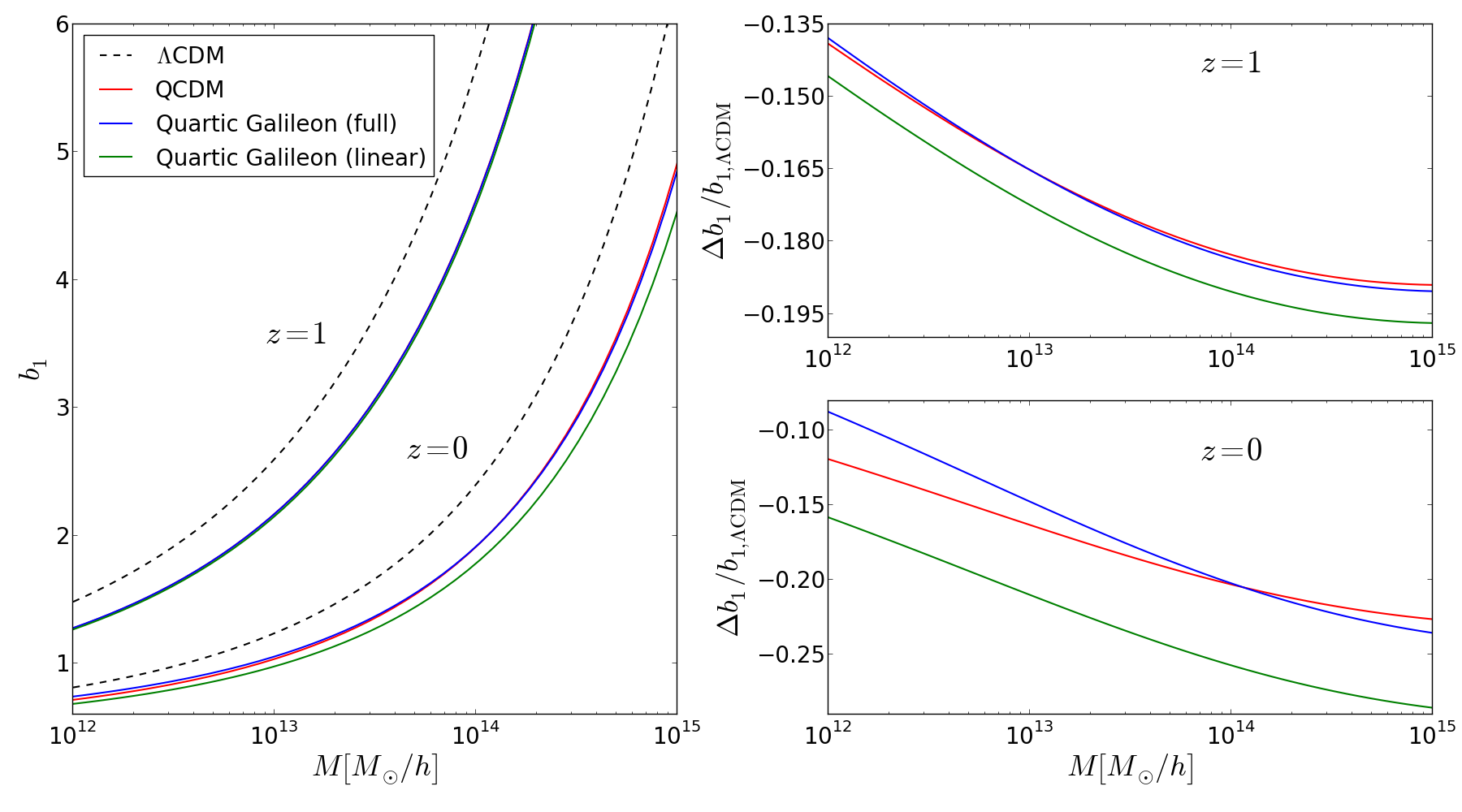}
	\caption{The left panel shows the linear halo bias parameter of Eq.~(\ref{eq:linear-bias}) for the $\Lambda$CDM (dashed black), QCDM (solid red), linear force Quartic model (solid green) and full Quartic Galileon model (solid blue), for two different redshifts $z = 1$ and $z = 0$. The two panels on the right show the difference relative to $\Lambda$CDM for the two redshifts.}
\label{fig:bias}\end{figure*}

Figure \ref{fig:bias} shows the linear halo bias of Eq.~(\ref{eq:linear-bias}) for the models listed in Table \ref{table:models}. The left panel shows the standard qualitative behaviour that high-mass halos cluster more ($b_1 > 1$, biased halos) and low-mass halos cluster less ($b_1 < 1$, anti-biased halos), with respect to the underlying linear dark matter distribution. The mass $M^*$ that separates these two regimes is determined by $S(M^*) = \delta_c^2$. This is a result of hierarchical structure formation which predicts that, in higher-density regions, low-mass halos are more likely to merge to form higher-mass halos. This results in an overabundance of the latter, and in a suppresion of the former. In this paper, we are more interested in the differences between models in this qualitative picture, which are determined by two factors. The first one is the different dynamics of the collapse, and is encapsulated in the different values of $\delta_c$. In particular, larger values of $\delta_c$ lead to higher bias at all mass scales (c.f.~Eq.~(\ref{eq:linear-bias})). The second factor is the different dynamics of the linear evolution of the regions where the halos are forming, and is described by the term $g(z)$ in Eq.~(\ref{eq:linear-bias}). Larger values of $g(z)$ increase the bias for $M > M^*$ ($\delta_c^2/S > 1$), but decrease it for $M < M^*$ ($\delta_c^2/S < 1$).

Following these considerations, the bias is generally smaller in the three variants of the Quartic Galileon model because of the lower value of $\delta_c$ compared to $\Lambda$CDM (c.f.~Fig.~\ref{fig:dc}). Moreover, $g(z)$ is also smaller in the Quartic model variants than in $\Lambda$CDM, which is why the differences become more pronounced (more negative in the right panels of Fig.~\ref{fig:bias}) with increasing mass. Note that, at the low-mass end of the panels, the changes in $\delta_c$ and $g(z)$ in the Quartic Galileon model variants with respect to $\Lambda$CDM shift the bias in opposite directions. However, the bias is still smaller in any of the Quartic model variants for low-mass halos, which shows that the changes in $\delta_c$ play the dominant role over $g(z)$ in determining the differences between these models and $\Lambda$CDM. The linear force Quartic Galileon model is that where the halos are less biased at all mass scales because it is the model where gravity is strongest (lowest $\delta_c$ value). One also notes that the difference between the linear force model and QCDM becomes slighlty more pronounced with halo mass, since $g(z)$ is smaller in the former compared to the latter. The case of the full Galileon model is perhaps the most interesting one due to the $\delta$-dependence in $G_{\rm eff}$. The dashed black and solid red lines in Fig.~\ref{fig:Geffmap}, show, respectively, the trajectories in $a-\delta$ space of a linear overdensity that has $\delta= 0.01$ and of a halo that collapses today. One sees that at late times, $a \gtrsim 0.5$, the spherical halo feels an overall weaker gravity compared to QCDM (larger $\delta_c$), but that the larger region, where the density is small, feels an overall stronger gravity (smaller $g(z)$) compared to QCDM. As a result, in light of the changes driven by $\delta_c$ and $g(z)$, one has that at the high-mass end, these effects shift the linear bias in opposite directions, and the net result is an approximate cancellation, w.r.t.~QCDM. On the other hand, at lower mass scales, the changes in $\delta_c$ and $g(z)$ both shift the bias upwards, which therefore becomes larger in the full Quartic model compared to QCDM. In particular, at the lowest mass scales shown, the bias in the full Quartic Galileon approaches that of the $\Lambda$CDM model.

Similarly to what we have seen in the previous subsections, the bias of halos that form at $z \gtrsim 1$ ($a \lesssim 0.5$) tend to become the same in the three variants of the Quartic Galileon model, because at these early times the three models are undistinguishable.

\section{Conclusions}\label{sec:conclusion}

We have studied the nonlinear growth of large scale structure in the Quartic and Quintic Galileon gravity models. For this, we have derived the nonlinear Einstein and Galileon field equations assuming spherical symmetry, the quasi-static approximation and the weak field approximation. Using these equations, we studied the spherical collapse of matter overdensities and used the excursion set formalism to predict the halo mass function and halo bias.

In these models, the spatial gradient, $\varphi,_{\chi}/\chi$, of the Galileon field contributes to the fifth force and its value is obtained by solving a nonlinear algebraic equation, Eq.~(\ref{eq:eom_sph_alg}). In the case of the Quintic Galileon model, we demonstrated that, if $\delta$ is above order unity, then the field equations do not admit physical solutions. This is because the branch of solutions for $\varphi,_{\chi}/\chi$ that vanishes when $\delta = 0$ (which characterizes the physical behaviour) becomes a complex root of Eq.~(\ref{eq:eom_sph_alg}) (c.f.~Fig.~\ref{fig:quintic-branches}). Evidently, the impossibility of evaluating the fifth force at these densities prevents the study of the spherical collapse in the Quintic model. We have argued that it is unlikely that relaxing the quasi-static approximation can help to solve this problem. In the case of the Quartic model, we showed that the physical solutions exist on sufficiently high density regions, but do not exist at late times if the density is sufficiently low (c.f.~Fig.~\ref{fig:Geffmap}). However, here the case is likely to be related to the breakdown of the approximations adopted, which, indeed, are not expected to hold in very low density regions. Fortunately, for the halo spherical collapse study we presented, such low densities are not important and the problem is irrelevant.

We have seen that at early times ($z \gtrsim 1$, or $a \lesssim 0.5$) in the Quartic model, the effective gravitational strength is $G_{\rm eff}/G \approx 1$ (c.f.~Fig.~\ref{fig:Geffmap}). With time, $G_{\rm eff}/G$ increases if the density is small ($\delta \ll 1$), and at the present day one has $G_{\rm eff}/G \approx 1.4$. On the other hand, if the density is of order unity or above, the value of $G_{\rm eff}/G$ decreases, and at the present day it is $G_{\rm eff}/G \approx 0.6$ for $\delta \gg 1$. Thus, the modifications to gravity are not completely screened on high densities. The reason for this is that the Galileon field contributes to $G_{\rm eff}/G$ not only through its spatial gradients, but also through the background time evolution that multiplies the standard gravity terms (c.f.~Eqs.~(\ref{eq:poisson}), (\ref{eq:slip}) and (\ref{eq:eom_sph})). The latter will still be present, even at high densities where the Vainshtein screening successfully suppresses the spatial gradients. 

A negative fifth force at high densities (small length scales) can have interesting observational consequences. For instance, the first results from N-body simulations of the Quartic Galileon model have been recently presented in \cite{Li:2013tda}. There, it was shown that the gravitational potential of the halos becomes shallower, which favours lower concentration, provided that the average velocity of dark matter particles does not decrease too much \cite{Li:2013tda}. Also, given the same kinematical data, i.e. galaxy rotation curves or galaxy velocity dispersion in clusters, or the same X-ray or Sunyaev-Zeldovich signals from difuse gas in clusters, then a lower value of the gravitational strength would result in a higher estimated dynamical mass for dark matter halos (see e.g. \cite{Schmidt:2010jr}). However, despite all this interesting phenomenology, it is important to bear in mind that the weaker and time-varying gravitational strength in the Quartic model is in fact putting the model into huge tension with the local Solar System tests of gravity. In this paper, we focused on the cosmological properties of this model, but if it turns out that the time variation of $G_{\rm eff}$ on local scales cannot be "cured", then the model is observationally ruled out. In our view, such a "cure" is also not likely to come from the relaxation of the quasi-static or weak-field approximations.

We have seen that the way the Galileon modifies the dynamics of the spherical collapse of overdensities is sensitive to the density of the halo, but not to its size or mass. In other words, the critical density for collapse, $\delta_c$, which determines the height of the barrier in the excursion sets, is "flat". Our results show that $\delta_c$ becomes smaller when one changes from the $\Lambda$CDM to the QCDM model. This is mostly because of the higher matter density in the latter, which makes the halos collapse faster. In the linear force Quartic model, the fifth force is non-negligible and positive for $a \gtrsim 0.5$, which further decreases the critical density $\delta_c$ because it further boosts the collapse of the halos. In the case of the full Quartic model, the spherical halos feel a negative fifth force in the collapsing stages for $a \gtrsim 0.5$ (c.f.~Fig~\ref{fig:Geffmap}), which makes $\delta_c$ larger than in QCDM. 

Using the excursion set theory formalism, we have computed the mass function for the Quartic Galileon model. We have seen that, at $z = 0$, all the variants of the Quartic Galileon model we studied predict more high-mass halos than $\Lambda$CDM, but fewer low-mass halos (c.~f.~Fig.~\ref{fig:mass-function}). This is mainly due to the higher matter density in the Galileon models, which enhances structure formation, and thus makes it easier for smaller halos to merge into more massive ones. In the case of the linear force model, the enhanced gravitational strength leads to more halos at the high-mass end. On the other hand, in the full Quartic Galileon model, the halos that collapse at $z = 0$ feel overall a weaker gravity, and therefore, the model predicts fewer high-mass halos compared to QCDM. In the full model, the fact that the screening mechanism cannot suppress all the modifications to gravity, compensates the boosting effect on halo formation driven by the higher matter density. This brings the model predictions slightly closer to those in $\Lambda$CDM. However, the number of high-mass halos in the full Quartic Galileon model is still considerably larger than in $\Lambda$CDM, which shows that the different matter density plays the dominant role in determining the different halo abundances in these two models.

Our results for the mass function in the Quartic Galileon model can be interpreted in the context of observations that have been claimed to be in tension with standard $\Lambda$CDM. In particular, X-ray and lensing measurements have detected galaxy clusters that seem to be too massive and to have formed too early, compared to what one would expect from $\Lambda$CDM \cite{Enqvist:2010bg, Jee:2011az, Hoyle:2010ce, Holz:2010ck}. In addition, the detection in the CMB of the ISW signal associated with super clusters (hot spots) and super voids (cold spots) has also been claimed to suggest that clusters were more massive and voids emptier in the past, and that they existed in higher number than in the $\Lambda$CDM expectation \cite{Granett:2008ju, Papai:2010gd, Nadathur:2011iu, Flender:2012wu}. However, \cite{Hotchkiss:2011ms, Harrison:2011ep, Hoyle:2011pj, Waizmann:2012wx, Waizmann:2011xu} have shown that the tension with the LCDM model is ameliorated by using an appropriate estimate of the statistics of rare structures, such as massive clusters or large voids (see also \cite{Yaryura11052011}). Such apparently unusual objects become more common in models with standard gravity if the initial distribution of density fluctuations is non-Gaussian \cite{Matarrese:2000iz, Sefusatti:2006eu}. The main difficulty with this solution is that the values of $f_{\rm NL}$ required to explain these observations are generally too large to be compatible with other CMB constraints (see however \cite{Trindade:2013lga}). On the other hand, our qualitative results for the mass function of the Quartic Galileon model show that the enhanced rate of structure formation (driven mainly by the higher matter density) helps to produce more high-mass objects relative to $\Lambda$CDM, whilst at the same time being compatible with the CMB temperature power spectrum.

We have also studied the halo bias in these models. In particular, we saw that the bias in the variants of the Quartic Galileon model is generally smaller than in $\Lambda$CDM, and that the differences become more pronounced with increasing halo mass. Within these variants, the bias is the lowest in the linear force model at all mass scales. In the case of the full Quartic model, at $z = 0$, the bias is larger for low-halo masses compared to QCDM, but the two predictions become comparable at higher halo masses (c.f. Fig.~\ref{fig:bias}). In the Galileon model, the bias of dark matter halos is determined by the interplay of the different values of $\delta_c$ and $g(z)$ (c.f.~Eq.~(\ref{eq:linear-bias})). While larger values of $\delta_c$ can only lead to higher bias, larger values of $g(z)$, (which corresponds to slower clustering of the underlying dark matter field), make high-mass halos more biased, but low-mass halos less biased. For instance, in the case of the full Quartic model, for high-mass halos, the weaker gravity felt by the halos and the faster evolution of the underlying matter density relative to QCDM compensate each other, which is why the two models predict roughly the same bias. However, for low-mass halos, the faster evolution of the linear density field will also push the linear bias to higher values, and consequently, the bias is noticeably larger. Our results for the halo bias show that the changes introduced by the modified gravitational strength (measured by the differences between the Quartic model variants) are sub-dominant over those introduced by the higher matter density w.r.t.~$\Lambda$CDM (measured by the differences between QCDM and $\Lambda$CDM). 

From the point of view of the linear matter power spectrum of Fig.~\ref{fig:w-h-cl-pk}, the lower halo bias in the Quartic Galileon model means that the amount by which its linear theory curve should be shifted upwards is smaller compared to $\Lambda$CDM. We note that the higher matter density in the Galileon model, which results in a higher clustering amplitude, also contributes significantly to the lower linear bias prediction. Moreover, the data in the bottom right panel of Fig.~\ref{fig:w-h-cl-pk} shows the host halo power spectrum of LRGs, which are thought to typically reside in halos with an effective mass $\approx 10^{14} M_{\odot}/h$ \cite{Wake:2008mf, Zheng:2008np, Sawangwit:2009bg}. Our linear bias results show that these high-halo masses are precisely those for which the differences between the Quartic Galileon and $\Lambda$CDM models are more pronounced. At these halo masses, the bias in the Quartic model can be a few tenths of percent smaller relative to $\Lambda$CDM. However, judging from Fig.~\ref{fig:w-h-cl-pk}, this may not be enough to fully ease the tension of the model with galaxy clustering data. As we mentioned before, our study is mostly qualitative and we shall leave more quantitative results for future work. In particular, it would be interesting to interpret the different linear halo bias in the Galileon model in light of current empirical halo ocupation models \cite{Zheng:2008np}, and halo-weighting schemes to reconstruct the mass distribution from galaxy surveys \cite{Cai:2010gz, Hamaus:2010im}.

In conclusion, our work shows that the modifications of gravity that arise in the Quartic Galileon gravity model can have interesting and testable predictions for the large scale structure in the Universe. In the present paper, our goal was to present a simplified study in order to get some first impressions on the phenomenology of the model and to help plan future studies. For instance, the first N-body simulations of the Quartic Galileon model were presented in \cite{Li:2013tda}, and it woud be interesting to use the excursion set methodology presented here to develop a halo model \cite{Cooray:2002dia} for Galileon gravity, and see how it compares with the results from high-resolution simulations (see e.g.~\cite{Schmidt:2009yj}). In principle, our equations can also be used to compute the halo mergers trees with less work than by using N-body simulations, which could be used to study galaxy formation in the Galileon model. Such studies would be important to help devise and interpret the results of future large-scale galaxy surveys, the goals of which include testing the laws of gravity on cosmological scales.

\begin{acknowledgments}

We thank Yan-Chuan Cai and Antony Lewis for comments on an earlier draft of this paper, and Shaun Cole for helpful discussions. AB is supported by FCT-Portugal through grant SFRH/BD/75791/2011. BL is supported by the Royal Astronomical Society and Durham University. This work has been partially supported by the European Union FP7  ITN INVISIBLES (Marie Curie Actions, PITN- GA-2011- 289442) and STFC.

\end{acknowledgments}

\appendix

\section{Time-dependent coefficients of the spherically symmetric nonlinear equations}

The coefficients $A_i$, $B_i$, $C_i$ in Eqs.~(\ref{eq:poisson}), (\ref{eq:slip}) and (\ref{eq:eom_sph}), are given, respectively, by

\bq
A_1 &=& -2c_3\xi\varphi' -12c_4\xi^2\varphi' - 15c_5\xi^3\varphi'\\
A_2 &=& 6c_4\xi\varphi' + 12c_5\xi^2\varphi' \\
A_3 &=& -4c_5\xi\varphi' \\
A_4 &=& 2 - 3c_4\xi^2\varphi'^2 - 6c_5\xi^3\varphi'^2\\
A_5 &=& 6c_5\xi^2\varphi'^2
\eq

\bq
B_0 &=& -2 - c_4\xi^2\varphi'^2 + 3c_5\xi^3\varphi''\varphi'\\
B_1 &=& 4c_4\left(-\xi^2\varphi' - \frac{3}{2}\xi^2\varphi''\right) \nonumber \\
&&+ 6c_5\left(-\frac{3}{2}\xi^3\varphi'' - \xi^3\varphi'\right)\\
B_2 &=& 2c_4\xi\varphi' + 6c_5\xi^2\varphi''\\
B_3 &=& -2 + 3c_4\xi^2\varphi'^2 + 6c_5\xi^3\varphi'^2\\
B_4 &=& 6c_5\xi^2\varphi'^2 
\eq

\bq
C_1 &=& -c_2 - 2c_3\left(4\xi + \xi\frac{\varphi''}{\varphi'}\right) -c_4\left(26\xi^2 + 6\xi^2\frac{\varphi''}{\varphi'}\right)  \nonumber \\
&& - 6c_5\left(4\xi^3 + \xi^3\frac{\varphi''}{\varphi}\right) \\
C_2 &=& 4c_3 + 6c_4\left(2\xi + \xi\frac{\varphi''}{\varphi'}\right) \nonumber \\
&&+ 6c_5\left(2\xi^2 + \xi^2\frac{\varphi''}{\varphi'}\right)\\
C_3 &=& -4c_4 - 4c_5\xi\frac{\varphi''}{\varphi'} \\
C_4 &=& 2c_4\left(3\xi^2\varphi'' + 2\xi^2\varphi'\right) \nonumber \\ 
&&+ 3c_5\left(3\xi^3\varphi'' + 2\xi^3\varphi'\right)\\
C_5 &=& -2c_3\xi\varphi' - 12c_4\xi^2\varphi' - 15c_5\xi^3\varphi' \\
C_6 &=& -4c_4\xi\varphi' - 12c_5\xi^2\varphi''\\
C_7 &=& 12c_4\xi\varphi' + 24c_5\xi^2\varphi'\\
C_8 &=& -12c_5\xi\varphi'\\
C_9 &=& -6c_5\xi^2\varphi'^2
\eq
where we have assumed the tracker background solution.

\bibliography{quartic-excursion.bib}

%Merlin.mbs v4.21 2009-07-09.
\begin{thebibliography}{10}%
\makeatletter
\providecommand \@ifxundefined [1]{%
 \ifx #1\undefined \expandafter \@firstoftwo
 \else \expandafter \@secondoftwo
\fi
}%
\providecommand \@ifnum [1]{%
 \ifnum #1\expandafter \@firstoftwo
 \else \expandafter \@secondoftwo
\fi
}%
\providecommand \enquote [1]{``#1''}%
\providecommand \bibnamefont  [1]{#1}%
\providecommand \bibfnamefont [1]{#1}%
\providecommand \citenamefont [1]{#1}%
\providecommand\href[0]{\@sanitize\@href}%
\providecommand\@href[1]{\endgroup\@@startlink{#1}\endgroup\@@href}%
\providecommand\@@href[1]{#1\@@endlink}%
\providecommand \@sanitize [0]{\begingroup\catcode`\&12\catcode`\#12\relax}%
\@ifxundefined \pdfoutput {\@firstoftwo}{%
 \@ifnum{\z@=\pdfoutput}{\@firstoftwo}{\@secondoftwo}%
}{%
 \providecommand\@@startlink[1]{\leavevmode\special{html:<a href="#1">}}%
 \providecommand\@@endlink[0]{\special{html:</a>}}%
}{%
 \providecommand\@@startlink[1]{%
  \leavevmode
  \pdfstartlink
   attr{/Border[0 0 1 ]/H/I/C[0 1 1]}%
   user{/Subtype/Link/A<</Type/Action/S/URI/URI(#1)>>}%
  \relax
 }%
 \providecommand\@@endlink[0]{\pdfendlink}%
}%
\providecommand \url  [0]{\begingroup\@sanitize \@url }%
\providecommand \@url [1]{\endgroup\@href {#1}{\urlprefix}}%
\providecommand \urlprefix [0]{URL }%
\providecommand \Eprint[0]{\href }%
\@ifxundefined \urlstyle {%
  \providecommand \doi [1]{doi:\discretionary{}{}{}#1}%
}{%
  \providecommand \doi [0]{doi:\discretionary{}{}{}\begingroup
  \urlstyle{rm}\Url }%
}%
\providecommand \doibase [0]{http://dx.doi.org/}%
\providecommand \Doi[1]{\href{\doibase#1}}%
\providecommand \bibAnnote [3]{%
  \BibitemShut{#1}%
  \begin{quotation}\noindent
    \textsc{Key:}\ #2\\\textsc{Annotation:}\ #3%
  \end{quotation}%
}%
\providecommand \bibAnnoteFile [2]{%
  \IfFileExists{#2}{\bibAnnote {#1} {#2} {\input{#2}}}{}%
}%
\providecommand \typeout [0]{\immediate \write \m@ne }%
\providecommand \selectlanguage [0]{\@gobble}%
\providecommand \bibinfo [0]{\@secondoftwo}%
\providecommand \bibfield [0]{\@secondoftwo}%
\providecommand \translation [1]{[#1]}%
\providecommand \BibitemOpen[0]{}%
\providecommand \bibitemStop [0]{}%
\providecommand \bibitemNoStop [0]{.\EOS\space}%
\providecommand \EOS [0]{\spacefactor3000\relax}%
\providecommand \BibitemShut [1]{\csname bibitem#1\endcsname}%
%</preamble>
\bibitem{Hinshaw:2012fq}%
  \BibitemOpen
  \bibfield{author}{%
  \bibinfo {author} {\bibfnamefont{G.}~\bibnamefont{Hinshaw}}, \bibinfo
  {author} {\bibfnamefont{D.}~\bibnamefont{Larson}}, \bibinfo {author}
  {\bibfnamefont{E.}~\bibnamefont{Komatsu}}, \bibinfo {author}
  {\bibfnamefont{D.}~\bibnamefont{Spergel}}, \bibinfo {author}
  {\bibfnamefont{C.}~\bibnamefont{Bennett}}, \emph{et~al.}}%
   (\bibinfo {year} {2012}),\
  \Eprint{http://arxiv.org/abs/1212.5226}{arXiv:1212.5226 [astro-ph.CO]}%
  \bibAnnoteFile{NoStop}{Hinshaw:2012fq}%
%%CITATION = ARXIV:1212.5226;%%
\bibitem{Ade:2013zuv}%
  \BibitemOpen
  \bibfield{author}{%
  \bibinfo {author} {\bibfnamefont{P.}~\bibnamefont{Ade}} \emph{et~al.}
  (\bibinfo {collaboration} {Planck Collaboration})}%
   (\bibinfo {year} {2013}),\
  \Eprint{http://arxiv.org/abs/1303.5076}{arXiv:1303.5076 [astro-ph.CO]}%
  \bibAnnoteFile{NoStop}{Ade:2013zuv}%
%%CITATION = ARXIV:1303.5076;%%
\bibitem{Guy:2010bc}%
  \BibitemOpen
  \bibfield{author}{%
  \bibinfo {author} {\bibfnamefont{J.}~\bibnamefont{Guy}}, \bibinfo {author}
  {\bibfnamefont{M.}~\bibnamefont{Sullivan}}, \bibinfo {author}
  {\bibfnamefont{A.}~\bibnamefont{Conley}}, \bibinfo {author}
  {\bibfnamefont{N.}~\bibnamefont{Regnault}}, \bibinfo {author}
  {\bibfnamefont{P.}~\bibnamefont{Astier}}, \emph{et~al.},\ }%
  \bibfield{journal}{%
  \Doi{10.1051/0004-6361/201014468}{\bibinfo {journal} {Astron.Astrophys.}}\ }%
  \textbf{\bibinfo {volume} {523}},\ \bibinfo {pages} {A7} (\bibinfo {year}
  {2010}),\ \Eprint{http://arxiv.org/abs/1010.4743}{arXiv:1010.4743
  [astro-ph.CO]}%
  \bibAnnoteFile{NoStop}{Guy:2010bc}%
%%CITATION = ARXIV:1010.4743;%%
\bibitem{Suzuki:2011hu}%
  \BibitemOpen
  \bibfield{author}{%
  \bibinfo {author} {\bibfnamefont{N.}~\bibnamefont{Suzuki}}, \bibinfo {author}
  {\bibfnamefont{D.}~\bibnamefont{Rubin}}, \bibinfo {author}
  {\bibfnamefont{C.}~\bibnamefont{Lidman}}, \bibinfo {author}
  {\bibfnamefont{G.}~\bibnamefont{Aldering}}, \bibinfo {author}
  {\bibfnamefont{R.}~\bibnamefont{Amanullah}}, \emph{et~al.},\ }%
  \bibfield{journal}{%
  \Doi{10.1088/0004-637X/746/1/85}{\bibinfo {journal} {Astrophys.J.}}\ }%
  \textbf{\bibinfo {volume} {746}},\ \bibinfo {pages} {85} (\bibinfo {year}
  {2012}),\ \Eprint{http://arxiv.org/abs/1105.3470}{arXiv:1105.3470
  [astro-ph.CO]}%
  \bibAnnoteFile{NoStop}{Suzuki:2011hu}%
%%CITATION = ARXIV:1105.3470;%%
\bibitem{Efstathiou:2001cw}%
  \BibitemOpen
  \bibfield{author}{%
  \bibinfo {author} {\bibfnamefont{G.}~\bibnamefont{Efstathiou}} \emph{et~al.}
  (\bibinfo {collaboration} {2dFGRS Collaboration}),\ }%
  \bibfield{journal}{%
  \bibinfo {journal} {Mon.Not.Roy.Astron.Soc.}\ }%
  \textbf{\bibinfo {volume} {330}},\ \bibinfo {pages} {L29} (\bibinfo {year}
  {2002}),\
  \Eprint{http://arxiv.org/abs/astro-ph/0109152}{arXiv:astro-ph/0109152
  [astro-ph]}%
  \bibAnnoteFile{NoStop}{Efstathiou:2001cw}%
%%CITATION = ASTRO-PH/0109152;%%
\bibitem{Sánchez11122009}%
  \BibitemOpen
  \bibfield{author}{%
  \bibinfo {author} {\bibfnamefont{A.~G.}\ \bibnamefont{Sanchez}}, \bibinfo
  {author} {\bibfnamefont{M.}~\bibnamefont{Crocce}}, \bibinfo {author}
  {\bibfnamefont{A.}~\bibnamefont{Cabre}}, \bibinfo {author}
  {\bibfnamefont{C.~M.}\ \bibnamefont{Baugh}},\ and\ \bibinfo {author}
  {\bibfnamefont{E.}~\bibnamefont{Gaztanaga}},\ }%
  \bibfield{journal}{%
  \Doi{10.1111/j.1365-2966.2009.15572.x}{\bibinfo {journal} {Monthly Notices of
  the Royal Astronomical Society}}\ }%
  \textbf{\bibinfo {volume} {400}},\ \bibinfo {pages} {1643} (\bibinfo {year}
  {2009})%
  \bibAnnoteFile{NoStop}{Sánchez11122009}%
\bibitem{Percival:2009xn}%
  \BibitemOpen
  \bibfield{author}{%
  \bibinfo {author} {\bibfnamefont{W.~J.}\ \bibnamefont{Percival}}
  \emph{et~al.} (\bibinfo {collaboration} {SDSS Collaboration}),\ }%
  \bibfield{journal}{%
  \Doi{10.1111/j.1365-2966.2009.15812.x}{\bibinfo {journal}
  {Mon.Not.Roy.Astron.Soc.}}\ }%
  \textbf{\bibinfo {volume} {401}},\ \bibinfo {pages} {2148} (\bibinfo {year}
  {2010}),\ \Eprint{http://arxiv.org/abs/0907.1660}{arXiv:0907.1660
  [astro-ph.CO]}%
  \bibAnnoteFile{NoStop}{Percival:2009xn}%
%%CITATION = ARXIV:0907.1660;%%
\bibitem{Beutler:2011hx}%
  \BibitemOpen
  \bibfield{author}{%
  \bibinfo {author} {\bibfnamefont{F.}~\bibnamefont{Beutler}}, \bibinfo
  {author} {\bibfnamefont{C.}~\bibnamefont{Blake}}, \bibinfo {author}
  {\bibfnamefont{M.}~\bibnamefont{Colless}}, \bibinfo {author}
  {\bibfnamefont{D.~H.}\ \bibnamefont{Jones}}, \bibinfo {author}
  {\bibfnamefont{L.}~\bibnamefont{Staveley-Smith}}, \emph{et~al.},\ }%
  \bibfield{journal}{%
  \Doi{10.1111/j.1365-2966.2011.19250.x}{\bibinfo {journal}
  {Mon.Not.Roy.Astron.Soc.}}\ }%
  \textbf{\bibinfo {volume} {416}},\ \bibinfo {pages} {3017} (\bibinfo {year}
  {2011}),\ \Eprint{http://arxiv.org/abs/1106.3366}{arXiv:1106.3366
  [astro-ph.CO]}%
  \bibAnnoteFile{NoStop}{Beutler:2011hx}%
%%CITATION = ARXIV:1106.3366;%%
\bibitem{Reid:2012sw}%
  \BibitemOpen
  \bibfield{author}{%
  \bibinfo {author} {\bibfnamefont{B.~A.}\ \bibnamefont{Reid}}, \bibinfo
  {author} {\bibfnamefont{L.}~\bibnamefont{Samushia}}, \bibinfo {author}
  {\bibfnamefont{M.}~\bibnamefont{White}}, \bibinfo {author}
  {\bibfnamefont{W.~J.}\ \bibnamefont{Percival}}, \bibinfo {author}
  {\bibfnamefont{M.}~\bibnamefont{Manera}}, \emph{et~al.}}%
   (\bibinfo {year} {2012}),\
  \Eprint{http://arxiv.org/abs/1203.6641}{arXiv:1203.6641 [astro-ph.CO]}%
  \bibAnnoteFile{NoStop}{Reid:2012sw}%
%%CITATION = ARXIV:1203.6641;%%
\bibitem{Reid:2009xm}%
  \BibitemOpen
  \bibfield{author}{%
  \bibinfo {author} {\bibfnamefont{B.~A.}\ \bibnamefont{Reid}}, \bibinfo
  {author} {\bibfnamefont{W.~J.}\ \bibnamefont{Percival}}, \bibinfo {author}
  {\bibfnamefont{D.~J.}\ \bibnamefont{Eisenstein}}, \bibinfo {author}
  {\bibfnamefont{L.}~\bibnamefont{Verde}}, \bibinfo {author}
  {\bibfnamefont{D.~N.}\ \bibnamefont{Spergel}}, \emph{et~al.},\ }%
  \bibfield{journal}{%
  \Doi{10.1111/j.1365-2966.2010.16276.x}{\bibinfo {journal}
  {Mon.Not.Roy.Astron.Soc.}}\ }%
  \textbf{\bibinfo {volume} {404}},\ \bibinfo {pages} {60} (\bibinfo {year}
  {2010}),\ \Eprint{http://arxiv.org/abs/0907.1659}{arXiv:0907.1659
  [astro-ph.CO]}%
  \bibAnnoteFile{NoStop}{Reid:2009xm}%
%%CITATION = ARXIV:0907.1659;%%
\bibitem{Anderson:2012sa}%
  \BibitemOpen
  \bibfield{author}{%
  \bibinfo {author} {\bibfnamefont{L.}~\bibnamefont{Anderson}}, \bibinfo
  {author} {\bibfnamefont{E.}~\bibnamefont{Aubourg}}, \bibinfo {author}
  {\bibfnamefont{S.}~\bibnamefont{Bailey}}, \bibinfo {author}
  {\bibfnamefont{D.}~\bibnamefont{Bizyaev}}, \bibinfo {author}
  {\bibfnamefont{M.}~\bibnamefont{Blanton}}, \emph{et~al.},\ }%
  \bibfield{journal}{%
  \Doi{10.1093/mnras/sts084}{\bibinfo {journal} {Mon.Not.Roy.Astron.Soc.}}\ }%
  \textbf{\bibinfo {volume} {428}},\ \bibinfo {pages} {1036} (\bibinfo {year}
  {2013}),\ \Eprint{http://arxiv.org/abs/1203.6594}{arXiv:1203.6594
  [astro-ph.CO]}%
  \bibAnnoteFile{NoStop}{Anderson:2012sa}%
%%CITATION = ARXIV:1203.6594;%%
\bibitem{Sanchez:2012sg}%
  \BibitemOpen
  \bibfield{author}{%
  \bibinfo {author} {\bibfnamefont{A.~G.}\ \bibnamefont{Sanchez}}, \bibinfo
  {author} {\bibfnamefont{C.}~\bibnamefont{Scoccola}}, \bibinfo {author}
  {\bibfnamefont{A.}~\bibnamefont{Ross}}, \bibinfo {author}
  {\bibfnamefont{W.}~\bibnamefont{Percival}}, \bibinfo {author}
  {\bibfnamefont{M.}~\bibnamefont{Manera}}, \emph{et~al.}}%
   (\bibinfo {year} {2012}),\
  \Eprint{http://arxiv.org/abs/1203.6616}{arXiv:1203.6616 [astro-ph.CO]}%
  \bibAnnoteFile{NoStop}{Sanchez:2012sg}%
%%CITATION = ARXIV:1203.6616;%%
\bibitem{Clifton:2011jh}%
  \BibitemOpen
  \bibfield{author}{%
  \bibinfo {author} {\bibfnamefont{T.}~\bibnamefont{Clifton}}, \bibinfo
  {author} {\bibfnamefont{P.~G.}\ \bibnamefont{Ferreira}}, \bibinfo {author}
  {\bibfnamefont{A.}~\bibnamefont{Padilla}},\ and\ \bibinfo {author}
  {\bibfnamefont{C.}~\bibnamefont{Skordis}},\ }%
  \bibfield{journal}{%
  \Doi{10.1016/j.physrep.2012.01.001}{\bibinfo {journal} {Phys.Rept.}}\ }%
  \textbf{\bibinfo {volume} {513}},\ \bibinfo {pages} {1} (\bibinfo {year}
  {2012}),\ \Eprint{http://arxiv.org/abs/1106.2476}{arXiv:1106.2476
  [astro-ph.CO]}%
  \bibAnnoteFile{NoStop}{Clifton:2011jh}%
%%CITATION = ARXIV:1106.2476;%%
\bibitem{PhysRevD.79.064036}%
  \BibitemOpen
  \bibfield{author}{%
  \bibinfo {author} {\bibfnamefont{A.}~\bibnamefont{Nicolis}}, \bibinfo
  {author} {\bibfnamefont{R.}~\bibnamefont{Rattazzi}},\ and\ \bibinfo {author}
  {\bibfnamefont{E.}~\bibnamefont{Trincherini}},\ }%
  \bibfield{journal}{%
  \Doi{10.1103/PhysRevD.79.064036}{\bibinfo {journal} {Phys. Rev. D}}\ }%
  \textbf{\bibinfo {volume} {79}},\ \bibinfo {pages} {064036} (\bibinfo {year}
  {2009})%
  \bibAnnoteFile{NoStop}{PhysRevD.79.064036}%
\bibitem{PhysRevD.79.084003}%
  \BibitemOpen
  \bibfield{author}{%
  \bibinfo {author} {\bibfnamefont{C.}~\bibnamefont{Deffayet}}, \bibinfo
  {author} {\bibfnamefont{G.}~\bibnamefont{Esposito-Far\`ese}},\ and\ \bibinfo
  {author} {\bibfnamefont{A.}~\bibnamefont{Vikman}},\ }%
  \bibfield{journal}{%
  \Doi{10.1103/PhysRevD.79.084003}{\bibinfo {journal} {Phys. Rev. D}}\ }%
  \textbf{\bibinfo {volume} {79}},\ \bibinfo {pages} {084003} (\bibinfo {year}
  {2009})%
  \bibAnnoteFile{NoStop}{PhysRevD.79.084003}%
\bibitem{Deffayet:2009mn}%
  \BibitemOpen
  \bibfield{author}{%
  \bibinfo {author} {\bibfnamefont{C.}~\bibnamefont{Deffayet}}, \bibinfo
  {author} {\bibfnamefont{S.}~\bibnamefont{Deser}},\ and\ \bibinfo {author}
  {\bibfnamefont{G.}~\bibnamefont{Esposito-Farese}},\ }%
  \bibfield{journal}{%
  \Doi{10.1103/PhysRevD.80.064015}{\bibinfo {journal} {Phys.Rev.}}\ }%
  \textbf{\bibinfo {volume} {D80}},\ \bibinfo {pages} {064015} (\bibinfo {year}
  {2009}),\ \Eprint{http://arxiv.org/abs/0906.1967}{arXiv:0906.1967 [gr-qc]}%
  \bibAnnoteFile{NoStop}{Deffayet:2009mn}%
%%CITATION = ARXIV:0906.1967;%%
\bibitem{Barreira:2012kk}%
  \BibitemOpen
  \bibfield{author}{%
  \bibinfo {author} {\bibfnamefont{A.}~\bibnamefont{Barreira}}, \bibinfo
  {author} {\bibfnamefont{B.}~\bibnamefont{Li}}, \bibinfo {author}
  {\bibfnamefont{C.~M.}\ \bibnamefont{Baugh}},\ and\ \bibinfo {author}
  {\bibfnamefont{S.}~\bibnamefont{Pascoli}},\ }%
  \bibfield{journal}{%
  \Doi{10.1103/PhysRevD.86.124016}{\bibinfo {journal} {Phys.Rev.}}\ }%
  \textbf{\bibinfo {volume} {D86}},\ \bibinfo {pages} {124016} (\bibinfo {year}
  {2012}),\ \Eprint{http://arxiv.org/abs/1208.0600}{arXiv:1208.0600
  [astro-ph.CO]}%
  \bibAnnoteFile{NoStop}{Barreira:2012kk}%
%%CITATION = ARXIV:1208.0600;%%
\bibitem{Barreira:2013jma}%
  \BibitemOpen
  \bibfield{author}{%
  \bibinfo {author} {\bibfnamefont{A.}~\bibnamefont{Barreira}}, \bibinfo
  {author} {\bibfnamefont{B.}~\bibnamefont{Li}}, \bibinfo {author}
  {\bibfnamefont{A.}~\bibnamefont{Sanchez}}, \bibinfo {author}
  {\bibfnamefont{C.~M.}\ \bibnamefont{Baugh}},\ and\ \bibinfo {author}
  {\bibfnamefont{S.}~\bibnamefont{Pascoli}}}%
   (\bibinfo {year} {2013}),\
  \Eprint{http://arxiv.org/abs/1302.6241}{arXiv:1302.6241 [astro-ph.CO]}%
  \bibAnnoteFile{NoStop}{Barreira:2013jma}%
%%CITATION = ARXIV:1302.6241;%%
\bibitem{Gannouji:2010au}%
  \BibitemOpen
  \bibfield{author}{%
  \bibinfo {author} {\bibfnamefont{R.}~\bibnamefont{Gannouji}}\ and\ \bibinfo
  {author} {\bibfnamefont{M.}~\bibnamefont{Sami}},\ }%
  \bibfield{journal}{%
  \Doi{10.1103/PhysRevD.82.024011}{\bibinfo {journal} {Phys.Rev.}}\ }%
  \textbf{\bibinfo {volume} {D82}},\ \bibinfo {pages} {024011} (\bibinfo {year}
  {2010}),\ \Eprint{http://arxiv.org/abs/1004.2808}{arXiv:1004.2808 [gr-qc]}%
  \bibAnnoteFile{NoStop}{Gannouji:2010au}%
%%CITATION = ARXIV:1004.2808;%%
\bibitem{PhysRevD.80.024037}%
  \BibitemOpen
  \bibfield{author}{%
  \bibinfo {author} {\bibfnamefont{N.}~\bibnamefont{Chow}}\ and\ \bibinfo
  {author} {\bibfnamefont{J.}~\bibnamefont{Khoury}},\ }%
  \bibfield{journal}{%
  \Doi{10.1103/PhysRevD.80.024037}{\bibinfo {journal} {Phys. Rev. D}}\ }%
  \textbf{\bibinfo {volume} {80}},\ \bibinfo {pages} {024037} (\bibinfo {year}
  {2009})%
  \bibAnnoteFile{NoStop}{PhysRevD.80.024037}%
\bibitem{DeFelice:2010pv}%
  \BibitemOpen
  \bibfield{author}{%
  \bibinfo {author} {\bibfnamefont{A.}~\bibnamefont{De~Felice}}\ and\ \bibinfo
  {author} {\bibfnamefont{S.}~\bibnamefont{Tsujikawa}},\ }%
  \bibfield{journal}{%
  \Doi{10.1103/PhysRevLett.105.111301}{\bibinfo {journal} {Phys.Rev.Lett.}}\ }%
  \textbf{\bibinfo {volume} {105}},\ \bibinfo {pages} {111301} (\bibinfo {year}
  {2010}),\ \Eprint{http://arxiv.org/abs/1007.2700}{arXiv:1007.2700
  [astro-ph.CO]}%
  \bibAnnoteFile{NoStop}{DeFelice:2010pv}%
%%CITATION = ARXIV:1007.2700;%%
\bibitem{Nesseris:2010pc}%
  \BibitemOpen
  \bibfield{author}{%
  \bibinfo {author} {\bibfnamefont{S.}~\bibnamefont{Nesseris}}, \bibinfo
  {author} {\bibfnamefont{A.}~\bibnamefont{De~Felice}},\ and\ \bibinfo {author}
  {\bibfnamefont{S.}~\bibnamefont{Tsujikawa}},\ }%
  \bibfield{journal}{%
  \Doi{10.1103/PhysRevD.82.124054}{\bibinfo {journal} {Phys.Rev.}}\ }%
  \textbf{\bibinfo {volume} {D82}},\ \bibinfo {pages} {124054} (\bibinfo {year}
  {2010}),\ \Eprint{http://arxiv.org/abs/1010.0407}{arXiv:1010.0407
  [astro-ph.CO]}%
  \bibAnnoteFile{NoStop}{Nesseris:2010pc}%
%%CITATION = ARXIV:1010.0407;%%
\bibitem{Appleby:2011aa}%
  \BibitemOpen
  \bibfield{author}{%
  \bibinfo {author} {\bibfnamefont{S.~A.}\ \bibnamefont{Appleby}}\ and\
  \bibinfo {author} {\bibfnamefont{E.~V.}\ \bibnamefont{Linder}},\ }%
  \bibfield{journal}{%
  \bibinfo {journal} {JCAP}\ }%
  \textbf{\bibinfo {volume} {1203}},\ \bibinfo {pages} {043} (\bibinfo {year}
  {2012}),\ \Eprint{http://arxiv.org/abs/1112.1981}{arXiv:1112.1981
  [astro-ph.CO]}%
  \bibAnnoteFile{NoStop}{Appleby:2011aa}%
%%CITATION = ARXIV:1112.1981;%%
\bibitem{PhysRevD.82.103015}%
  \BibitemOpen
  \bibfield{author}{%
  \bibinfo {author} {\bibfnamefont{A.}~\bibnamefont{Ali}}, \bibinfo {author}
  {\bibfnamefont{R.}~\bibnamefont{Gannouji}},\ and\ \bibinfo {author}
  {\bibfnamefont{M.}~\bibnamefont{Sami}},\ }%
  \bibfield{journal}{%
  \Doi{10.1103/PhysRevD.82.103015}{\bibinfo {journal} {Phys. Rev. D}}\ }%
  \textbf{\bibinfo {volume} {82}},\ \bibinfo {pages} {103015} (\bibinfo {year}
  {2010})%
  \bibAnnoteFile{NoStop}{PhysRevD.82.103015}%
\bibitem{Neveu:2013mfa}%
  \BibitemOpen
  \bibfield{author}{%
  \bibinfo {author} {\bibfnamefont{J.}~\bibnamefont{Neveu}}, \bibinfo {author}
  {\bibfnamefont{V.}~\bibnamefont{Ruhlmann-Kleider}}, \bibinfo {author}
  {\bibfnamefont{A.}~\bibnamefont{Conley}}, \bibinfo {author}
  {\bibfnamefont{N.}~\bibnamefont{Palanque-Delabrouille}}, \bibinfo {author}
  {\bibfnamefont{P.}~\bibnamefont{Astier}}, \emph{et~al.}}%
   (\bibinfo {year} {2013}),\
  \Eprint{http://arxiv.org/abs/1302.2786}{arXiv:1302.2786 [gr-qc]}%
  \bibAnnoteFile{NoStop}{Neveu:2013mfa}%
%%CITATION = ARXIV:1302.2786;%%
\bibitem{Appleby:2012ba}%
  \BibitemOpen
  \bibfield{author}{%
  \bibinfo {author} {\bibfnamefont{S.~A.}\ \bibnamefont{Appleby}}\ and\
  \bibinfo {author} {\bibfnamefont{E.~V.}\ \bibnamefont{Linder}}}%
   (\bibinfo {year} {2012}),\
  \Eprint{http://arxiv.org/abs/1204.4314}{arXiv:1204.4314 [astro-ph.CO]}%
  \bibAnnoteFile{NoStop}{Appleby:2012ba}%
%%CITATION = ARXIV:1204.4314;%%
\bibitem{Okada:2012mn}%
  \BibitemOpen
  \bibfield{author}{%
  \bibinfo {author} {\bibfnamefont{H.}~\bibnamefont{Okada}}, \bibinfo {author}
  {\bibfnamefont{T.}~\bibnamefont{Totani}},\ and\ \bibinfo {author}
  {\bibfnamefont{S.}~\bibnamefont{Tsujikawa}}}%
   (\bibinfo {year} {2012}),\
  \Eprint{http://arxiv.org/abs/1208.4681}{arXiv:1208.4681 [astro-ph.CO]}%
  \bibAnnoteFile{NoStop}{Okada:2012mn}%
%%CITATION = ARXIV:1208.4681;%%
\bibitem{Bartolo:2013ws}%
  \BibitemOpen
  \bibfield{author}{%
  \bibinfo {author} {\bibfnamefont{N.}~\bibnamefont{Bartolo}}, \bibinfo
  {author} {\bibfnamefont{E.}~\bibnamefont{Bellini}}, \bibinfo {author}
  {\bibfnamefont{D.}~\bibnamefont{Bertacca}},\ and\ \bibinfo {author}
  {\bibfnamefont{S.}~\bibnamefont{Matarrese}}}%
   (\bibinfo {year} {2013}),\
  \Eprint{http://arxiv.org/abs/1301.4831}{arXiv:1301.4831 [astro-ph.CO]}%
  \bibAnnoteFile{NoStop}{Bartolo:2013ws}%
%%CITATION = ARXIV:1301.4831;%%
\bibitem{Woodard:2006nt}%
  \BibitemOpen
  \bibfield{author}{%
  \bibinfo {author} {\bibfnamefont{R.~P.}\ \bibnamefont{Woodard}},\ }%
  \bibfield{journal}{%
  \Doi{10.1007/978-3-540-71013-4_14}{\bibinfo {journal} {Lect.Notes Phys.}}\ }%
  \textbf{\bibinfo {volume} {720}},\ \bibinfo {pages} {403} (\bibinfo {year}
  {2007}),\
  \Eprint{http://arxiv.org/abs/astro-ph/0601672}{arXiv:astro-ph/0601672
  [astro-ph]}%
  \bibAnnoteFile{NoStop}{Woodard:2006nt}%
%%CITATION = ASTRO-PH/0601672;%%
\bibitem{Horndeski:1974wa}%
  \BibitemOpen
  \bibfield{author}{%
  \bibinfo {author} {\bibfnamefont{G.~W.}\ \bibnamefont{Horndeski}},\ }%
  \bibfield{journal}{%
  \Doi{10.1007/BF01807638}{\bibinfo {journal} {Int.J.Theor.Phys.}}\ }%
  \textbf{\bibinfo {volume} {10}},\ \bibinfo {pages} {363} (\bibinfo {year}
  {1974})%
  \bibAnnoteFile{NoStop}{Horndeski:1974wa}%
%%CITATION = IJTPB,10,363;%%
\bibitem{1990PhRvL..64..123D}%
  \BibitemOpen
  \bibfield{author}{%
  \bibinfo {author} {\bibfnamefont{T.}~\bibnamefont{{Damour}}}, \bibinfo
  {author} {\bibfnamefont{G.~W.}\ \bibnamefont{{Gibbons}}},\ and\ \bibinfo
  {author} {\bibfnamefont{C.}~\bibnamefont{{Gundlach}}},\ }%
  \bibfield{journal}{%
  \Doi{10.1103/PhysRevLett.64.123}{\bibinfo {journal} {Phys.~Rev.~Lett.}}\ }%
  \textbf{\bibinfo {volume} {64}},\ \bibinfo {pages} {123} (\bibinfo {year}
  {1990})%
  \bibAnnoteFile{NoStop}{1990PhRvL..64..123D}%
\bibitem{1999PhRvL..83.3585B}%
  \BibitemOpen
  \bibfield{author}{%
  \bibinfo {author} {\bibfnamefont{S.}~\bibnamefont{{Bae{\ss}ler}}}, \bibinfo
  {author} {\bibfnamefont{B.~R.}\ \bibnamefont{{Heckel}}}, \bibinfo {author}
  {\bibfnamefont{E.~G.}\ \bibnamefont{{Adelberger}}}, \bibinfo {author}
  {\bibfnamefont{J.~H.}\ \bibnamefont{{Gundlach}}}, \bibinfo {author}
  {\bibfnamefont{U.}~\bibnamefont{{Schmidt}}},\ and\ \bibinfo {author}
  {\bibfnamefont{H.~E.}\ \bibnamefont{{Swanson}}},\ }%
  \bibfield{journal}{%
  \Doi{10.1103/PhysRevLett.83.3585}{\bibinfo {journal} {Phys.~Rev.~Lett.}}\ }%
  \textbf{\bibinfo {volume} {83}},\ \bibinfo {pages} {3585} (\bibinfo {year}
  {1999})%
  \bibAnnoteFile{NoStop}{1999PhRvL..83.3585B}%
\bibitem{Will:2005va}%
  \BibitemOpen
  \bibfield{author}{%
  \bibinfo {author} {\bibfnamefont{C.~M.}\ \bibnamefont{Will}},\ }%
  \bibfield{journal}{%
  \bibinfo {journal} {Living Rev.Rel.}\ }%
  \textbf{\bibinfo {volume} {9}},\ \bibinfo {pages} {3} (\bibinfo {year}
  {2006}),\ \Eprint{http://arxiv.org/abs/gr-qc/0510072}{arXiv:gr-qc/0510072
  [gr-qc]}%
  \bibAnnoteFile{NoStop}{Will:2005va}%
%%CITATION = GR-QC/0510072;%%
\bibitem{Kapner:2006si}%
  \BibitemOpen
  \bibfield{author}{%
  \bibinfo {author} {\bibfnamefont{D.}~\bibnamefont{Kapner}}, \bibinfo {author}
  {\bibfnamefont{T.}~\bibnamefont{Cook}}, \bibinfo {author}
  {\bibfnamefont{E.}~\bibnamefont{Adelberger}}, \bibinfo {author}
  {\bibfnamefont{J.}~\bibnamefont{Gundlach}}, \bibinfo {author}
  {\bibfnamefont{B.~R.}\ \bibnamefont{Heckel}}, \emph{et~al.},\ }%
  \bibfield{journal}{%
  \Doi{10.1103/PhysRevLett.98.021101}{\bibinfo {journal} {Phys.Rev.Lett.}}\ }%
  \textbf{\bibinfo {volume} {98}},\ \bibinfo {pages} {021101} (\bibinfo {year}
  {2007}),\ \Eprint{http://arxiv.org/abs/hep-ph/0611184}{arXiv:hep-ph/0611184
  [hep-ph]}%
  \bibAnnoteFile{NoStop}{Kapner:2006si}%
%%CITATION = HEP-PH/0611184;%%
\bibitem{Vainshtein1972393}%
  \BibitemOpen
  \bibfield{author}{%
  \bibinfo {author} {\bibfnamefont{A.}~\bibnamefont{Vainshtein}},\ }%
  \bibfield{journal}{%
  \Doi{10.1016/0370-2693(72)90147-5}{\bibinfo {journal} {Phys.~Lett.~B}}\ }%
  \textbf{\bibinfo {volume} {39}},\ \bibinfo {pages} {393 } (\bibinfo {year}
  {1972}),\ ISSN \bibinfo {issn} {0370-2693}%
  \bibAnnoteFile{NoStop}{Vainshtein1972393}%
\bibitem{Babichev:2013usa}%
  \BibitemOpen
  \bibfield{author}{%
  \bibinfo {author} {\bibfnamefont{E.}~\bibnamefont{Babichev}}\ and\ \bibinfo
  {author} {\bibfnamefont{C.}~\bibnamefont{Deffayet}}}%
   (\bibinfo {year} {2013}),\
  \Eprint{http://arxiv.org/abs/1304.7240}{arXiv:1304.7240 [gr-qc]}%
  \bibAnnoteFile{NoStop}{Babichev:2013usa}%
%%CITATION = ARXIV:1304.7240;%%
\bibitem{Koyama:2013paa}%
  \BibitemOpen
  \bibfield{author}{%
  \bibinfo {author} {\bibfnamefont{K.}~\bibnamefont{Koyama}}, \bibinfo {author}
  {\bibfnamefont{G.}~\bibnamefont{Niz}},\ and\ \bibinfo {author}
  {\bibfnamefont{G.}~\bibnamefont{Tasinato}}}%
   (\bibinfo {year} {2013}),\
  \Eprint{http://arxiv.org/abs/1305.0279}{arXiv:1305.0279 [hep-th]}%
  \bibAnnoteFile{NoStop}{Koyama:2013paa}%
%%CITATION = ARXIV:1305.0279;%%
\bibitem{camb_notes}%
  \BibitemOpen
  \bibfield{author}{%
  \bibinfo {author} {\bibfnamefont{A.}~\bibnamefont{Lewis}},\ }%
  \bibinfo {note} {\url{http://camb.info/}}%
  \bibAnnoteFile{NoStop}{camb_notes}%
\bibitem{Lewis:2002ah}%
  \BibitemOpen
  \bibfield{author}{%
  \bibinfo {author} {\bibfnamefont{A.}~\bibnamefont{Lewis}}\ and\ \bibinfo
  {author} {\bibfnamefont{S.}~\bibnamefont{Bridle}},\ }%
  \bibfield{journal}{%
  \Doi{10.1103/PhysRevD.66.103511}{\bibinfo {journal} {Phys.Rev.}}\ }%
  \textbf{\bibinfo {volume} {D66}},\ \bibinfo {pages} {103511} (\bibinfo {year}
  {2002}),\
  \Eprint{http://arxiv.org/abs/astro-ph/0205436}{arXiv:astro-ph/0205436
  [astro-ph]}%
  \bibAnnoteFile{NoStop}{Lewis:2002ah}%
%%CITATION = ASTRO-PH/0205436;%%
\bibitem{Barreira:2013eea}%
  \BibitemOpen
  \bibfield{author}{%
  \bibinfo {author} {\bibfnamefont{A.}~\bibnamefont{Barreira}}, \bibinfo
  {author} {\bibfnamefont{B.}~\bibnamefont{Li}}, \bibinfo {author}
  {\bibfnamefont{W.~A.}\ \bibnamefont{Hellwing}}, \bibinfo {author}
  {\bibfnamefont{C.~M.}\ \bibnamefont{Baugh}},\ and\ \bibinfo {author}
  {\bibfnamefont{S.}~\bibnamefont{Pascoli}},\ }%
  \bibfield{journal}{%
  \bibinfo {journal} {JCAP}\ }%
  \textbf{\bibinfo {volume} {2013}},\ \bibinfo {pages} {027} (\bibinfo {year}
  {2013}),\ \Eprint{http://arxiv.org/abs/1306.3219}{arXiv:1306.3219
  [astro-ph.CO]}%
  \bibAnnoteFile{NoStop}{Barreira:2013eea}%
%%CITATION = ARXIV:1306.3219;%%
\bibitem{Li:2011vk}%
  \BibitemOpen
  \bibfield{author}{%
  \bibinfo {author} {\bibfnamefont{B.}~\bibnamefont{Li}}, \bibinfo {author}
  {\bibfnamefont{G.-B.}\ \bibnamefont{Zhao}}, \bibinfo {author}
  {\bibfnamefont{R.}~\bibnamefont{Teyssier}},\ and\ \bibinfo {author}
  {\bibfnamefont{K.}~\bibnamefont{Koyama}},\ }%
  \bibfield{journal}{%
  \Doi{10.1088/1475-7516/2012/01/051}{\bibinfo {journal} {JCAP}}\ }%
  \textbf{\bibinfo {volume} {1201}},\ \bibinfo {pages} {051} (\bibinfo {year}
  {2012}),\ \Eprint{http://arxiv.org/abs/1110.1379}{arXiv:1110.1379
  [astro-ph.CO]}%
  \bibAnnoteFile{NoStop}{Li:2011vk}%
%%CITATION = ARXIV:1110.1379;%%
\bibitem{Li:2013nua}%
  \BibitemOpen
  \bibfield{author}{%
  \bibinfo {author} {\bibfnamefont{B.}~\bibnamefont{Li}}, \bibinfo {author}
  {\bibfnamefont{G.-B.}\ \bibnamefont{Zhao}},\ and\ \bibinfo {author}
  {\bibfnamefont{K.}~\bibnamefont{Koyama}},\ }%
  \bibfield{journal}{%
  \Doi{10.1088/1475-7516/2013/05/023}{\bibinfo {journal} {JCAP}}\ }%
  \textbf{\bibinfo {volume} {1305}},\ \bibinfo {pages} {023} (\bibinfo {year}
  {2013}),\ \Eprint{http://arxiv.org/abs/1303.0008}{arXiv:1303.0008
  [astro-ph.CO]}%
  \bibAnnoteFile{NoStop}{Li:2013nua}%
%%CITATION = ARXIV:1303.0008;%%
\bibitem{Li:2013tda}%
  \BibitemOpen
  \bibfield{author}{%
  \bibinfo {author} {\bibfnamefont{B.}~\bibnamefont{Li}}, \bibinfo {author}
  {\bibfnamefont{A.}~\bibnamefont{Barreira}}, \bibinfo {author}
  {\bibfnamefont{C.~M.}\ \bibnamefont{Baugh}}, \bibinfo {author}
  {\bibfnamefont{W.~A.}\ \bibnamefont{Hellwing}}, \bibinfo {author}
  {\bibfnamefont{K.}~\bibnamefont{Koyama}}, \emph{et~al.}}%
   (\bibinfo {year} {2013}),\
  \Eprint{http://arxiv.org/abs/1308.3491}{arXiv:1308.3491 [astro-ph.CO]}%
  \bibAnnoteFile{NoStop}{Li:2013tda}%
%%CITATION = ARXIV:1308.3491;%%
\bibitem{1991ApJ...379..440B}%
  \BibitemOpen
  \bibfield{author}{%
  \bibinfo {author} {\bibfnamefont{J.~R.}\ \bibnamefont{{Bond}}}, \bibinfo
  {author} {\bibfnamefont{S.}~\bibnamefont{{Cole}}}, \bibinfo {author}
  {\bibfnamefont{G.}~\bibnamefont{{Efstathiou}}},\ and\ \bibinfo {author}
  {\bibfnamefont{N.}~\bibnamefont{{Kaiser}}},\ }%
  \bibfield{journal}{%
  \Doi{10.1086/170520}{\bibinfo {journal} {\apj}}\ }%
  \textbf{\bibinfo {volume} {379}},\ \bibinfo {pages} {440} (\bibinfo {month}
  {Oct.}\ \bibinfo {year} {1991})%
  \bibAnnoteFile{NoStop}{1991ApJ...379..440B}%
\bibitem{Deffayet:2010qz}%
  \BibitemOpen
  \bibfield{author}{%
  \bibinfo {author} {\bibfnamefont{C.}~\bibnamefont{Deffayet}}, \bibinfo
  {author} {\bibfnamefont{O.}~\bibnamefont{Pujolas}}, \bibinfo {author}
  {\bibfnamefont{I.}~\bibnamefont{Sawicki}},\ and\ \bibinfo {author}
  {\bibfnamefont{A.}~\bibnamefont{Vikman}},\ }%
  \bibfield{journal}{%
  \Doi{10.1088/1475-7516/2010/10/026}{\bibinfo {journal} {JCAP}}\ }%
  \textbf{\bibinfo {volume} {1010}},\ \bibinfo {pages} {026} (\bibinfo {year}
  {2010}),\ \Eprint{http://arxiv.org/abs/1008.0048}{arXiv:1008.0048 [hep-th]}%
  \bibAnnoteFile{NoStop}{Deffayet:2010qz}%
%%CITATION = ARXIV:1008.0048;%%
\bibitem{Babichev:2012re}%
  \BibitemOpen
  \bibfield{author}{%
  \bibinfo {author} {\bibfnamefont{E.}~\bibnamefont{Babichev}}\ and\ \bibinfo
  {author} {\bibfnamefont{G.}~\bibnamefont{Esposito-Farèse}},\ }%
  \bibfield{journal}{%
  \Doi{10.1103/PhysRevD.87.044032}{\bibinfo {journal} {Phys.Rev.}}\ }%
  \textbf{\bibinfo {volume} {D87}},\ \bibinfo {pages} {044032} (\bibinfo {year}
  {2013}),\ \Eprint{http://arxiv.org/abs/1212.1394}{arXiv:1212.1394 [gr-qc]}%
  \bibAnnoteFile{NoStop}{Babichev:2012re}%
%%CITATION = ARXIV:1212.1394;%%
\bibitem{DeFelice:2010as}%
  \BibitemOpen
  \bibfield{author}{%
  \bibinfo {author} {\bibfnamefont{A.}~\bibnamefont{De~Felice}}, \bibinfo
  {author} {\bibfnamefont{R.}~\bibnamefont{Kase}},\ and\ \bibinfo {author}
  {\bibfnamefont{S.}~\bibnamefont{Tsujikawa}},\ }%
  \bibfield{journal}{%
  \Doi{10.1103/PhysRevD.83.043515}{\bibinfo {journal} {Phys.Rev.}}\ }%
  \textbf{\bibinfo {volume} {D83}},\ \bibinfo {pages} {043515} (\bibinfo {year}
  {2011}),\ \Eprint{http://arxiv.org/abs/1011.6132}{arXiv:1011.6132
  [astro-ph.CO]}%
  \bibAnnoteFile{NoStop}{DeFelice:2010as}%
%%CITATION = ARXIV:1011.6132;%%
\bibitem{Bellini:2012qn}%
  \BibitemOpen
  \bibfield{author}{%
  \bibinfo {author} {\bibfnamefont{E.}~\bibnamefont{Bellini}}, \bibinfo
  {author} {\bibfnamefont{N.}~\bibnamefont{Bartolo}},\ and\ \bibinfo {author}
  {\bibfnamefont{S.}~\bibnamefont{Matarrese}},\ }%
  \bibfield{journal}{%
  \Doi{10.1088/1475-7516/2012/06/019}{\bibinfo {journal} {JCAP}}\ }%
  \textbf{\bibinfo {volume} {1206}},\ \bibinfo {pages} {019} (\bibinfo {year}
  {2012}),\ \Eprint{http://arxiv.org/abs/1202.2712}{arXiv:1202.2712
  [astro-ph.CO]}%
  \bibAnnoteFile{NoStop}{Bellini:2012qn}%
%%CITATION = ARXIV:1202.2712;%%
\bibitem{Babichev:2011iz}%
  \BibitemOpen
  \bibfield{author}{%
  \bibinfo {author} {\bibfnamefont{E.}~\bibnamefont{Babichev}}, \bibinfo
  {author} {\bibfnamefont{C.}~\bibnamefont{Deffayet}},\ and\ \bibinfo {author}
  {\bibfnamefont{G.}~\bibnamefont{Esposito-Farese}},\ }%
  \bibfield{journal}{%
  \Doi{10.1103/PhysRevLett.107.251102}{\bibinfo {journal} {Phys.Rev.Lett.}}\ }%
  \textbf{\bibinfo {volume} {107}},\ \bibinfo {pages} {251102} (\bibinfo {year}
  {2011}),\ \Eprint{http://arxiv.org/abs/1107.1569}{arXiv:1107.1569 [gr-qc]}%
  \bibAnnoteFile{NoStop}{Babichev:2011iz}%
%%CITATION = ARXIV:1107.1569;%%
\bibitem{Kimura:2011dc}%
  \BibitemOpen
  \bibfield{author}{%
  \bibinfo {author} {\bibfnamefont{R.}~\bibnamefont{Kimura}}, \bibinfo {author}
  {\bibfnamefont{T.}~\bibnamefont{Kobayashi}},\ and\ \bibinfo {author}
  {\bibfnamefont{K.}~\bibnamefont{Yamamoto}},\ }%
  \bibfield{journal}{%
  \Doi{10.1103/PhysRevD.85.024023}{\bibinfo {journal} {Phys.Rev.}}\ }%
  \textbf{\bibinfo {volume} {D85}},\ \bibinfo {pages} {024023} (\bibinfo {year}
  {2012}),\ \Eprint{http://arxiv.org/abs/1111.6749}{arXiv:1111.6749
  [astro-ph.CO]}%
  \bibAnnoteFile{NoStop}{Kimura:2011dc}%
%%CITATION = ARXIV:1111.6749;%%
\bibitem{Williams:2004qba}%
  \BibitemOpen
  \bibfield{author}{%
  \bibinfo {author} {\bibfnamefont{J.~G.}\ \bibnamefont{Williams}}, \bibinfo
  {author} {\bibfnamefont{S.~G.}\ \bibnamefont{Turyshev}},\ and\ \bibinfo
  {author} {\bibfnamefont{D.~H.}\ \bibnamefont{Boggs}},\ }%
  \bibfield{journal}{%
  \Doi{10.1103/PhysRevLett.93.261101}{\bibinfo {journal} {Phys.Rev.Lett.}}\ }%
  \textbf{\bibinfo {volume} {93}},\ \bibinfo {pages} {261101} (\bibinfo {year}
  {2004}),\ \Eprint{http://arxiv.org/abs/gr-qc/0411113}{arXiv:gr-qc/0411113
  [gr-qc]}%
  \bibAnnoteFile{NoStop}{Williams:2004qba}%
%%CITATION = GR-QC/0411113;%%
\bibitem{Zentner:2006vw}%
  \BibitemOpen
  \bibfield{author}{%
  \bibinfo {author} {\bibfnamefont{A.~R.}\ \bibnamefont{Zentner}},\ }%
  \bibfield{journal}{%
  \Doi{10.1142/S0218271807010511}{\bibinfo {journal} {Int.J.Mod.Phys.}}\ }%
  \textbf{\bibinfo {volume} {D16}},\ \bibinfo {pages} {763} (\bibinfo {year}
  {2007}),\
  \Eprint{http://arxiv.org/abs/astro-ph/0611454}{arXiv:astro-ph/0611454
  [astro-ph]}%
  \bibAnnoteFile{NoStop}{Zentner:2006vw}%
%%CITATION = ASTRO-PH/0611454;%%
\bibitem{Gaztanaga:2000vw}%
  \BibitemOpen
  \bibfield{author}{%
  \bibinfo {author} {\bibfnamefont{E.}~\bibnamefont{Gaztanaga}}\ and\ \bibinfo
  {author} {\bibfnamefont{J.~A.}\ \bibnamefont{Lobo}},\ }%
  \bibfield{journal}{%
  \Doi{10.1086/318684}{\bibinfo {journal} {Astrophys.J.}}\ }%
  \textbf{\bibinfo {volume} {548}},\ \bibinfo {pages} {47} (\bibinfo {year}
  {2001}),\
  \Eprint{http://arxiv.org/abs/astro-ph/0003129}{arXiv:astro-ph/0003129
  [astro-ph]}%
  \bibAnnoteFile{NoStop}{Gaztanaga:2000vw}%
%%CITATION = ASTRO-PH/0003129;%%
\bibitem{Schaefer:2007nf}%
  \BibitemOpen
  \bibfield{author}{%
  \bibinfo {author} {\bibfnamefont{B.~M.}\ \bibnamefont{Schaefer}}\ and\
  \bibinfo {author} {\bibfnamefont{K.}~\bibnamefont{Koyama}},\ }%
  \bibfield{journal}{%
  \Doi{10.1111/j.1365-2966.2008.12841.x}{\bibinfo {journal}
  {Mon.Not.Roy.Astron.Soc.}}\ }%
  \textbf{\bibinfo {volume} {385}},\ \bibinfo {pages} {411} (\bibinfo {year}
  {2008}),\ \Eprint{http://arxiv.org/abs/0711.3129}{arXiv:0711.3129
  [astro-ph]}%
  \bibAnnoteFile{NoStop}{Schaefer:2007nf}%
%%CITATION = ARXIV:0711.3129;%%
\bibitem{Martino:2008ae}%
  \BibitemOpen
  \bibfield{author}{%
  \bibinfo {author} {\bibfnamefont{M.~C.}\ \bibnamefont{Martino}}, \bibinfo
  {author} {\bibfnamefont{H.~F.}\ \bibnamefont{Stabenau}},\ and\ \bibinfo
  {author} {\bibfnamefont{R.~K.}\ \bibnamefont{Sheth}},\ }%
  \bibfield{journal}{%
  \Doi{10.1103/PhysRevD.79.084013}{\bibinfo {journal} {Phys.Rev.}}\ }%
  \textbf{\bibinfo {volume} {D79}},\ \bibinfo {pages} {084013} (\bibinfo {year}
  {2009}),\ \Eprint{http://arxiv.org/abs/0812.0200}{arXiv:0812.0200
  [astro-ph]}%
  \bibAnnoteFile{NoStop}{Martino:2008ae}%
%%CITATION = ARXIV:0812.0200;%%
\bibitem{Li:2011qda}%
  \BibitemOpen
  \bibfield{author}{%
  \bibinfo {author} {\bibfnamefont{B.}~\bibnamefont{Li}}\ and\ \bibinfo
  {author} {\bibfnamefont{G.}~\bibnamefont{Efstathiou}},\ }%
  \bibfield{journal}{%
  \Doi{10.1111/j.1365-2966.2011.20404.x}{\bibinfo {journal}
  {Mon.Not.Roy.Astron.Soc.}}\ }%
  \textbf{\bibinfo {volume} {421}},\ \bibinfo {pages} {1431} (\bibinfo {year}
  {2012}),\ \Eprint{http://arxiv.org/abs/1110.6440}{arXiv:1110.6440
  [astro-ph.CO]}%
  \bibAnnoteFile{NoStop}{Li:2011qda}%
%%CITATION = ARXIV:1110.6440;%%
\bibitem{Borisov:2011fu}%
  \BibitemOpen
  \bibfield{author}{%
  \bibinfo {author} {\bibfnamefont{A.}~\bibnamefont{Borisov}}, \bibinfo
  {author} {\bibfnamefont{B.}~\bibnamefont{Jain}},\ and\ \bibinfo {author}
  {\bibfnamefont{P.}~\bibnamefont{Zhang}},\ }%
  \bibfield{journal}{%
  \Doi{10.1103/PhysRevD.85.063518}{\bibinfo {journal} {Phys.Rev.}}\ }%
  \textbf{\bibinfo {volume} {D85}},\ \bibinfo {pages} {063518} (\bibinfo {year}
  {2012}),\ \Eprint{http://arxiv.org/abs/1102.4839}{arXiv:1102.4839
  [astro-ph.CO]}%
  \bibAnnoteFile{NoStop}{Borisov:2011fu}%
%%CITATION = ARXIV:1102.4839;%%
\bibitem{Lam:2012fa}%
  \BibitemOpen
  \bibfield{author}{%
  \bibinfo {author} {\bibfnamefont{T.~Y.}\ \bibnamefont{Lam}}\ and\ \bibinfo
  {author} {\bibfnamefont{B.}~\bibnamefont{Li}},\ }%
  \bibfield{journal}{%
  \Doi{10.1111/j.1365-2966.2012.21746.x}{\bibinfo {journal}
  {Mon.Not.Roy.Astron.Soc.}}\ }%
  \textbf{\bibinfo {volume} {426}},\ \bibinfo {pages} {3260} (\bibinfo {year}
  {2012}),\ \Eprint{http://arxiv.org/abs/1205.0059}{arXiv:1205.0059
  [astro-ph.CO]}%
  \bibAnnoteFile{NoStop}{Lam:2012fa}%
%%CITATION = ARXIV:1205.0059;%%
\bibitem{Li:2012ez}%
  \BibitemOpen
  \bibfield{author}{%
  \bibinfo {author} {\bibfnamefont{B.}~\bibnamefont{Li}}\ and\ \bibinfo
  {author} {\bibfnamefont{T.~Y.}\ \bibnamefont{Lam}},\ }%
  \bibfield{journal}{%
  \bibinfo {journal} {MNRAS, 425,}\ }%
  \textbf{\bibinfo {volume} {730}} (\bibinfo {year} {2012}),\ \doi{\bibinfo
  {doi} {10.1111/j.1365-2966.2012.21592.x}},\
  \Eprint{http://arxiv.org/abs/1205.0058}{arXiv:1205.0058 [astro-ph.CO]}%
  \bibAnnoteFile{NoStop}{Li:2012ez}%
%%CITATION = ARXIV:1205.0058;%%
\bibitem{Clampitt:2012ub}%
  \BibitemOpen
  \bibfield{author}{%
  \bibinfo {author} {\bibfnamefont{J.}~\bibnamefont{Clampitt}}\ and\ \bibinfo
  {author} {\bibfnamefont{Y.-C.}\ \bibnamefont{Cai}},\ }%
  \bibfield{journal}{%
  \Doi{10.1093/mnras/stt219}{\bibinfo {journal} {2013}}\ }%
  \textbf{\bibinfo {volume} {431}},\ \bibinfo {pages} {749C} (\bibinfo {year}
  {MNRAS}),\ \Eprint{http://arxiv.org/abs/1212.2216}{arXiv:1212.2216
  [astro-ph.CO]}%
  \bibAnnoteFile{NoStop}{Clampitt:2012ub}%
%%CITATION = ARXIV:1212.2216;%%
\bibitem{Lombriser:2013wta}%
  \BibitemOpen
  \bibfield{author}{%
  \bibinfo {author} {\bibfnamefont{L.}~\bibnamefont{Lombriser}}, \bibinfo
  {author} {\bibfnamefont{B.}~\bibnamefont{Li}}, \bibinfo {author}
  {\bibfnamefont{K.}~\bibnamefont{Koyama}},\ and\ \bibinfo {author}
  {\bibfnamefont{G.-B.}\ \bibnamefont{Zhao}},\ }%
  \bibfield{journal}{%
  \bibinfo {journal} {Phys. Rev. D 87,}\ }%
  \textbf{\bibinfo {volume} {123511}} (\bibinfo {year} {2013}),\ \doi{\bibinfo
  {doi} {10.1103/PhysRevD.87.123511}},\
  \Eprint{http://arxiv.org/abs/1304.6395}{arXiv:1304.6395 [astro-ph.CO]}%
  \bibAnnoteFile{NoStop}{Lombriser:2013wta}%
%%CITATION = ARXIV:1304.6395;%%
\bibitem{Kopp:2013lea}%
  \BibitemOpen
  \bibfield{author}{%
  \bibinfo {author} {\bibfnamefont{M.}~\bibnamefont{Kopp}}, \bibinfo {author}
  {\bibfnamefont{S.~A.}\ \bibnamefont{Appleby}}, \bibinfo {author}
  {\bibfnamefont{I.}~\bibnamefont{Achitouv}},\ and\ \bibinfo {author}
  {\bibfnamefont{J.}~\bibnamefont{Weller}}}%
   (\bibinfo {year} {2013}),\
  \Eprint{http://arxiv.org/abs/1306.3233}{arXiv:1306.3233 [astro-ph.CO]}%
  \bibAnnoteFile{NoStop}{Kopp:2013lea}%
%%CITATION = ARXIV:1306.3233;%%
\bibitem{Bardeen:1985tr}%
  \BibitemOpen
  \bibfield{author}{%
  \bibinfo {author} {\bibfnamefont{J.~M.}\ \bibnamefont{Bardeen}}, \bibinfo
  {author} {\bibfnamefont{J.}~\bibnamefont{Bond}}, \bibinfo {author}
  {\bibfnamefont{N.}~\bibnamefont{Kaiser}},\ and\ \bibinfo {author}
  {\bibfnamefont{A.}~\bibnamefont{Szalay}},\ }%
  \bibfield{journal}{%
  \Doi{10.1086/164143}{\bibinfo {journal} {Astrophys.J.}}\ }%
  \textbf{\bibinfo {volume} {304}},\ \bibinfo {pages} {15} (\bibinfo {year}
  {1986})%
  \bibAnnoteFile{NoStop}{Bardeen:1985tr}%
%%CITATION = ASJOA,304,15;%%
\bibitem{1974ApJ...187..425P}%
  \BibitemOpen
  \bibfield{author}{%
  \bibinfo {author} {\bibfnamefont{W.~H.}\ \bibnamefont{{Press}}}\ and\
  \bibinfo {author} {\bibfnamefont{P.}~\bibnamefont{{Schechter}}},\ }%
  \bibfield{journal}{%
  \Doi{10.1086/152650}{\bibinfo {journal} {\apj}}\ }%
  \textbf{\bibinfo {volume} {187}},\ \bibinfo {pages} {425} (\bibinfo {month}
  {Feb.}\ \bibinfo {year} {1974})%
  \bibAnnoteFile{NoStop}{1974ApJ...187..425P}%
\bibitem{Parfrey:2010uy}%
  \BibitemOpen
  \bibfield{author}{%
  \bibinfo {author} {\bibfnamefont{K.}~\bibnamefont{Parfrey}}, \bibinfo
  {author} {\bibfnamefont{L.}~\bibnamefont{Hui}},\ and\ \bibinfo {author}
  {\bibfnamefont{R.~K.}\ \bibnamefont{Sheth}},\ }%
  \bibfield{journal}{%
  \Doi{10.1103/PhysRevD.83.063511}{\bibinfo {journal} {Phys.Rev.}}\ }%
  \textbf{\bibinfo {volume} {D83}},\ \bibinfo {pages} {063511} (\bibinfo {year}
  {2011}),\ \Eprint{http://arxiv.org/abs/1012.1335}{arXiv:1012.1335
  [astro-ph.CO]}%
  \bibAnnoteFile{NoStop}{Parfrey:2010uy}%
%%CITATION = ARXIV:1012.1335;%%
\bibitem{Mo:1995cs}%
  \BibitemOpen
  \bibfield{author}{%
  \bibinfo {author} {\bibfnamefont{H.}~\bibnamefont{Mo}}\ and\ \bibinfo
  {author} {\bibfnamefont{S.~D.}\ \bibnamefont{White}},\ }%
  \bibfield{journal}{%
  \bibinfo {journal} {Mon.Not.Roy.Astron.Soc.}\ }%
  \textbf{\bibinfo {volume} {282}},\ \bibinfo {pages} {347} (\bibinfo {year}
  {1996}),\
  \Eprint{http://arxiv.org/abs/astro-ph/9512127}{arXiv:astro-ph/9512127
  [astro-ph]}%
  \bibAnnoteFile{NoStop}{Mo:1995cs}%
%%CITATION = ASTRO-PH/9512127;%%
\bibitem{Fry:1992vr}%
  \BibitemOpen
  \bibfield{author}{%
  \bibinfo {author} {\bibfnamefont{J.~N.}\ \bibnamefont{Fry}}\ and\ \bibinfo
  {author} {\bibfnamefont{E.}~\bibnamefont{Gaztanaga}},\ }%
  \bibfield{journal}{%
  \Doi{10.1086/173015}{\bibinfo {journal} {Astrophys.J.}}\ }%
  \textbf{\bibinfo {volume} {413}},\ \bibinfo {pages} {447} (\bibinfo {year}
  {1993}),\
  \Eprint{http://arxiv.org/abs/astro-ph/9302009}{arXiv:astro-ph/9302009
  [astro-ph]}%
  \bibAnnoteFile{NoStop}{Fry:1992vr}%
%%CITATION = ASTRO-PH/9302009;%%
\bibitem{Sheth:1999mn}%
  \BibitemOpen
  \bibfield{author}{%
  \bibinfo {author} {\bibfnamefont{R.~K.}\ \bibnamefont{Sheth}}\ and\ \bibinfo
  {author} {\bibfnamefont{G.}~\bibnamefont{Tormen}},\ }%
  \bibfield{journal}{%
  \Doi{10.1046/j.1365-8711.1999.02692.x}{\bibinfo {journal}
  {Mon.Not.Roy.Astron.Soc.}}\ }%
  \textbf{\bibinfo {volume} {308}},\ \bibinfo {pages} {119} (\bibinfo {year}
  {1999}),\
  \Eprint{http://arxiv.org/abs/astro-ph/9901122}{arXiv:astro-ph/9901122
  [astro-ph]}%
  \bibAnnoteFile{NoStop}{Sheth:1999mn}%
%%CITATION = ASTRO-PH/9901122;%%
\bibitem{Sheth:1999su}%
  \BibitemOpen
  \bibfield{author}{%
  \bibinfo {author} {\bibfnamefont{R.~K.}\ \bibnamefont{Sheth}}, \bibinfo
  {author} {\bibfnamefont{H.}~\bibnamefont{Mo}},\ and\ \bibinfo {author}
  {\bibfnamefont{G.}~\bibnamefont{Tormen}},\ }%
  \bibfield{journal}{%
  \Doi{10.1046/j.1365-8711.2001.04006.x}{\bibinfo {journal}
  {Mon.Not.Roy.Astron.Soc.}}\ }%
  \textbf{\bibinfo {volume} {323}},\ \bibinfo {pages} {1} (\bibinfo {year}
  {2001}),\
  \Eprint{http://arxiv.org/abs/astro-ph/9907024}{arXiv:astro-ph/9907024
  [astro-ph]}%
  \bibAnnoteFile{NoStop}{Sheth:1999su}%
%%CITATION = ASTRO-PH/9907024;%%
\bibitem{Sheth:2001dp}%
  \BibitemOpen
  \bibfield{author}{%
  \bibinfo {author} {\bibfnamefont{R.~K.}\ \bibnamefont{Sheth}}\ and\ \bibinfo
  {author} {\bibfnamefont{G.}~\bibnamefont{Tormen}},\ }%
  \bibfield{journal}{%
  \Doi{10.1046/j.1365-8711.2002.04950.x}{\bibinfo {journal}
  {Mon.Not.Roy.Astron.Soc.}}\ }%
  \textbf{\bibinfo {volume} {329}},\ \bibinfo {pages} {61} (\bibinfo {year}
  {2002}),\
  \Eprint{http://arxiv.org/abs/astro-ph/0105113}{arXiv:astro-ph/0105113
  [astro-ph]}%
  \bibAnnoteFile{NoStop}{Sheth:2001dp}%
%%CITATION = ASTRO-PH/0105113;%%
\bibitem{Schmidt:2010jr}%
  \BibitemOpen
  \bibfield{author}{%
  \bibinfo {author} {\bibfnamefont{F.}~\bibnamefont{Schmidt}},\ }%
  \bibfield{journal}{%
  \Doi{10.1103/PhysRevD.81.103002}{\bibinfo {journal} {Phys.Rev.}}\ }%
  \textbf{\bibinfo {volume} {D81}},\ \bibinfo {pages} {103002} (\bibinfo {year}
  {2010}),\ \Eprint{http://arxiv.org/abs/1003.0409}{arXiv:1003.0409
  [astro-ph.CO]}%
  \bibAnnoteFile{NoStop}{Schmidt:2010jr}%
%%CITATION = ARXIV:1003.0409;%%
\bibitem{Enqvist:2010bg}%
  \BibitemOpen
  \bibfield{author}{%
  \bibinfo {author} {\bibfnamefont{K.}~\bibnamefont{Enqvist}}, \bibinfo
  {author} {\bibfnamefont{S.}~\bibnamefont{Hotchkiss}},\ and\ \bibinfo {author}
  {\bibfnamefont{O.}~\bibnamefont{Taanila}},\ }%
  \bibfield{journal}{%
  \Doi{10.1088/1475-7516/2011/04/017}{\bibinfo {journal} {JCAP}}\ }%
  \textbf{\bibinfo {volume} {1104}},\ \bibinfo {pages} {017} (\bibinfo {year}
  {2011}),\ \Eprint{http://arxiv.org/abs/1012.2732}{arXiv:1012.2732
  [astro-ph.CO]}%
  \bibAnnoteFile{NoStop}{Enqvist:2010bg}%
%%CITATION = ARXIV:1012.2732;%%
\bibitem{Jee:2011az}%
  \BibitemOpen
  \bibfield{author}{%
  \bibinfo {author} {\bibfnamefont{M.}~\bibnamefont{Jee}}, \bibinfo {author}
  {\bibfnamefont{K.}~\bibnamefont{Dawson}}, \bibinfo {author}
  {\bibfnamefont{H.}~\bibnamefont{Hoekstra}}, \bibinfo {author}
  {\bibfnamefont{S.}~\bibnamefont{Perlmutter}}, \bibinfo {author}
  {\bibfnamefont{P.}~\bibnamefont{Rosati}}, \emph{et~al.},\ }%
  \bibfield{journal}{%
  \Doi{10.1088/0004-637X/737/2/59}{\bibinfo {journal} {Astrophys.J.}}\ }%
  \textbf{\bibinfo {volume} {737}},\ \bibinfo {pages} {59} (\bibinfo {year}
  {2011}),\ \Eprint{http://arxiv.org/abs/1105.3186}{arXiv:1105.3186
  [astro-ph.CO]}%
  \bibAnnoteFile{NoStop}{Jee:2011az}%
%%CITATION = ARXIV:1105.3186;%%
\bibitem{Hoyle:2010ce}%
  \BibitemOpen
  \bibfield{author}{%
  \bibinfo {author} {\bibfnamefont{B.}~\bibnamefont{Hoyle}}, \bibinfo {author}
  {\bibfnamefont{R.}~\bibnamefont{Jimenez}},\ and\ \bibinfo {author}
  {\bibfnamefont{L.}~\bibnamefont{Verde}},\ }%
  \bibfield{journal}{%
  \Doi{10.1103/PhysRevD.83.103502}{\bibinfo {journal} {Phys.Rev.}}\ }%
  \textbf{\bibinfo {volume} {D83}},\ \bibinfo {pages} {103502} (\bibinfo {year}
  {2011}),\ \Eprint{http://arxiv.org/abs/1009.3884}{arXiv:1009.3884
  [astro-ph.CO]}%
  \bibAnnoteFile{NoStop}{Hoyle:2010ce}%
%%CITATION = ARXIV:1009.3884;%%
\bibitem{Holz:2010ck}%
  \BibitemOpen
  \bibfield{author}{%
  \bibinfo {author} {\bibfnamefont{D.~E.}\ \bibnamefont{Holz}}\ and\ \bibinfo
  {author} {\bibfnamefont{S.}~\bibnamefont{Perlmutter}}}%
   (\bibinfo {year} {2010}),\ \doi{\bibinfo {doi}
  {10.1088/2041-8205/755/2/L36}},\
  \Eprint{http://arxiv.org/abs/1004.5349}{arXiv:1004.5349 [astro-ph.CO]}%
  \bibAnnoteFile{NoStop}{Holz:2010ck}%
%%CITATION = ARXIV:1004.5349;%%
\bibitem{Granett:2008ju}%
  \BibitemOpen
  \bibfield{author}{%
  \bibinfo {author} {\bibfnamefont{B.~R.}\ \bibnamefont{Granett}}, \bibinfo
  {author} {\bibfnamefont{M.~C.}\ \bibnamefont{Neyrinck}},\ and\ \bibinfo
  {author} {\bibfnamefont{I.}~\bibnamefont{Szapudi}},\ }%
  \bibfield{journal}{%
  \Doi{10.1086/591670}{\bibinfo {journal} {Astrophys.J.}}\ }%
  \textbf{\bibinfo {volume} {683}},\ \bibinfo {pages} {L99} (\bibinfo {year}
  {2008}),\ \Eprint{http://arxiv.org/abs/0805.3695}{arXiv:0805.3695
  [astro-ph]}%
  \bibAnnoteFile{NoStop}{Granett:2008ju}%
%%CITATION = ARXIV:0805.3695;%%
\bibitem{Papai:2010gd}%
  \BibitemOpen
  \bibfield{author}{%
  \bibinfo {author} {\bibfnamefont{P.}~\bibnamefont{Papai}}, \bibinfo {author}
  {\bibfnamefont{I.}~\bibnamefont{Szapudi}},\ and\ \bibinfo {author}
  {\bibfnamefont{B.~R.}\ \bibnamefont{Granett}},\ }%
  \bibfield{journal}{%
  \Doi{10.1088/0004-637X/732/1/27}{\bibinfo {journal} {Astrophys.J.}}\ }%
  \textbf{\bibinfo {volume} {732}},\ \bibinfo {pages} {27} (\bibinfo {year}
  {2011}),\ \Eprint{http://arxiv.org/abs/1012.3750}{arXiv:1012.3750
  [astro-ph.CO]}%
  \bibAnnoteFile{NoStop}{Papai:2010gd}%
%%CITATION = ARXIV:1012.3750;%%
\bibitem{Nadathur:2011iu}%
  \BibitemOpen
  \bibfield{author}{%
  \bibinfo {author} {\bibfnamefont{S.}~\bibnamefont{Nadathur}}, \bibinfo
  {author} {\bibfnamefont{S.}~\bibnamefont{Hotchkiss}},\ and\ \bibinfo {author}
  {\bibfnamefont{S.}~\bibnamefont{Sarkar}},\ }%
  \bibfield{journal}{%
  \Doi{10.1088/1475-7516/2012/06/042}{\bibinfo {journal} {JCAP}}\ }%
  \textbf{\bibinfo {volume} {1206}},\ \bibinfo {pages} {042} (\bibinfo {year}
  {2012}),\ \Eprint{http://arxiv.org/abs/1109.4126}{arXiv:1109.4126
  [astro-ph.CO]}%
  \bibAnnoteFile{NoStop}{Nadathur:2011iu}%
%%CITATION = ARXIV:1109.4126;%%
\bibitem{Flender:2012wu}%
  \BibitemOpen
  \bibfield{author}{%
  \bibinfo {author} {\bibfnamefont{S.}~\bibnamefont{Flender}}, \bibinfo
  {author} {\bibfnamefont{S.}~\bibnamefont{Hotchkiss}},\ and\ \bibinfo {author}
  {\bibfnamefont{S.}~\bibnamefont{Nadathur}},\ }%
  \bibfield{journal}{%
  \Doi{10.1088/1475-7516/2013/02/013}{\bibinfo {journal} {JCAP}}\ }%
  \textbf{\bibinfo {volume} {1302}},\ \bibinfo {pages} {013} (\bibinfo {year}
  {2013}),\ \Eprint{http://arxiv.org/abs/1212.0776}{arXiv:1212.0776
  [astro-ph.CO]}%
  \bibAnnoteFile{NoStop}{Flender:2012wu}%
%%CITATION = ARXIV:1212.0776;%%
\bibitem{Hotchkiss:2011ms}%
  \BibitemOpen
  \bibfield{author}{%
  \bibinfo {author} {\bibfnamefont{S.}~\bibnamefont{Hotchkiss}},\ }%
  \bibfield{journal}{%
  \Doi{10.1088/1475-7516/2011/07/004}{\bibinfo {journal} {JCAP}}\ }%
  \textbf{\bibinfo {volume} {1107}},\ \bibinfo {pages} {004} (\bibinfo {year}
  {2011}),\ \Eprint{http://arxiv.org/abs/1105.3630}{arXiv:1105.3630
  [astro-ph.CO]}%
  \bibAnnoteFile{NoStop}{Hotchkiss:2011ms}%
%%CITATION = ARXIV:1105.3630;%%
\bibitem{Harrison:2011ep}%
  \BibitemOpen
  \bibfield{author}{%
  \bibinfo {author} {\bibfnamefont{I.}~\bibnamefont{Harrison}}\ and\ \bibinfo
  {author} {\bibfnamefont{P.}~\bibnamefont{Coles}},\ }%
  \bibfield{journal}{%
  \bibinfo {journal} {Mon.Not.Roy.Astron.Soc.}\ }%
  \textbf{\bibinfo {volume} {421}},\ \bibinfo {pages} {L19} (\bibinfo {year}
  {2012}),\ \Eprint{http://arxiv.org/abs/1111.1184}{arXiv:1111.1184
  [astro-ph.CO]}%
  \bibAnnoteFile{NoStop}{Harrison:2011ep}%
%%CITATION = ARXIV:1111.1184;%%
\bibitem{Hoyle:2011pj}%
  \BibitemOpen
  \bibfield{author}{%
  \bibinfo {author} {\bibfnamefont{B.}~\bibnamefont{Hoyle}}, \bibinfo {author}
  {\bibfnamefont{R.}~\bibnamefont{Jimenez}}, \bibinfo {author}
  {\bibfnamefont{L.}~\bibnamefont{Verde}},\ and\ \bibinfo {author}
  {\bibfnamefont{S.}~\bibnamefont{Hotchkiss}},\ }%
  \bibfield{journal}{%
  \Doi{10.1088/1475-7516/2012/02/009}{\bibinfo {journal} {JCAP}}\ }%
  \textbf{\bibinfo {volume} {1202}},\ \bibinfo {pages} {009} (\bibinfo {year}
  {2012}),\ \Eprint{http://arxiv.org/abs/1108.5458}{arXiv:1108.5458
  [astro-ph.CO]}%
  \bibAnnoteFile{NoStop}{Hoyle:2011pj}%
%%CITATION = ARXIV:1108.5458;%%
\bibitem{Waizmann:2012wx}%
  \BibitemOpen
  \bibfield{author}{%
  \bibinfo {author} {\bibfnamefont{J.-C.}\ \bibnamefont{Waizmann}}, \bibinfo
  {author} {\bibfnamefont{S.}~\bibnamefont{Ettori}},\ and\ \bibinfo {author}
  {\bibfnamefont{M.}~\bibnamefont{Bartelmann}}}%
   (\bibinfo {year} {2012}),\
  \Eprint{http://arxiv.org/abs/1210.6021}{arXiv:1210.6021 [astro-ph.CO]}%
  \bibAnnoteFile{NoStop}{Waizmann:2012wx}%
%%CITATION = ARXIV:1210.6021;%%
\bibitem{Waizmann:2011xu}%
  \BibitemOpen
  \bibfield{author}{%
  \bibinfo {author} {\bibfnamefont{J.-C.}\ \bibnamefont{Waizmann}}, \bibinfo
  {author} {\bibfnamefont{S.}~\bibnamefont{Ettori}},\ and\ \bibinfo {author}
  {\bibfnamefont{L.}~\bibnamefont{Moscardini}},\ }%
  \bibfield{journal}{%
  \Doi{10.1111/j.1365-2966.2011.20171.x}{\bibinfo {journal}
  {Mon.Not.Roy.Astron.Soc.}}\ }%
  \textbf{\bibinfo {volume} {420}},\ \bibinfo {pages} {1754} (\bibinfo {year}
  {2012}),\ \Eprint{http://arxiv.org/abs/1109.4820}{arXiv:1109.4820
  [astro-ph.CO]}%
  \bibAnnoteFile{NoStop}{Waizmann:2011xu}%
%%CITATION = ARXIV:1109.4820;%%
\bibitem{Yaryura11052011}%
  \BibitemOpen
  \bibfield{author}{%
  \bibinfo {author} {\bibfnamefont{C.~Y.}\ \bibnamefont{Yaryura}}, \bibinfo
  {author} {\bibfnamefont{C.~M.}\ \bibnamefont{Baugh}},\ and\ \bibinfo {author}
  {\bibfnamefont{R.~E.}\ \bibnamefont{Angulo}},\ }%
  \bibfield{journal}{%
  \Doi{10.1111/j.1365-2966.2011.18233.x}{\bibinfo {journal} {Monthly Notices of
  the Royal Astronomical Society}}\ }%
  \textbf{\bibinfo {volume} {413}},\ \bibinfo {pages} {1311} (\bibinfo {year}
  {2011}),\
  \Eprint{http://arxiv.org/abs/http://mnras.oxfordjournals.org/content/413/2/1%
311.full.pdf+html}{http://mnras.oxfordjournals.org/content/413/2/1311.full.pdf%
+html},\ \url{http://mnras.oxfordjournals.org/content/413/2/1311.abstract}%
  \bibAnnoteFile{NoStop}{Yaryura11052011}%
\bibitem{Matarrese:2000iz}%
  \BibitemOpen
  \bibfield{author}{%
  \bibinfo {author} {\bibfnamefont{S.}~\bibnamefont{Matarrese}}, \bibinfo
  {author} {\bibfnamefont{L.}~\bibnamefont{Verde}},\ and\ \bibinfo {author}
  {\bibfnamefont{R.}~\bibnamefont{Jimenez}},\ }%
  \bibfield{journal}{%
  \Doi{10.1086/309412}{\bibinfo {journal} {Astrophys.J.}}\ }%
  \textbf{\bibinfo {volume} {541}},\ \bibinfo {pages} {10} (\bibinfo {year}
  {2000}),\
  \Eprint{http://arxiv.org/abs/astro-ph/0001366}{arXiv:astro-ph/0001366
  [astro-ph]}%
  \bibAnnoteFile{NoStop}{Matarrese:2000iz}%
%%CITATION = ASTRO-PH/0001366;%%
\bibitem{Sefusatti:2006eu}%
  \BibitemOpen
  \bibfield{author}{%
  \bibinfo {author} {\bibfnamefont{E.}~\bibnamefont{Sefusatti}}, \bibinfo
  {author} {\bibfnamefont{C.}~\bibnamefont{Vale}}, \bibinfo {author}
  {\bibfnamefont{K.}~\bibnamefont{Kadota}},\ and\ \bibinfo {author}
  {\bibfnamefont{J.}~\bibnamefont{Frieman}},\ }%
  \bibfield{journal}{%
  \Doi{10.1086/511331}{\bibinfo {journal} {Astrophys.J.}}\ }%
  \textbf{\bibinfo {volume} {658}},\ \bibinfo {pages} {669} (\bibinfo {year}
  {2007}),\
  \Eprint{http://arxiv.org/abs/astro-ph/0609124}{arXiv:astro-ph/0609124
  [astro-ph]}%
  \bibAnnoteFile{NoStop}{Sefusatti:2006eu}%
%%CITATION = ASTRO-PH/0609124;%%
\bibitem{Trindade:2013lga}%
  \BibitemOpen
  \bibfield{author}{%
  \bibinfo {author} {\bibfnamefont{A.}~\bibnamefont{Trindade}}, \bibinfo
  {author} {\bibfnamefont{P.}~\bibnamefont{Avelino}},\ and\ \bibinfo {author}
  {\bibfnamefont{P.}~\bibnamefont{Viana}}}%
   (\bibinfo {year} {2013}),\
  \Eprint{http://arxiv.org/abs/1304.4275}{arXiv:1304.4275 [astro-ph.CO]}%
  \bibAnnoteFile{NoStop}{Trindade:2013lga}%
%%CITATION = ARXIV:1304.4275;%%
\bibitem{Wake:2008mf}%
  \BibitemOpen
  \bibfield{author}{%
  \bibinfo {author} {\bibfnamefont{D.~A.}\ \bibnamefont{Wake}}, \bibinfo
  {author} {\bibfnamefont{R.~K.}\ \bibnamefont{Sheth}}, \bibinfo {author}
  {\bibfnamefont{R.~C.}\ \bibnamefont{Nichol}}, \bibinfo {author}
  {\bibfnamefont{C.~M.}\ \bibnamefont{Baugh}}, \bibinfo {author}
  {\bibfnamefont{J.}~\bibnamefont{Bland-Hawthorn}}, \emph{et~al.},\ }%
  \bibfield{journal}{%
  \Doi{10.1111/j.1365-2966.2008.13333.x}{\bibinfo {journal}
  {Mon.Not.Roy.Astron.Soc.}}\ }%
  \textbf{\bibinfo {volume} {387}},\ \bibinfo {pages} {1045} (\bibinfo {year}
  {2008}),\ \Eprint{http://arxiv.org/abs/0802.4288}{arXiv:0802.4288
  [astro-ph]}%
  \bibAnnoteFile{NoStop}{Wake:2008mf}%
%%CITATION = ARXIV:0802.4288;%%
\bibitem{Zheng:2008np}%
  \BibitemOpen
  \bibfield{author}{%
  \bibinfo {author} {\bibfnamefont{Z.}~\bibnamefont{Zheng}}, \bibinfo {author}
  {\bibfnamefont{I.}~\bibnamefont{Zehavi}}, \bibinfo {author}
  {\bibfnamefont{D.~J.}\ \bibnamefont{Eisenstein}}, \bibinfo {author}
  {\bibfnamefont{D.~H.}\ \bibnamefont{Weinberg}},\ and\ \bibinfo {author}
  {\bibfnamefont{Y.}~\bibnamefont{Jing}},\ }%
  \bibfield{journal}{%
  \Doi{10.1088/0004-637X/707/1/554}{\bibinfo {journal} {Astrophys.J.}}\ }%
  \textbf{\bibinfo {volume} {707}},\ \bibinfo {pages} {554} (\bibinfo {year}
  {2009}),\ \Eprint{http://arxiv.org/abs/0809.1868}{arXiv:0809.1868
  [astro-ph]}%
  \bibAnnoteFile{NoStop}{Zheng:2008np}%
%%CITATION = ARXIV:0809.1868;%%
\bibitem{Sawangwit:2009bg}%
  \BibitemOpen
  \bibfield{author}{%
  \bibinfo {author} {\bibfnamefont{U.}~\bibnamefont{Sawangwit}}, \bibinfo
  {author} {\bibfnamefont{T.}~\bibnamefont{Shanks}}, \bibinfo {author}
  {\bibfnamefont{F.}~\bibnamefont{Abdalla}}, \bibinfo {author}
  {\bibfnamefont{R.}~\bibnamefont{Cannon}}, \bibinfo {author}
  {\bibfnamefont{S.}~\bibnamefont{Croom}}, \emph{et~al.},\ }%
  \bibfield{journal}{%
  \Doi{10.1111/j.1365-2966.2011.19251.x}{\bibinfo {journal}
  {Mon.Not.Roy.Astron.Soc.}}\ }%
  \textbf{\bibinfo {volume} {416}},\ \bibinfo {pages} {3033} (\bibinfo {year}
  {2011}),\ \Eprint{http://arxiv.org/abs/0912.0511}{arXiv:0912.0511
  [astro-ph.CO]}%
  \bibAnnoteFile{NoStop}{Sawangwit:2009bg}%
%%CITATION = ARXIV:0912.0511;%%
\bibitem{Cai:2010gz}%
  \BibitemOpen
  \bibfield{author}{%
  \bibinfo {author} {\bibfnamefont{Y.-C.}\ \bibnamefont{Cai}}, \bibinfo
  {author} {\bibfnamefont{G.}~\bibnamefont{Bernstein}},\ and\ \bibinfo {author}
  {\bibfnamefont{R.~K.}\ \bibnamefont{Sheth}}}%
   (\bibinfo {year} {2010}),\
  \Eprint{http://arxiv.org/abs/1007.3500}{arXiv:1007.3500 [astro-ph.CO]}%
  \bibAnnoteFile{NoStop}{Cai:2010gz}%
%%CITATION = ARXIV:1007.3500;%%
\bibitem{Hamaus:2010im}%
  \BibitemOpen
  \bibfield{author}{%
  \bibinfo {author} {\bibfnamefont{N.}~\bibnamefont{Hamaus}}, \bibinfo {author}
  {\bibfnamefont{U.}~\bibnamefont{Seljak}}, \bibinfo {author}
  {\bibfnamefont{V.}~\bibnamefont{Desjacques}}, \bibinfo {author}
  {\bibfnamefont{R.~E.}\ \bibnamefont{Smith}},\ and\ \bibinfo {author}
  {\bibfnamefont{T.}~\bibnamefont{Baldauf}},\ }%
  \bibfield{journal}{%
  \Doi{10.1103/PhysRevD.82.043515}{\bibinfo {journal} {Phys.Rev.}}\ }%
  \textbf{\bibinfo {volume} {D82}},\ \bibinfo {pages} {043515} (\bibinfo {year}
  {2010}),\ \Eprint{http://arxiv.org/abs/1004.5377}{arXiv:1004.5377
  [astro-ph.CO]}%
  \bibAnnoteFile{NoStop}{Hamaus:2010im}%
%%CITATION = ARXIV:1004.5377;%%
\bibitem{Cooray:2002dia}%
  \BibitemOpen
  \bibfield{author}{%
  \bibinfo {author} {\bibfnamefont{A.}~\bibnamefont{Cooray}}\ and\ \bibinfo
  {author} {\bibfnamefont{R.~K.}\ \bibnamefont{Sheth}},\ }%
  \bibfield{journal}{%
  \Doi{10.1016/S0370-1573(02)00276-4}{\bibinfo {journal} {Phys.Rept.}}\ }%
  \textbf{\bibinfo {volume} {372}},\ \bibinfo {pages} {1} (\bibinfo {year}
  {2002}),\
  \Eprint{http://arxiv.org/abs/astro-ph/0206508}{arXiv:astro-ph/0206508
  [astro-ph]}%
  \bibAnnoteFile{NoStop}{Cooray:2002dia}%
%%CITATION = ASTRO-PH/0206508;%%
\bibitem{Schmidt:2009yj}%
  \BibitemOpen
  \bibfield{author}{%
  \bibinfo {author} {\bibfnamefont{F.}~\bibnamefont{Schmidt}}, \bibinfo
  {author} {\bibfnamefont{W.}~\bibnamefont{Hu}},\ and\ \bibinfo {author}
  {\bibfnamefont{M.}~\bibnamefont{Lima}},\ }%
  \bibfield{journal}{%
  \Doi{10.1103/PhysRevD.81.063005}{\bibinfo {journal} {Phys.Rev.}}\ }%
  \textbf{\bibinfo {volume} {D81}},\ \bibinfo {pages} {063005} (\bibinfo {year}
  {2010}),\ \Eprint{http://arxiv.org/abs/0911.5178}{arXiv:0911.5178
  [astro-ph.CO]}%
  \bibAnnoteFile{NoStop}{Schmidt:2009yj}%
%%CITATION = ARXIV:0911.5178;%%
\end{thebibliography}%

\end{document}